\documentclass[rmp,aps,amsfonts,superscriptaddress,preprintnumbers]{revtex4}

\usepackage{amssymb}
\usepackage{epsf}
\usepackage{epsfig}
\usepackage{psfrag}
\usepackage{rotating}
\usepackage{color}

\newcommand{\lsim}{
\mathrel{\hbox{\rlap{\hbox{\lower4pt\hbox{$\sim$}}}\hbox{$<$}}}}

\newcommand{\gsim}{
\mathrel{\hbox{\rlap{\hbox{\lower4pt\hbox{$\sim$}}}\hbox{$>$}}}}

\newcommand{\KLmm}{K_L \to \mu^+ \mu^-}

\newcommand{\BXdnn}{\bar{B} \to X_d \nu \bar{\nu}}
\newcommand{\BXsnn}{\bar{B} \to X_s \nu \bar{\nu}}

\newcommand{\Bdmm}{B_d \to \mu^+ \mu^-}
\newcommand{\Bsmm}{B_s \to \mu^+ \mu^-}

\newcommand{\BRKp}{{Br} (\kpn)}
\newcommand{\BRKL}{{Br} (\klpn)}
\newcommand{\BRKm}{{Br} (\KLmm)_{\rm SD}}
\newcommand{\BRXd}{{Br} (\BXdnn)}
\newcommand{\BRXs}{{Br} (\BXsnn)}
\newcommand{\BRBd}{Br (\Bdmm)}
\newcommand{\BRBs}{{Br} (\Bsmm)}

\newcommand{\vcb}{|V_{cb}|}
\newcommand{\vtd}{|V_{td}|}
\newcommand{\vub}{|V_{ub}/V_{cb}|}

\def\eps{\varepsilon}
\def\epe{\varepsilon'/\varepsilon}
\newcommand{\gev}{\, {\rm GeV}}
\newcommand{\mev}{\, {\rm MeV}}

\newcommand{\mt}{m_{\rm t}}

\newcommand{\mc}{m_{\rm c}}

\newcommand{\mw}{M_{\rm W}}

\newcommand{\imlt}{\IM\lambda_t}
\newcommand{\relt}{\RE\lambda_t}
\newcommand{\relc}{\RE\lambda_c}
\newcommand{\beqa}{\begin{eqnarray}}
\newcommand{\eeqa}{\end{eqnarray}}

\newcommand{\be}{\begin{equation}}
\newcommand{\ee}{\end{equation}}
\newcommand{\bi}{\begin{itemize}}
\newcommand{\ei}{\end{itemize}}
\newcommand{\ord}{{\cal O}}

\newcommand{\RE}{{\rm Re}}
\newcommand{\IM}{{\rm Im}}

\def\kpn{K^+\rightarrow\pi^+\nu\bar\nu}

\def\klpn{K_{\rm L}\rightarrow\pi^0\nu\bar\nu}

\def\kspn{K_{\rm S}\rightarrow\pi^0\nu\bar\nu}

\begin{document}
\title{Waiting for precise measurements of $\kpn$ and $\klpn$}

\author{Andrzej J. Buras}
\affiliation{Physik Department, Technische Universit\"at M\"unchen,
D-85748 Garching, Germany}
\author{Felix Schwab}
\affiliation{Max-Planck-Institut f{\"u}r Physik -- Werner-Heisenberg-Institut,
 D-80805 Munich, Germany} 
\affiliation{Physik Department, Technische Universit\"at M\"unchen,
D-85748 Garching, Germany}
\author{Selma Uhlig}
\affiliation{Physik Department, Technische Universit\"at M\"unchen,
D-85748 Garching, Germany}

\begin{abstract}
\noindent
 In view of future plans for accurate measurements of the 
theoretically clean branching ratios $Br(\kpn)$ and $Br(\klpn)$, that should
take place in the next decade, we collect the relevant formulae for quantities of
interest and analyze their theoretical and parametric uncertainties. 
We point out that in addition to the angle $\beta$ 
in the unitarity triangle (UT) also the angle $\gamma$ can in principle
be determined from these decays with respectable 
precision and emphasize in this context the importance of the recent NNLO QCD calculation of 
the charm contribution to $\kpn$ and of the improved estimate of the long distance contribution by
means of chiral perturbation theory.
In addition to known expressions we present several new ones that should allow 
transparent tests of the Standard Model (SM) and of its extensions. While our 
presentation is centered around the SM, we also discuss  models with
minimal flavour violation and scenarios with new complex phases in 
decay amplitudes and meson mixing. 
We give a brief review of existing results within
specific extensions of the SM, in particular the Littlest Higgs Model with
T-parity, $Z^\prime$ models, the MSSM and a model with one universal extra
dimension. 
We derive a new ``golden" relation between $B$ and $K$ systems that involves 
$(\beta,\gamma)$ and $Br(\klpn)$ and investigate the virtues of 
$(R_t,\beta)$, $(R_b,\gamma)$, $(\beta,\gamma)$ and $(\bar\eta,\gamma)$ 
strategies for the UT in the context of $K\to\pi\nu\bar\nu$ decays with the 
goal of testing the SM and its extensions.
\end{abstract}



\maketitle
\tableofcontents



%
%
%
\section{Introduction}\label{sec:intro}
\setcounter{equation}{0}

The rare decays of $K$ and $B$ mesons play an important role in the search 
for the underlying flavour dynamics and in particular in the search for the 
origin of CP violation
\cite{Buras:2005xt,Buras:2003jf,Buras:2004sc,Buchalla:1995vs,Buras:1998ra,Fleischer:2002ys,Fleischer:2004xw,Ali:2003te,Hurth:2003vb,Nir:2001ge,Buchalla:2003ux,Isidori:2005xm}. 
Among the many $K$ and $B$ decays, the rare 
decays $\kpn$ and 
$\klpn$ are very special as 
their branching ratios can be computed to an exceptionally high degree of 
precision that is not matched by any other loop induced decay of mesons. In
particular the 
theoretical uncertainties in the prominent decays like 
$B\to X_s\mu^+\mu^-$ and $B_{s}\to\mu^+\mu^-$ amount 
typically to $\pm 10\% $ or larger at the level of the branching ratio,
although progress in the calculation of the branching ratio 
of $B\to X_s\gamma$ at
the NNLO level shows that in this case an error below 10$\%$ is in principle possible
\cite{Becher:2006pu,Misiak:2006zs}. 
On the other hand the 
corresponding uncertainties in $\klpn$ amount to 1-2$\%$ 
\cite{Buchalla:1992zm,Buchalla:1993bv,Misiak:1999yg,Buchalla:1998ba}. In the case 
of $\kpn$, the 
presence of the internal charm contributions in the relevant $Z^0$ penguin 
and box diagrams contained the theoretical perturbative uncertainty of $\pm 7\%$ at the NLO 
level \cite{Buchalla:1998ba,Buchalla:1993wq}, but 
this uncertainty has been recently reduced down to $\pm 1-2\%$ through 
a complete NNLO calculation \cite{BUGOHA,Buras:2006gb}.

The reason for the exceptional theoretical cleanness of 
$\kpn$ and $\klpn$  \cite{Littenberg:1989ix} is the fact 
that the required hadronic matrix elements can be extracted, including isospin 
breaking corrections \cite{Marciano:1996wy,Mescia:2007kn}, from the leading semileptonic 
decay $K^+\to\pi^ 0e^+\nu$. Moreover, 
extensive studies of other long-distance contributions
\cite{Rein:1989tr,Hagelin:1989wt,Lu:1994ww,Fajfer:1996tc,Geng:1996kd, Ecker:1987hd,Falk:2000nm,Buchalla:1998ux}
and of higher order electroweak effects \cite{Buchalla:1997kz} 
have shown that they can safely be neglected  in $\klpn$ and are small in $\kpn$. 
In particular, the most recent improved calculation of long distance contributions to $\kpn$ results
in an enhancement of the relevant branching ratio by $6\pm 2 \%$. Further
progress in calculating these contributions is in principle possible with the
help of lattice QCD \cite{Isidori:2005tv}.
Some recent reviews on $K\to\pi\nu\bar\nu$ can be found 
in \cite{Buras:2005xt,Buras:2003jf,Buras:2004sc,Isidori:2003ij,Isidori:2005xm}.

We are fortunate that, while the decay $\kpn$  
is CP conserving and depends sensitively on the underlying flavour dynamics, 
its partner $\klpn$ is 
purely CP violating within the Standard Model (SM) and most of its extensions
and consequently depends also on the mechanism of CP violation. 
Moreover, the combination of these two decays allows to eliminate 
the parametric uncertainties due to the 
CKM element $\vcb$ and $m_t$ in the determination of the angle $\beta$ in the unitarity
triangle (UT) or equivalently of the phase of the CKM element $V_{td}$ 
\cite{Buchalla:1994tr,Buchalla:1996fp}. 
The resulting theoretical uncertainty in $\sin 2\beta$  
 is comparable to the 
one present   
in the mixing induced CP asymmetry $a_{\psi K_S}$ and  with the measurements 
of both branching ratios at the $\pm 10\%$ and $\pm 5\%$ level, $\sin2\beta$ 
could be determined 
with $\pm 0.08$ and $\pm 0.04$ precision, respectively. This independent 
determination of $\sin 2\beta$ with a very small theoretical error 
offers a 
powerful test of the SM and of its simplest extensions in which the flavour 
and 
CP violation are governed by the CKM matrix, the so-called MFV (minimal 
flavour violation) models \cite{Buras:2005xt,Buras:2003jf,Buras:2004sc,Buras:2000dm,D'Ambrosio:2002ex}. 
Indeed, in $K\to\pi\nu\bar\nu$ the phase $\beta$
 originates in $Z^0$ penguin 
diagrams ($\Delta S=1$), whereas in the case of $a_{\psi K_S}$ in the 
$B^0_d-\bar B^0_d$ box diagrams ($\Delta B=2$). 
Any ``non-minimal" contributions to $Z^0$ penguin diagrams and/or  box 
$B^0_d-\bar B^0_d$ diagrams would then be signaled by the violation of 
the MFV ``golden" relation \cite{Buchalla:1994tr}
\be\label{R7}
(\sin 2\beta)_{\pi\nu\bar\nu}=(\sin 2\beta)_{\psi K_S}.
\ee

Now, strictly speaking, according to the common classification of 
different types of CP violation 
\cite{Buras:2005xt,Buras:2003jf,Buras:2004sc,Fleischer:2002ys,Fleischer:2004xw,Ali:2003te,Hurth:2003vb,Nir:2001ge,Buchalla:2003ux}, both the asymmetry 
$a_{\psi K_S}$ and a non-vanishing rate for $\klpn$ in the SM and in most 
of its extensions signal the CP violation in the interference of mixing and
decay. However, as the CP violation in mixing (indirect CP violation)
in $K$ decays is governed by the small parameter $\varepsilon_K$, one 
can show \cite{Littenberg:1989ix,Buchalla:1996fp,Grossman:1997sk} that the observation of $Br(\klpn)$ 
at the level of $10^{-11}$ and higher is a manifestation of a large direct 
CP violation with the indirect one contributing less than $\sim 1\%$ to 
the branching ratio. The great potential of $\klpn$ in testing the physics beyond the SM has been
summarized in \cite{Bryman:2005xp}.

Additionally, this large direct CP violation can be directly measured
without essentially any hadronic uncertainties, due to the
presence of the $\nu\bar\nu$ in the final state. This should be contrasted
with the very popular studies of direct CP violation in non-leptonic two--body
$B$ decays
\cite{Buras:2005xt,Buras:2003jf,Buras:2004sc,Fleischer:2002ys,Fleischer:2004xw,Ali:2003te,Hurth:2003vb,Nir:2001ge,Buchalla:2003ux}, that are subject to significant hadronic 
uncertainties. In particular, the extraction of weak phases requires
generally rather involved strategies
using often certain assumptions about the strong dynamics
\cite{Harrison:1998yr,Ball:2000ba,Anikeev:2001rk}. 
Only a handful of strategies, which we will briefly review in Section \ref{sec:decays}, 
allow direct determinations of weak phases from non-leptonic $B$ decays
without practically any hadronic uncertainties.

Returning to (\ref{R7}), an important consequence of this relation is the 
following one \cite{Buras:2001af}: 
for a given 
$\sin2\beta$ extracted from $a_{\psi K_S}$, the measurement of $Br(\kpn)$
 determines up to a two-fold 
ambiguity the value of $Br(\klpn)$, independent of any new parameters present in the 
MFV models. Consequently, measuring $Br(\klpn)$ will either select one of the 
possible values or rule 
out all MFV models. Recent analyses of the MFV models indicate that one of 
these values is very unlikely \cite{Bobeth:2005ck, Haisch:2007ia}.
A spectacular violation of the relation (\ref{R7})
  is found in the 
context of new physics scenarios with enhanced $Z^0$ penguins carrying a new 
CP-violating phase
 \cite{Buras:2003dj,Buras:2004ub, Nir:1997tf,Buras:1997ij,Colangelo:1998pm,Buras:1998ed,Buras:1999da,
        Buchalla:2000sk,Atwood:2003tg}. An explicit realization of such a
	scenario is the Littlest Higgs Model with T-parity \cite{Blanke:2006eb}
	which we will
	discuss in Section \ref{sec:models}.

Another important virtue of $\kpn$ is a theoretically clean determination of 
$\vtd$ or equivalently of the length $R_t$ in the unitarity 
triangle. This 
determination is only subject to theoretical uncertainties in the charm sector, 
that amount after the recent NNLO calculation to $\pm 1-2\%$. The remaining 
parametric uncertainties in the 
determination of $\vtd$ related to $\vcb$ and $m_t$ should be soon reduced
to the 1-2$\%$ level. Finally, the decay $\klpn$ offers the cleanest 
determination of the Jarlskog 
CP-invariant $J_{CP}$ \cite{Buchalla:1996fp} or equivalently of the area of the 
unrescaled unitarity 
triangle that cannot be matched by any $B$ decay. With the improved precision 
on $m_t$ and $\vcb$, also a precise measurement of the height $\bar\eta$ 
of the 
unitarity 
triangle becomes possible.

The clean determinations of $\sin 2\beta$, $\vtd$, $R_t$, $J_{CP}$, and 
of the UT in general, as well as the test of 
the MFV relation (\ref{R7}) and generally of the physics beyond the SM, 
put these two decays in the class 
of ``golden decays", essentially on the 
same level as the determination of $\sin 2\beta$ through the asymmetry 
$a_{\psi K_S}$  and 
certain clean strategies for the determination of the angle $\gamma$ in the
UT \cite{Buras:2005xt,Buras:2003jf,Buras:2004sc,Fleischer:2002ys,Fleischer:2004xw,Ali:2003te,Hurth:2003vb,Nir:2001ge,Buchalla:2003ux},  that will be 
available at LHC \cite{Ball:2000ba}. We will discuss briefly
the latter in Section \ref{sec:decays}. 
Therefore precise measurements of $Br(\kpn)$ and $Br(\klpn)$ 
are of utmost importance and should be aimed for, even when realizing that 
the determination of the branching ratios in question with an accuracy 
of $5\%$ is extremely challenging.

With the NNLO calculation \cite{BUGOHA} at hand the branching ratios of
$\kpn$ and $\klpn$ within the SM can be predicted as
{ \begin{equation}\label{SMkp0}
Br(\kpn)_{\rm SM}=
(8.1 \pm 1.1 )\cdot 10^{-11},
\ee
\be\label{SMkl0}
 Br(\klpn)_{\rm SM}=
(2.6 \pm 0.3)\cdot 10^{-11} .
\end{equation}}
This is an accuracy of $\pm 14\%$ and $\pm 12\%$, 
respectively. We will demonstrate that
further progress on the determination of the CKM parameters coming in 
the next
few years dominantly  from BaBar, Belle, 
Tevatron and later from LHC as well as the improved determination of $m_c$ relevant for $\kpn$, 
should allow eventually 
predictions for $Br(\kpn)$ and $Br(\klpn)$ with the uncertainties of 
$\pm 5\%$ or better.
This
accuracy cannot be matched by any other rare decay branching ratio 
in the field of meson decays. \\

On the experimental side the AGS E787 collaboration at Brookhaven was the 
first to observe the decay $\kpn$ \cite{Adler:1997am,Adler:2000by}. 
The resulting branching ratio based on two 
events and published in 2002 was \cite{Adler:2001xv,Adler:2004hp}
\be\label{EXP0}
Br(\kpn)=
(15.7^{+17.5}_{-8.2})\cdot 10^{-11} \qquad (2002).
\ee
In 2004, a new $\kpn$ experiment, AGS E949 \cite{Anisimovsky:2004hr}, 
released its first results 
 that are based on the 2002 running. One additional event has been 
observed. Including the result of AGS E787 the present branching ratio reads
\be\label{EXP1}
Br(\kpn)=
(14.7^{+13.0}_{-8.9})\cdot 10^{-11} \qquad (2004).
\ee
It is not clear, at present, how this result will be improved in the coming 
years as the activities of AGS E949
and the efforts at Fermilab around the CKM 
experiment \cite{CKMEXP} have unfortunately been terminated. On the other hand, 
the corresponding efforts at CERN around the NA48
collaboration \cite{NA48EXP} and at JPARC in Japan 
\cite{JPAR} could provide additional 50-100 events at the beginning of the next decade.

The situation is different for $\klpn$. 
The older upper bound on its branching ratio from KTeV \cite{E391KL}, 
$Br(\klpn)<2.9 \cdot 10^{-7}$ has recently been improved to  
{\be\label{EXP2}
Br(\klpn)<2.1\cdot 10^{-7},
\end{equation}}
by E391 Experiment at KEK { \cite{Ahn:2006uf}}. 
While this is about four orders of magnitude above the SM expectation, the prospects for 
an improved measurement of $\klpn$ appear almost better than for 
$\kpn$ from the present perspective.

Indeed, a $\klpn$ experiment at KEK, E391a \cite{E391}
should in its first stage improve the bound in (\ref{EXP2}) by  three orders 
of magnitude. While this 
is insufficient to reach the SM level, a few events could be observed if 
$Br(\klpn)$ turned out to be by one order of magnitude larger due to 
new physics contributions. 

While a very interesting experiment at Brookhaven, KOPIO \cite{Littenberg:2002um,Bryman:2002dv}, 
that was supposed to in due time 
provide 40-60 events of $\klpn$ at the SM level has unfortunately not been approved to run at Brookhaven, the ideas
presented in this proposal can hopefully be realized one day. 
Finally, the second stage of 
the E391 experiment could, using the high intensity 50 GeV/c proton beam from
JPARC \cite{JPAR}, provide roughly 1000 SM events of 
$\klpn$, which would be truly fantastic! Perspectives of a search for $\klpn$
at a $\Phi$-factory have been discussed in \cite{Bossi:1998it}.
Further reviews on experimental prospects for $K\to\pi\nu\bar\nu$ 
can be found in \cite{Belyaev:2001kz,Diwan:2002zq,Barker:2000gd}.

Parallel to these efforts, during the coming years we will certainly witness
unprecedented tests of the CKM picture of flavour and CP violation in 
$B$ decays that will
be available at SLAC, KEK, Tevatron and 
at CERN. The most prominent of these tests will involve the CP violation in the $B^0_s-\bar
B^0_s$ mixing and a number of clean strategies
for the determination of the angles $\gamma$ and $\beta$ in the  
UT that will involve $B^\pm$, $B^0_d$ and $B^0_s$ two-body non-leptonic
decays.

These efforts will be accompanied by the studies of CP
violation in decays like $B\to\pi\pi$, $B\to\pi K$  and $B\to KK$, 
that in spite of being
less theoretically clean than the quantities considered in the present review,
will certainly contribute to the tests of the CKM paradigm \cite{Cabibbo:1963yz,Kobayashi:1973fv}. 
In addition, rare 
decays like $B\to X_s\gamma$, 
$B\to X_{s,d}\mu^+\mu^-$, $B_{s,d}\to\mu^+\mu^-$, 
$B\to X_{s,d}\nu\bar\nu$, $B\to \tau \bar \nu$, $K_L\to\pi^0 e^+e^-$ and $K_L\to\pi^0 \mu^+\mu^-$
will play an important role.

In 1994, two detailed analyses of $\kpn$, $\klpn$, $B^0_s-\bar B^0_s$ 
mixing and of CP asymmetries in $B$ decays have been presented in the 
anticipation of future precise measurements of several theoretically clean
observables, that could be used for a determination of the CKM matrix and of
the unitarity triangle within the SM \cite{Buras:1994ec,Buras:1994rj}. 
These analyses were very speculative
as in 1994 even the top quark mass was unknown, none of the observables
listed above have been measured and the CKM elements $\vcb$ and 
$|V_{ub}|$ were rather poorly known.

During the last thirteen years an impressive progress has taken place: 
the top quark mass, the angle $\beta$ in the UT and the  $B^0_s-\bar B_s^0$ mixing mass difference $\Delta M_s$ have been precisely
measured
and three events of $\kpn$ have been observed. We are still 
waiting for the observation of $\klpn$ and 
a precise direct measurement of the angle $\gamma$ in the UT from tree level
decays, but now we are rather
confident that we will be awarded already in the next decade.

This progress makes it possible to considerably improve the analyses of 
\cite{Buras:1994ec,Buras:1994rj} within the SM and to generalize them to its simplest 
extensions. This is one of the goals of our review. We will see that the 
decays $\kpn$ and $\klpn$, as in 1994, play an important role in these 
investigations. 

In this context we would like to emphasize that new physics contributions in
$\kpn$ and $\klpn$, 
in essentially all extensions of the SM,\footnote{Exceptions will be briefly
discussed in Section \ref{sec:models}.} can be parametrized in a model-independent manner 
by just two parameters \cite{Buras:1997ij}, the magnitude of the short distance
function $X$ \cite{Buras:2005xt,Buras:2003jf,Buras:2004sc} and its complex phase: 
\be\label{NX}
X=|X|e^{i\theta_X}
\ee
with $|X|=X(x_t)$ and $\theta_X=0$ in the SM. The important virtues 
of the $K\to\pi\nu\bar\nu$ system here are the following ones
\begin{itemize}
\item
$|X|$ and $\theta_X$ can be extracted from $Br(\klpn)$ and $Br(\kpn)$ 
without any hadronic uncertainties,
\item
As in many extensions of the SM, the function $X$ is governed by the 
$Z^0$ penguins with top quark and new particle 
exchanges\footnote{Box diagrams seem to be relevant only in the SM and can be
calculated with high accuracy.}, the determination of the function $X$ is
actually the determination of the $Z^0$ penguins that enter other decays.
\item
The theoretical cleanness of this determination cannot be matched by any other
decay. For instance, the decays like $B\to X_{s,d}\mu^+\mu^-$ and 
$B_{s,d}\to\mu^+\mu^-$, that can also be used for this purpose, are subject 
to theoretical uncertainties of $\pm 10\%$ or more.
\end{itemize}

Already at this stage we would like to emphasize
that the clean 
theoretical character of these decays remains valid in essentially all 
extensions of the SM, whereas this is generally not the case for non-leptonic 
two-body B decays used to determine the CKM parameters through CP 
asymmetries and/or other strategies. While several mixing induced CP 
asymmetries  in non-leptonic B decays within the SM are essentially free 
from hadronic uncertainties, as the latter cancel out due to the dominance 
of a single CKM amplitude, this is often not the case in extensions of the SM 
in which the amplitudes receive new contributions with different weak phases 
implying no cancellation of hadronic 
uncertainties in the relevant observables.  
A classic example of this situation, 
as stressed in \cite{Ciuchini:2002pd}, is the 
mixing induced CP asymmetry in $B^0_d(\bar B^0_d)\to \phi K_S$ decays that 
within the SM measures the 
angle $\beta$ in the UT with very small hadronic uncertainties. 
As soon as the 
extensions of the SM are considered in which new operators and new weak 
phases are present, the mixing induced asymmetry $a_{\phi K_S}$ suffers 
from potential 
hadronic uncertainties that make the determination of the relevant 
parameters problematic unless the hadronic matrix elements can be 
calculated with sufficient precision. This is evident from the many papers 
on the anomaly in $B^0_d(\bar B^0_d)\to \phi K_S$ decays of which the subset
is given in \cite{Ciuchini:2002pd,Fleischer:2001pc,Hiller:2002ci,Datta:2002nr,Raidal:2002ph,Grossman:2003qp,Khalil:2003ng}.

The goal of the present review is to collect the relevant formulae for 
the decays $\kpn$ and $\klpn$  and
to investigate their theoretical and parametric uncertainties. In
addition to known expressions we derive new ones that should allow transparent
tests of the SM and of its extensions. While our 
presentation is centered around the SM, we also discuss  models with
MFV and scenarios with new complex phases in particular the Littlest Higgs Model with
T-parity, the MSSM, $Z^\prime$ models and a model with one universal extra
dimension.
We also give a brief review of other models.
Moreover, we 
investigate the interplay between the $K\to\pi\nu\bar\nu$ complex , the 
$B^0_{d,s}-\bar B^0_{d,s}$ mass differences $\Delta M_{d,s}$ and the angles
$\beta$ and $\gamma$ in the unitarity triangle that can be measured precisely
in two body $B$ decays one day.

Our review is organized as follows. Sections \ref{sec:basic} and \ref{sec:phen} can be considered as a 
compendium of formulae for the decays $\kpn$ and $\klpn$ within the SM. We
also give there the formulae for the CKM factors and the UT 
that are relevant for us. In particular in Section \ref{sec:phen} we investigate
the interplay between $K\to\pi\nu\bar\nu$, the 
mass differences $\Delta M_{d,s}$ and the angles
$\beta$ and $\gamma$. In Section \ref{sec:num} a detailed numerical analysis of the
formulae of Sections \ref{sec:basic} and \ref{sec:phen} is presented. Section
\ref{sec:guide} is a short guide to subsequent sections in which we review $K \to \pi \nu
\bar \nu$ in various extensions of the SM. First in Section \ref {sec:MFV} we indicate how 
the discussion of previous sections is generalized to the class of the MFV 
models. In Section \ref{sec:newphys} our discussion is further generalized to three scenarios
involving new complex phases: a scenario with new physics entering only 
$Z^0$ penguins, a scenario with new physics entering only $B^0_d-\bar B^0_d$ 
mixing and a hybrid 
scenario in which both $Z^0$ penguins and $B^0_d-\bar B^0_d$ 
mixing are affected by new physics.
Here we derive a number of expressions that were not presented in the
literature so far and illustrate how the new phases, and other new physics 
parameters can be determined by means of the $(R_b,\gamma)$ strategy 
\cite{Buras:2002yj} and the related reference unitarity triangle \cite{Goto:1995hj,Cohen:1996sq,Barenboim:1999in,Grossman:1997dd}.
While the discussion of Section \ref{sec:newphys} is practically model independent within three scenarios considered
we give
in Section \ref{sec:models} a brief review of the existing results for both decays 
within
specific extensions of the SM, like Little Higgs, $Z^\prime$ and supersymmetric models, models with 
extra dimensions, models with lepton-flavour mixing and other selected models
considered in the literature.
In Section \ref{sec:decays} we compare the $K\to\pi\nu\bar\nu$ decays with other 
$K$ and $B$ decays used for the determination of the CKM phases and of the UT
with respect to the theoretical cleanness.
 In Section \ref{sec:longdist} we describe briefly the long distance contributions 
that are taken into account in the numerical analyses.
Finally, in Section \ref{sec:concl} we summarize our results and give 
a brief outlook for the future.

\section{Basic Formulae}\label{sec:basic}
\setcounter{equation}{0}
\subsection{Preliminaries}
In this section we will collect the formulae for the branching ratios for 
the decays $\kpn$ and $\klpn$ that constitute the basis for the rest 
of our review. We will also give the values of the relevant parameters as 
well as recall the formulae related to the CKM matrix and the unitarity 
triangle that are relevant for our review. 
Clearly, many formulae listed below have been presented previously in the
literature, in particular in \cite{Buras:2005xt,Buras:2003jf,Buras:2004sc,Buchalla:1995vs,Buras:1998ra,Buchalla:1998ba,Buchalla:1996fp,Buras:2002yj, Battaglia:2003in}. 
Still the collection 
of them at one place and the addition of new ones should 
be useful for future investigations.

\begin{figure}[hbt]
\vspace{0.10in}
\centerline{
\epsfysize=8cm
\epsffile{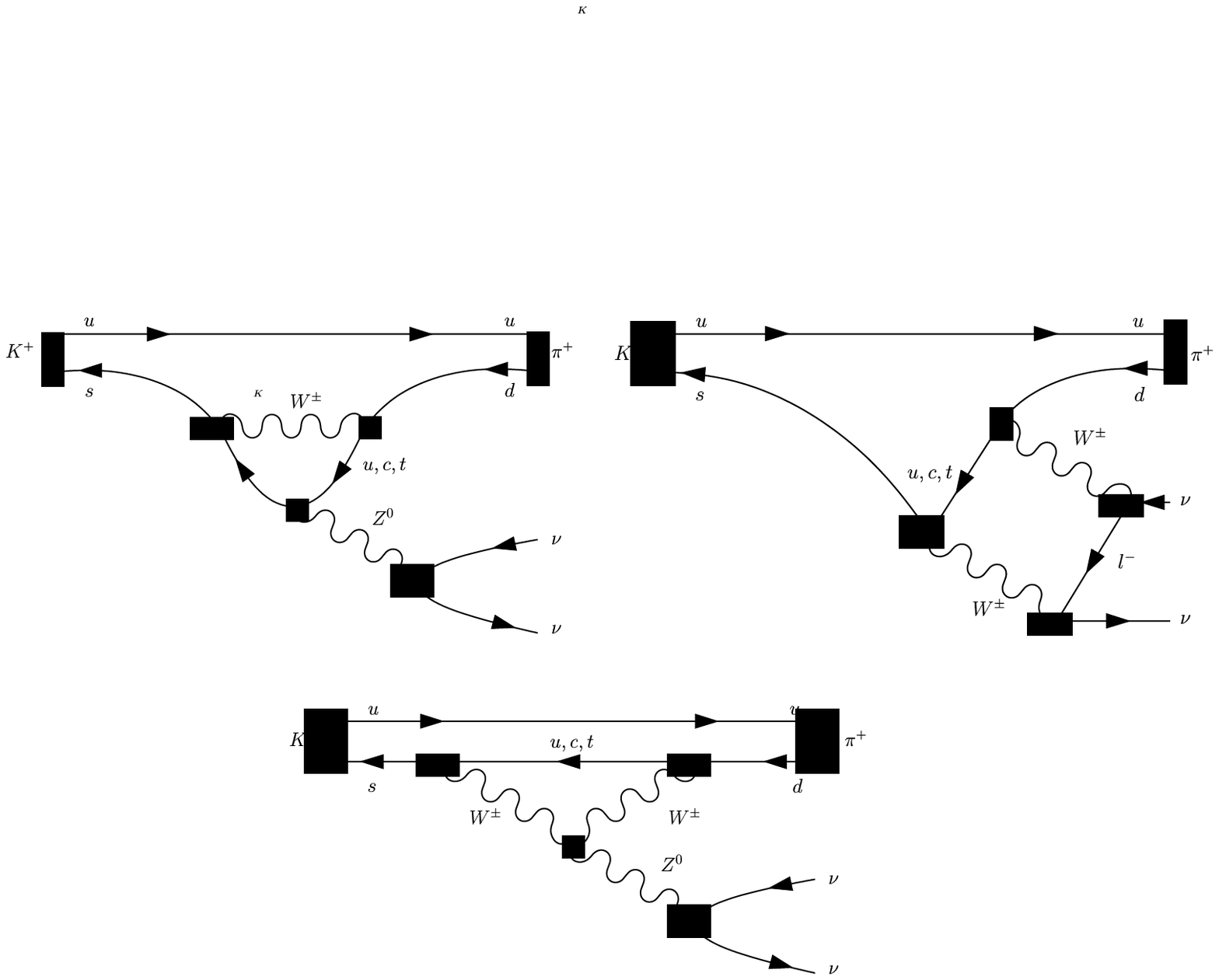}
}
\vspace{0.08in}
\caption{The penguin and box diagrams contributing to $\kpn$. For
  $\klpn$ only the spectator quark is changed from $u$ to $d$.}\label{fig:diagrams}
\end{figure}

The effective Hamiltonian relevant for $\kpn$  and $\klpn$ decays can 
be written in the 
SM as follows \cite{Buchalla:1998ba,Buchalla:1993wq}
\begin{equation}\label{hkpn} 
{\cal H}^{\rm SM}_{\rm eff}={G_{\rm F} \over{\sqrt 2}}{\alpha\over 2\pi 
\sin^2\theta_{\rm w}}
 \sum_{l=e,\mu,\tau}\left( V^{\ast}_{cs}V_{cd} X^l_{\rm NL}+
V^{\ast}_{ts}V_{td} X(x_t)\right)
 (\bar sd)_{V-A}(\bar\nu_l\nu_l)_{V-A} \, 
\end{equation}
with all symbols defined below.
It is obtained from the relevant $Z^0$ penguin  and box diagrams with 
the up, charm and top quark exchanges shown in Fig. {\ref{fig:diagrams}} and includes 
QCD corrections at the NLO level
\cite{Buchalla:1992zm,Buchalla:1993bv,Misiak:1999yg,Buchalla:1998ba,Buchalla:1993wq}
and the NNLO calculated recently \cite{BUGOHA, Buras:2006gb}. 
The presence of up quark 
contributions is only needed for the GIM mechanism to work but 
otherwise only the internal charm and top contributions matter.
The relevance of these contributions in each decay is spelled out below.

The index $l$ = $e$, $\mu$, $\tau$ denotes the lepton flavour.
The dependence on the charged lepton mass resulting from the box diagrams
is negligible for the top contribution. In the charm sector this is the
case only for the electron and the muon but not for the $\tau$-lepton.
In what follows we give the branching ratios that follow from (\ref{hkpn}).

\boldmath
\subsection{$\kpn$}
\unboldmath
The branching ratio for $K^+\to\pi^+\nu\bar\nu$ in the SM is dominated
by $Z^0$ penguin diagrams with a significant contribution from the box 
diagrams. Summing over three neutrino flavours, it can be written
as follows \cite{Buchalla:1998ba, Mescia:2007kn}  
\begin{equation}\label{bkpnn}
Br(K^+\to\pi^+\nu\bar\nu)=\kappa_+ (1+\Delta_{EM})\cdot
\left[\left(\frac{{\rm Im}\lambda_t}{\lambda^5}X(x_t)\right)^2+
\left(\frac{{\rm Re}\lambda_c}{\lambda}P_c(X)+
\frac{{\rm Re}\lambda_t}{\lambda^5}X(x_t)\right)^2\right],
\end{equation}
\begin{equation}\label{kapp}
\kappa_+={ (5.173\pm 0.025 )\cdot 10^{-11}
\left[\frac{\lambda}{0.225}\right]^8}.
\end{equation}

\noindent
An explicit derivation of (\ref{bkpnn}) can be found in \cite{Buras:1998ra}.
Here $x_t=m^2_t/M^2_W$, $\lambda_i=V^*_{is}V_{id}$ are the CKM factors
discussed below and $\kappa_+$ summarizes all the remaining factors following
from (\ref{hkpn}), in particular the relevant hadronic matrix elements that can
be extracted from leading semi-leptonic decays of $K^+$, $K_L$ and $K_S$ mesons.
The original calculation of these matrix elements \cite{Marciano:1996wy} has
recently been significantly improved by Mescia and Smith \cite{Mescia:2007kn},
where details can be found, in particular $\Delta_{EM}$ amounts to $-0.3$ \% which we will
neglect in what follows.
In obtaining the numerical value in (\ref{kapp}) \cite{Mescia:2007kn} the
values \cite{PDBook} 
\be\label{INPUT}
\sin^2\theta_w=0.231, \qquad \alpha=\frac{1}{127.9}, 
\ee
given in the $\overline{MS}$ scheme have been used.  
Their errors are below $0.1\%$ and can be neglected.
There is an issue related to $\sin^2\theta_w$ that although very well
measured in a given renormalization scheme, is a scheme dependent quantity
with the scheme dependence only removed by considering higher order 
electroweak effects in $K\to \pi \nu\bar\nu$.
An analysis of such effects in the large $m_t$ 
limit \cite{Buchalla:1997kz} shows that in principle they could introduce 
a $\pm 5\%$ correction in the $K\to \pi \nu\bar\nu$ branching ratios but 
with the $\overline{MS}$ definition of $\sin^2\theta_w$, these higher order 
electroweak corrections are found below $2\%$ and can also be safely 
neglected. Similar comments apply to $\alpha$. This pattern of higher 
order electroweak corrections is also found in the 
$B_{d,s}^0-\bar B_{d,s}^0$ mixing \cite{Gambino:1998rt}. Yet, in the future, a
complete analysis of two-loop
electroweak contributions to $K\to \pi \bar \nu \nu$ would certainly be of interest.

The apparent large sensitivity of $Br(\kpn)$ to $\lambda$ is spurious as 
$P_c(X)\sim \lambda^{-4}$ and the dependence on $\lambda$ in (\ref{kapp}) 
cancels the one in (\ref{bkpnn}) to a large extent. However, basically for
aesthetic reasons it is useful to write first these formulae as given above. In
doing this it is essential to keep track of the $\lambda$ dependence as it is
hidden in $P_c(X)$ (see (\ref{p0k})) and changing $\lambda$ while 
keeping $P_c(X)$ 
fixed would give wrong results.
For later purposes we will also introduce
\be\label{kapbar}
\bar\kappa_+=\frac{\kappa_+}{\lambda^8}={ (7.87\pm 0.04 ) }\cdot 10^{-6}.
\ee

The function $X(x_t)$ relevant for the top part is given by
\begin{equation}\label{xx9} 
X(x_t)=X_0(x_t)+\frac{\alpha_s(m_t)}{4\pi} X_1(x_t) 
=\eta_X\cdot X_0(x_t), \qquad\quad \eta_X=0.995,
\end{equation}
where
\begin{equation}\label{xx0} 
X_0(x_t)={x_t\over 8}\left[ -{2+x_t\over 1-x_t}+
{3x_t-6\over (1-x_t)^2}\ln x_t\right] 
\end{equation}
describes the contribution of $Z^0$ penguin diagrams and box diagrams 
without the QCD corrections \cite{Inami:1980fz,Buchalla:1990qz} and the second term
stands for  the QCD correction \cite{Buchalla:1992zm,Buchalla:1993bv,Misiak:1999yg,Buchalla:1998ba} with
\begin{eqnarray}\label{xx1}
X_1(x_t)=&-&{29x_t-x_t^2-4x_t^3\over 3(1-x_t)^2}
-{x_t+9x_t^2-x_t^3-x_t^4\over (1-x_t)^3}\ln x_t
\nonumber\\
&+&{8x_t+4x_t^2+x_t^3-x_t^4\over 2(1-x_t)^3}\ln^2 x_t
-{4x_t-x_t^3\over (1-x_t)^2}L_2(1-x_t)
\nonumber\\
&+&8x{\partial X_0(x_t)\over\partial x_t}\ln x_\mu
\end{eqnarray}
where $x_\mu=\mu_t^2/M^2_W$, $\mu_t=\ord(m_t)$ and
\begin{equation}\label{l2} 
L_2(1-x_t)=\int^{x_t}_1 dt {\ln t\over 1-t}.   
\end{equation}
The $\mu_t$-dependence in the last term in (\ref{xx1}) cancels to the
order considered the $\mu_t$-dependence of the leading term $X_0(x_t(\mu_t))$ in
 (\ref{xx9}).
The leftover $\mu_t$-dependence in $X(x_t)$ is below $1\%$.
The factor $\eta_X$ summarizes the NLO 
corrections represented by the second
term in (\ref{xx9}).
With $\mt\equiv \mt(\mt)$ the QCD factor $\eta_X$
is practically independent of $m_t$ and $\alpha_s(M_Z)$
and is very close to unity. Varying $\mt(\mt)$ from $150\gev$ to 
$180\gev$ changes $\eta_X$ by at most $0.1\%$.

The uncertainty in $X(x_t)$ is then fully dominated by the experimental error
in $m_t$. The $\overline{MS}$ top-quark mass \footnote{We thank M. Jamin
for discussions on this subject.},  including 
one- two- and three-loop contributions \cite{Melnikov:2000qh} and corresponding to the most recent 
{$m_t^{\rm pole}=(170.9\pm 1.1 \pm 1.5 )\gev$ }\cite{Group:2006qt} is given by
\be\label{mtop} 
{ \mt(\mt)=(161.0 \pm 1.7 )\gev.}
\ee
One finds then
\be\label{XT}
{X(x_t)=1.443 \pm 0.017 }.
\ee
$X(x_t)$ increases with $m_t$ roughly as $m_t^{1.15}$. After the LHC era 
the error on $m_t$ should decrease below $\pm 1\gev$, implying the error of 
$\pm 0.01$ in $X(x_t)$ that can be neglected for all practical purposes.


The parameter $P_c(X)$ summarizes the charm contribution
and is defined through
\be
{ P_c(X)=P_c^{\rm SD}(X)+\delta P_{c,u}}
\ee
{ with the long-distance contributions { $\delta P_{c,u}=0.04 \pm
0.02$}
\cite{Isidori:2005xm}. The short-distance part is given by}
\begin{equation}\label{p0k}
P_c^{\rm SD}(X)=\frac{1}{\lambda^4}\left[\frac{2}{3} X^e_{\rm NNL}+\frac{1}{3}
 X^\tau_{\rm NNL}\right]
\end{equation}
where
the functions
$X^l_{\rm NNL}$ result from the NLO calculation \cite{Buchalla:1998ba,Buchalla:1993wq} and 
NNLO \cite{BUGOHA,Buras:2006gb}. 
The index ``$l$" distinguishes between the charged lepton flavours in 
the box diagrams. This distinction is irrelevant in the top contribution 
due to $m_t\gg m_l$ but is relevant in the charm contribution as 
$m_{\tau}>m_c$.
The inclusion of NLO corrections reduced considerably the large
$\mu_c$ dependence
(with $\mu_c={\cal O}(m_c)$) present in the leading order expressions
for the charm contribution
 \cite{Vainshtein:1976eu,Ellis:1982ve,Dib:1989cc}.
Varying $\mu_c$ in the range $1\gev\le\mu_c\le 3\gev$ changes 
$X^l_{\rm NNL}$ by roughly $24\%$ at NLO to be compared to $56\%$ in the 
leading order. At NNLO, the $\mu_c$ dependence is further decreased as discussed in detail below.

The net effect of QCD corrections is to suppress the charm contribution by 
roughly $30\%$.
For our purposes we need only $P_c(X)$.
In table~\ref{tab:P0Kplus} we give its values 
for different $\alpha_s(M_Z)$ and $m_c\equiv m_c(m_c)$.
The chosen range for $m_c(m_c)$ is in the ballpark of the most recent 
estimates. For instance $m_c(m_c)=1.286(13),~1.29(7)(13)$ and $1.29(7)$ 
(all in $\gev$) have been found from $R^{e^+e^-}(s)$ \cite{Kuhn:2007vp}, 
quenched combined with dynamical lattice QCD \cite{Dougall:2005ev} and charmonium sum rules \cite{Hoang:2004xm}, 
respectively. Further references can be found in these papers and in 
\cite{Battaglia:2003in}.

Finally, in table~\ref{tab:mucdep} we show 
the dependence of $P_c(X)$ on $\alpha_s(M_Z)$ and $\mu_c$ at fixed 
$m_c(m_c)=1.30\gev$.

{%
\renewcommand{\arraystretch}{1.25}
\begin{table}[!t]
\begin{center}
\begin{tabular}{|c|c|c|c|c|c|c|c|}
\hline  
\multicolumn{1}{|c|}{} & \multicolumn{7}{c|}{$P_c(X)$} \\
\hline 
$\alpha_s (M_Z) \hspace{1.5mm} \backslash \hspace{1.5mm} m_c (m_c)
\hspace{1.5mm} [{\rm GeV}]$ & 1.15 & 1.20 & 1.25 & 1.30 & 1.35 & 1.40 &
1.45 \\
\hline 
0.115 & 0.307 & 0.336 & 0.366 & 0.397 & 0.430 & 0.463 & 0.497 \\
0.116 & 0.303 & 0.332 & 0.362 & 0.394 & 0.426 & 0.459 & 0.493 \\
0.117 & 0.300 & 0.329 & 0.359 & 0.390 & 0.422 & 0.455 & 0.489 \\
0.118 & 0.296 & 0.325 & 0.355 & 0.386 & 0.417 & 0.450 & 0.484 \\
0.119 & 0.292 & 0.321 & 0.350 & 0.381 & 0.413 & 0.446 & 0.480 \\
0.120 & 0.288 & 0.316 & 0.346 & 0.377 & 0.409 & 0.441 & 0.475 \\
0.121 & 0.283 & 0.312 & 0.342 & 0.372 & 0.404 & 0.437 & 0.470 \\
0.122 & 0.279 & 0.307 & 0.337 & 0.368 & 0.399 & 0.432 & 0.465 \\
0.123 & 0.274 & 0.303 & 0.332 & 0.363 & 0.394 & 0.426 & 0.460 \\
\hline 
\end{tabular} 
\vspace{2mm}
\caption{ The parameter $P_c(X)$ in NNLO approximation for various values
of $\alpha_s (M_Z)$ and $m_c (m_c)$ \cite{Buras:2006gb}. The numerical values for
$P_c(X)$ correspond to $\lambda = 0.2248$, 
$\mu_W = 80.0 \, {\rm GeV}$, $\mu_b = 5.0 \, {\rm GeV}$, and $\mu_c = 1.50 \,
{\rm GeV}$. }                  
\label{tab:P0Kplus}
\end{center}
\end{table}
}%

Restricting the three parameters involved to the ranges 
\be
1.15\gev\le\mc(m_c)\le 1.45\gev,
 \qquad
1.0\gev\le\mu_c\le 3.0\gev,
\ee
\be
0.115\le \alpha_s(M_Z)\le 0.123
\ee
one arrives at \cite{BUGOHA}
\be\label{PC}
{ P_c(X)^{\rm SD}=(0.375\pm 0.031_{m_c} \pm 0.009_{\mu_c} \pm 0.009_{\alpha_s})
\left(\frac{0.2248}{\lambda}\right)^4}
\ee
where the errors correspond to $m_c(m_c)$, $\mu_c$ and 
$\alpha_s(M_Z)$, respectively.
The uncertainty due to $m_c$ is 
significant. On the other hand, the uncertainty due to $\alpha_s$ is small. 
In principle one could add the errors in (\ref{PC}) linearly, which would 
result in an error of $\pm 0.049$.  
We think that this estimate would be too
conservative.
Adding the errors in quadrature gives $\pm 0.033$. This could be, on the
other hand, too optimistic, since the uncertainties are not
  statistically distributed. 
Therefore, as the final result for $P_c(X)$ we 
quote
\be\label{PCFIN}
{ P_c(X)=0.41\pm 0.05}
\ee
that we will use in the rest of our review.

{%
\renewcommand{\arraystretch}{1.25}
\begin{table}[!t]
\begin{center}
\begin{tabular}{|c|c|c|c|c|c|}
\hline  
\multicolumn{1}{|c|}{} & \multicolumn{5}{c|}{$P_c(X)$} \\
\hline 
$\alpha_s (M_Z) \hspace{1.5mm} \backslash \hspace{1.5mm} \mu_c
\hspace{1.5mm} [{\rm GeV}]$ & 1.0 & 1.5 & 2.0 & 2.5 & 3.0 \\
\hline 
0.115 & 0.393 & 0.397 & 0.395 & 0.392 & 0.388 \\
0.116 & 0.389 & 0.394 & 0.391 & 0.388 & 0.383 \\
0.117 & 0.384 & 0.390 & 0.387 & 0.383 & 0.379 \\
0.118 & 0.380 & 0.386 & 0.383 & 0.379 & 0.374 \\
0.119 & 0.375 & 0.381 & 0.379 & 0.374 & 0.369 \\
0.120 & 0.370 & 0.377 & 0.374 & 0.369 & 0.364 \\
0.121 & 0.365 & 0.372 & 0.369 & 0.364 & 0.359 \\
0.122 & 0.359 & 0.368 & 0.364 & 0.359 & 0.354 \\
0.123 & 0.353 & 0.363 & 0.359 & 0.354 & 0.348 \\
\hline 
\end{tabular} 
\vspace{2mm}
\caption{ The parameter $P_c(X)$ in NNLO approximation for various values
of $\alpha_s (M_Z)$ and $\mu_c$ \cite{Buras:2006gb}. The numerical values for $P_c(X)$
correspond to $\lambda = 0.2248$, $m_c
(m_c) = 1.30 \, {\rm GeV}$, $\mu_W = 80.0 \, {\rm GeV}$, and $\mu_b = 5.0 \,
{\rm GeV}$.}                 
\label{tab:mucdep}
\end{center}
\end{table}
}%

We expect that the reduction of the error in
$\alpha_s(M_Z)$ to $\pm 0.001$ will decrease the corresponding error 
to $0.005$, making it negligible.  
Concerning the error due to $m_c(m_c)$, it should be remarked that 
increasing the error in 
$\mc(\mc)$ to $\pm 70\mev$
would increase the first error in (\ref{PC}) to $0.047$, whereas 
its decrease to
$\pm 30\mev$ would decrease it to $0.020$. More generally we have to a good 
approximation
\be\label{FS}
\sigma (P_c(X))_{m_c}= \left[\frac{0.67}{\rm GeV}\right]
 \sigma(\mc(\mc)).
\ee

>From the present perspective, 
unless some important advances in the determination of $\mc(m_c)$ will be made,
it will be very difficult to decrease the error on $P_c(X)$ below 
$\pm 0.03$, although $\pm 0.02$ cannot be fully excluded. We will use this
information in our numerical analysis in Section \ref{sec:num}.

\boldmath
\subsection{$\klpn$}
\unboldmath
The neutrino pair produced by ${\cal H}^{\rm SM}_{\rm eff}$ 
in (\ref{hkpn}) is a CP eigenstate with positive eigenvalue. 
Consequently, within 
the approximation of keeping only operators of dimension six, as done
in (\ref{hkpn}), the decay $\klpn$ proceeds entirely through CP violation 
\cite{Littenberg:1989ix}. However, as pointed out in \cite{Buchalla:1998ux}, even in the SM 
there are CP-conserving contributions to $\klpn$, that are 
generated only by local operators of $d\ge 8$ or by long distance effects. 
Fortunately, these effects are by a factor of $10^5$ smaller than the 
leading CP-violating contribution and can be safely neglected \cite{Buchalla:1998ux}.
As we will discuss in Section \ref{sec:models},
the situation can be in principle very different beyond the SM.

The branching ratio for $\klpn$ in the SM is then fully dominated by the 
diagrams with internal top exchanges 
with the charm contribution well below $1\%$. It can be written then
as follows \cite{Buchalla:1996fp,Buchalla:1995vs,Buras:1998ra}
\begin{equation}\label{bklpn}
Br(K_L\to\pi^0\nu\bar\nu)=\kappa_L\cdot
\left(\frac{{\rm Im}\lambda_t}{\lambda^5}X(x_t)\right)^2
\end{equation}
\begin{equation}\label{kapl}
\kappa_L=
{ (2.231\pm 0.013)\cdot 10^{-10}\left[\frac{\lambda}{0.225}\right]^8}
\end{equation}
{ where we have
summed over three
neutrino flavours. 
An explicit derivation of (\ref{bklpn}) can be found 
in \cite{Buras:1998ra}.
Here $\kappa_L$ is the factor corresponding to $\kappa_+$ in (\ref{bkpnn}). The
original calculation of $\kappa_L$ \cite{Marciano:1996wy} has been recently
significantly improved by \cite{Mescia:2007kn}, where details can be found. }
Due to the absence of $P_c(X)$ in (\ref{bklpn}), $Br(K_L\to\pi^0\nu\bar\nu)$ 
has essentially no theoretical uncertainties and is only affected by
parametric uncertainties coming from $m_t$, $\imlt$ and $\kappa_L$. They 
should be decreased significantly in the coming years so that a precise 
prediction for $Br(K_L\to\pi^0\nu\bar\nu)$ should be available in this
decade. On the other hand, as discussed below, once this branching ratio has 
been measured, $\imlt$ can be in principle determined with exceptional 
precision not matched by any other decay \cite{Buchalla:1996fp}.

\boldmath
\subsection{$\kspn$}
\unboldmath
Next, mainly for completeness, we give the expression for $Br(\kspn)$, that, 
due to $\tau(K_S)\ll \tau(K_L)$, is suppressed by roughly 2 orders of
magnitude relative to $Br(\klpn)$. We have \cite{Bossi:1998it}
\begin{equation}\label{bspnn}
Br(\kspn)=\kappa_S\cdot
\left(\frac{{\rm Re}\lambda_c}{\lambda}P_c(X)+
\frac{{\rm Re}\lambda_t}{\lambda^5}X(x_t)\right)^2,
\end{equation}
\begin{equation}\label{kapps}
\kappa_S=
\kappa_L\frac{\tau(K_S)}{\tau(K_L)}
={ (3.91 \pm 0.02)\cdot 10^{-13}}
\left[\frac{\lambda}{0.2248}\right]^8.
\end{equation}
Introducing the 
``reduced'' branching ratio
\begin{equation}\label{b3}
B_3={Br(\kspn)\over \kappa_S}
\end{equation}
and analogous ratios $B_1$ and $B_2$ for $\kpn$ and $\klpn$ given in
(\ref{b1b2}) we find a simple relation between the three $K\to\pi\nu\bar\nu$ 
decays 
\be\label{COR3}
B_1=B_2+B_3.
\ee

We would like to emphasize that, while $Br(\klpn)$ being only sensitive to 
$\imlt$ provides a direct determination of $\bar\eta$, 
$Br(\kspn)$ being only sensitive to 
$\relt$ provides a direct determination of $\bar\varrho$. The latter
determination is not as clean as the one of $\bar\eta$ from $\klpn$ due to 
the presence of the charm contribution in (\ref{bspnn}). However, it is much
cleaner than the corresponding determination of $\bar\varrho$ from 
$K_L\to\mu^+\mu^-$. Unfortunately, the tiny branching ratio 
$Br(\kspn)\approx 5\cdot 10^{-13} $ will not allow this determination in a 
foreseeable future. Therefore we will not consider $\kspn$ in the rest of 
our review. Still one should not forget that the presence of another
theoretically clean observable would be very useful in testing the extensions
of the SM.
Interesting discussions of the complex $\klpn$ and $\kspn$ and its analogies
to the studies of $\epe$ can be found in \cite{Bossi:1998it,D'Ambrosio:1994wx}.

\subsection{CKM Parameters}\label{ssec:CKM}
\boldmath
\subsubsection{Unitarity Triangle, {$\imlt$} and {$\relt$}}
\begin{figure}[hbt]
\vspace{0.10in}
\centerline{
\epsfysize=1.7in
\epsffile{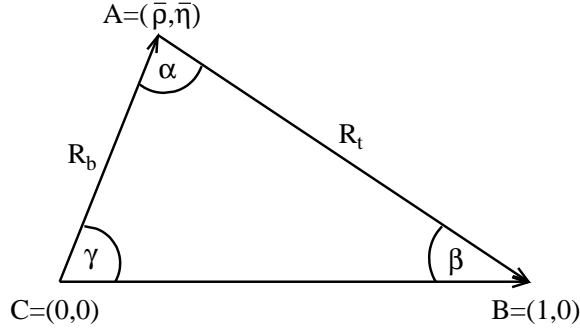}
}
\vspace{0.08in}
\caption{Unitarity Triangle.}\label{fig:utriangle}
\end{figure}

\unboldmath
Concerning the CKM parameters, we will use in our numerical analysis 
the Wolfenstein parametrization
\cite{Wolfenstein:1983yz}, generalized to include higher orders in $\lambda\equiv|V_{us}|$ 
\cite{Buras:1994ec}.
This turns out to be very useful in making the structure of various formulae 
transparent and gives results very close to the ones obtained by means of
the exact standard parametrization \cite{Hagiwara:2002fs,Chau:1984fp}. 
The basic parameters are then
\be\label{GW}
\lambda, \qquad A=\frac{\vcb}{\lambda^2}, \qquad  
\bar\varrho=\varrho (1-\frac{\lambda^2}{2}),
\qquad
\bar\eta=\eta (1-\frac{\lambda^2}{2})
\end{equation}
with $\varrho$ and $\eta$ being the usual Wolfenstein parameters \cite{Wolfenstein:1983yz}. 
The parameters 
$\bar\varrho$ and $\bar\eta$, introduced in \cite{Buras:1994ec}, are particularly 
useful as they describe the apex of the standard UT as
shown in Fig.~\ref{fig:utriangle}.
More details on the unitarity triangle and the generalized Wolfenstein 
parametrization can be found in \cite{Buras:2005xt,Buras:2003jf,Buras:2004sc,Buras:1994ec,Battaglia:2003in}. Below, we only recall 
certain expressions that we need in the course of our discussion.

Parallel to the use of the parameters in (\ref{GW}) it will turn out useful 
to express the CKM elements $V_{td}$ and $V_{ts}$ 
as follows \cite{Buras:2004ub}
\be\label{VTDVTS}
V_{td}=A R_t\lambda^3 e^{-i\beta},\quad V_{ts}=-|V_{ts}|e^{-i\beta_s},
\ee
with $\tan\beta_s\approx -\lambda^2 \bar\eta$.
The smallness of $\beta_s$ follows from the CKM phase conventions and the 
unitarity of the CKM matrix. Consequently it is valid beyond the SM
if three generation unitarity is assumed.
$R_t$ and $\beta$ are defined in Fig.~\ref{fig:utriangle}. 

We have then
\be\label{LAMT}
\lambda_t\equiv V_{ts}^*V_{td}=-\tilde r \lambda\vcb^2 R_t e^{-i\beta} 
e^{i\beta_s} \quad\mbox{with}\quad 
\tilde r=\left|\frac{V_{ts}}{V_{cb}}\right|
=\sqrt{1+\lambda^2(2\bar\varrho-1)}
\approx 0.985,
\ee
where in order to avoid high powers of $\lambda$ we expressed the parameter
 $A$ through $\vcb$. 
Consequently 
\be\label{IMRE}
\imlt=\tilde r \lambda\vcb^2 R_t\sin(\beta_{\rm eff}), \qquad 
\relt=-\tilde r \lambda\vcb^2 R_t\cos(\beta_{\rm eff})
\ee
with $\beta_{\rm eff}=\beta-\beta_s$.

Alternatively, using the parameters in (\ref{GW}),
one has \cite{Buras:1994ec}
\begin{equation}\label{IMREBLO}
 \imlt= \eta \lambda\vcb^2, \qquad
\relt= -(1-\frac{\lambda^2}{2}) \lambda\vcb^2 (1-\bar\varrho)
\ee
\begin{equation}\label{REC}
 \relc=-\lambda (1-\frac{\lambda^2}{2})~.
\end{equation}
The expressions for $\imlt$ and $\relc$ given here represent to an accuracy of
0.2\% the exact formulae obtained using the standard parametrization. 
The expression for $\relt$ in (\ref{IMREBLO})
deviates by at most $0.5\%$ from the exact formula in the
full range of parameters considered. 
After inserting the expressions (\ref{IMREBLO}) and (\ref{REC}) in the exact
formulae for quantities of interest a further expansion in $\lambda$
should not be made. 

\boldmath
\subsubsection{Leading Strategies for $(\bar\varrho,\bar\eta)$}
\unboldmath
Next, we have the following useful 
relations, that correspond to the best strategies for the determination of 
$(\bar\varrho,\bar\eta)$ considered in \cite{Buras:2002yj}:

{\bf \boldmath{$(R_t,\beta)$} Strategy}:

\be\label{S1}
\bar\varrho=1-R_t\cos\beta, \qquad \bar\eta=R_t\sin\beta
\ee
with $R_t$ determined through (\ref{Rt}) below and $\beta$ through 
$a_{\psi K_S}$. 
In this strategy, $R_b$ and $\gamma$ are given by
\be\label{VUBG}
R_b=\sqrt{1+R_t^2-2 R_t\cos\beta},\qquad
\cot\gamma=\frac{1-R_t\cos\beta}{R_t\sin\beta}.
\ee

{\bf \boldmath{($R_b,\gamma)$} Strategy}:

\be\label{S2}
\bar\varrho=R_b\cos\gamma, \qquad \bar\eta=R_b\sin\gamma
\ee
with $\gamma$ (see Fig.~\ref{fig:utriangle}), determined through clean 
strategies in tree dominated $B$-decays \cite{Buras:2005xt,Buras:2003jf,Buras:2004sc,Fleischer:2002ys,Fleischer:2004xw,Ali:2003te,Hurth:2003vb,Nir:2001ge,Buchalla:2003ux,Ball:2000ba,Anikeev:2001rk}.
In this strategy, $R_t$ and $\beta$ are given by
\be\label{VTDG}
R_t=\sqrt{1+R_b^2-2 R_b\cos\gamma},\qquad
\cot\beta=\frac{1-R_b\cos\gamma}{R_b\sin\gamma}.
\ee

{\bf \boldmath{$(\beta,\gamma)$} Strategy}:

Formulae in (\ref{S1}) and
\be\label{S3}
R_t=\frac{\sin\gamma}{\sin(\beta+\gamma)}
\ee
with $\beta$ and $\gamma$ determined through $a_{\psi K_S}$ and clean 
strategies for $\gamma$ as in (\ref{S2}).
In this strategy, the length $R_b$ and $\vub$ can be determined through
\be\label{RTVUB2}
R_b=\frac{\sin\beta}{\sin(\beta+\gamma)},\qquad
\left|\frac{V_{ub}}{V_{cb}}\right|=\left(\frac{\lambda}{1-\lambda^2/2}
\right) R_b.
\ee

{\bf \boldmath{$(\bar\eta,\gamma)$} Strategy}:

\be\label{S4}
\bar\varrho=\frac{\bar\eta}{\tan\gamma}
\ee
with $\bar\eta$ determined for instance through $Br(\klpn)$ as discussed 
in Section \ref{sec:phen} and $\gamma$ as in the two strategies above.

As demonstrated in \cite{Buras:2002yj}, the $(R_t,\beta)$ strategy is very useful now that the $B^0_{s}-\bar B^0_{s}$ mixing mass difference
$\Delta M_s$ has been measured. However, the remaining three strategies turn 
out to be more efficient in determining $(\bar\varrho,\bar\eta)$. 
The strategies $(\beta,\gamma)$ and $(\bar\eta,\gamma)$
are theoretically cleanest as $\beta$ and $\gamma$  
can be measured 
precisely in two body B decays one day and $\bar\eta$ can be extracted 
from $Br(\klpn)$ subject only to uncertainty in $\vcb$. Combining these
two strategies offers a precise determination of the CKM matrix 
including $\vcb$ and $|V_{ub}|$ \cite{Buras:1994rj}. 
On the other hand, these two strategies 
are subject to uncertainties
coming from new physics that can enter through $\beta$ and $\bar\eta$. 
The angle $\gamma$, the phase of $V_{ub}$, can be determined in principle 
without 
these uncertainties. 

The strategy $(R_b,\gamma)$, on the other hand, while
subject to hadronic uncertainties in the determination of $R_b$, is not
polluted by new physics contributions as, in addition to $\gamma$, also 
 $R_b$ can be determined from tree level decays. This strategy results 
in the so-called {\it reference unitarity triangle} (RUT) as proposed and discussed
in \cite{Goto:1995hj,Cohen:1996sq,Barenboim:1999in,Grossman:1997dd}. We will return to all these strategies in the course of 
our presentation.

\subsubsection{Constraints from the Standard Analysis of the UT}\label{sssec:CKMconstr}
Other useful expressions that represent the constraints from 
the CP-violating parameter $\varepsilon_K$ and  $\Delta M_{s,d}$, 
that parametrize the size of $B^0_{s,d}-\bar B^0_{s,d}$
mixings are as follows.

First we have
\begin{equation}
\eps_K=-C_{\eps} \hat B_K \IM\lambda_t \left\{
\lambda^4\RE\lambda_c P_c(\varepsilon)  +
\RE\lambda_t \eta^{QCD}_2 S_0(x_t) \right\} e^{i \pi/4}\,,
\label{eq:epsformula}
\end{equation}
where { $S_0(x_t)=2.27 \pm 0.04$} results from $\Delta S=2$ box diagrams and  
the numerical constant $C_\eps$ is given by ($M_W=80.4\gev$)
\begin{equation}
C_\eps = \frac{G_{\rm F}^2 F_K^2 m_K \mw^2}{6 \sqrt{2} \pi^2 \Delta M_K}
       = 3.837 \cdot 10^4 \, .
\label{eq:Ceps}
\end{equation}
Next \cite{Herrlich:1993yv,Herrlich:1995hh,Herrlich:1996vf,JaminNierste}, 
\be
P_c(\varepsilon)=\frac{\bar P_c(\varepsilon)}{\lambda^4}=
(0.29\pm0.07) \left[\frac{0.2248}{\lambda}\right]^4, 
\qquad 
\bar P_c(\varepsilon)=(7.3\pm1.7)\cdot 10^{-4},
\ee
 $\eta^{  QCD}_2=0.574\pm0.003$ 
\cite{Buras:1990fn,Buchalla:1995vs,Buras:1998ra} and 
$\hat B_K$ is 
a non-perturbative parameter.
In obtaining (\ref{eq:epsformula}) a small term amounting to at most
$5\%$ correction to $\eps_K$ has been neglected. This is justified
in view of other uncertainties, in particular those connected with
$\hat B_K$ but in the future should be taken into account \cite{Andriyash:2003ym}.

Comparing (\ref{eq:epsformula}) with the
experimental value for $\eps_K$ \cite{Hagiwara:2002fs}
\begin{equation}\label{eexp}
(\varepsilon_K)_{exp}
=(2.280\pm0.013)\cdot10^{-3}\;\exp{i\pi/4},
\end{equation}
one obtains a constraint on the UT that with the help of  
(\ref{IMREBLO}) and (\ref{REC}) can be cast into
\begin{equation}\label{100}
\bar\eta \left[(1-\bar\varrho) \vcb^2 \eta^{  QCD}_2 S_0(x_t)
+ \bar P_c(\varepsilon) \right] \vcb^2 \hat B_K = 1.184 \cdot 10^{-6}
\left[\frac{0.2248}{\lambda}\right]^{2}.
\end{equation}

Next, the constraint from $\Delta M_d$ implies
\begin{equation}\label{106}
 R_t= \frac{1}{\lambda}\frac{|V_{td}|}{\vcb} = 0.834 \cdot
\left[\frac{|V_{td}|}{7.75\cdot 10^{-3}} \right] 
\left[ \frac{0.0415}{\vcb} \right] \left[\frac{0.2248}{\lambda}\right],
\end{equation}
\begin{equation}\label{VT}
\vtd=
7.75\cdot 10^{-3}\left[ 
\frac{230\mev}{\sqrt{\hat B_{B_d}}F_{B_d}}\right]
\sqrt{\frac{\Delta M_d}{0.50/{\rm ps}}}  
\sqrt{\frac{0.55}{\eta_B^{  QCD}}}
\sqrt{\frac{2.40}{S_0(x_t)}}~.
\end{equation}
Here $\sqrt{\hat B_{B_d}}F_{B_d}$ is a non-perturbative parameter and 
$\eta_B^{  QCD}=0.551\pm0.003$ the QCD correction \cite{Buras:1990fn,Urban:1997gw}.

Finally, the simultaneous use of $\Delta M_{d}$ and $\Delta M_{s}$  gives
{\be\label{Rt}
R_t=0.935~\tilde r\left[\frac{\xi}{1.24}\right] 
\left[\frac{0.2248}{\lambda}\right]
\sqrt{\frac{17.8/ps}{\Delta M_s}} 
\sqrt{\frac{\Delta M_d}{0.50/ps}},
\qquad
\xi = 
\frac{\sqrt{\hat B_{B_s}}F_{B_s} }{ \sqrt{\hat B_{B_d}}F_{B_d}}
\ee}
with $\tilde r$ defined in (\ref{LAMT})
and $\xi$ standing for a nonperturbative parameter that is subject 
to smaller theoretical uncertainties than the individual 
$\sqrt{\hat B_{B_d}}F_{B_d}$ and $\sqrt{\hat B_{B_s}}F_{B_s}$.

The main uncertainties in these constraints originate in the theoretical 
uncertainties in  $\hat B_K$ and 
$\sqrt{\hat B_d}F_{B_d}$, { $\sqrt{\hat B_s}F_{B_s}$} and to a lesser extent in $\xi$ 
{ \cite{Hashimoto:2004hn,Dawson:2005zv}}: 
{ \be\label{paramet}
\hat B_K=0.79\pm0.04\pm 0.08, \quad  
\sqrt{\hat B_d}F_{B_d}=(214\pm 38)~{\rm MeV},
\quad
\sqrt{\hat B_s}F_{B_s}=(262\pm 35)~{\rm MeV},
\quad \xi=1.23\pm 0.06~.
\ee}
The QCD sum rules results for  
the parameters in question are similar and can be found in 
\cite{Battaglia:2003in}. 
Finally  \cite{Battaglia:2003in,Abulencia:2006ze}
 \be
\Delta M_d=(0.507\pm 0.005)/{\rm ps}, \qquad 
\Delta M_s=(17.77\pm 0.12)/{\rm ps}
\ee 

Extensive discussion of the formulae (\ref{eq:epsformula}), 
(\ref{100}), (\ref{VT}) and (\ref{Rt}) 
can be 
found in \cite{Battaglia:2003in}.
For our numerical analysis, we will use \cite{Bona:2005vz}
{ \be\label{CKM1}
\lambda=0.2258 \pm 0.0014,\quad
A=0.816\pm0.016,\quad \vcb=(41.6\pm 0.6)\cdot 10^{-3},
\ee}
{\be\label{CKM2}
 \left|\frac{V_{ub}}{V_{cb}}\right|=0.088\pm0.005,\qquad
R_b=0.38\pm0.01
\ee}
{ \be\label{CKM3}
\beta=(22.2\pm 0.9)^\circ,\quad \beta_s=-1^\circ
\ee }
with the value of $\beta$ following from the UTfit and slightly higher than the
one
determined from
measurements of the time-dependent CP asymmetry
$a_{\psi K_S}(t)$ that give 
\cite{Aubert:2002ic,Abe:2002px,Browder:2003ii,Barberio:2007cr}
 \be
(\sin 2\beta)_{\psi K_S}=0.675\pm 0.026~ \qquad \beta=(21.2 \pm 1.0)^°.
\label{ga}
\ee

\section{Phenomenological Applications in the SM}\label{sec:phen}
\setcounter{equation}{0}
\subsection{Preliminaries}
During the last ten years several analyses of $K\to\pi\nu\bar\nu$ 
decays within the SM were presented, in particular in
 \cite{Buras:2005xt,Buras:2003jf,Buras:2004sc,Buchalla:1998ba,D'Ambrosio:2001zh,Kettell:2002ep,Haisch:2005um,
Mescia:2007kn,Bona:2006ah,Charles:2004jd}. Moreover, 
correlations 
with other decays have been pointed out 
\cite{Buras:1998ed,Buras:1999da,Bergmann:2001pm,Bergmann:2000ak}. 
In this section we collect and update many of these formulae  
and  derive a number 
of useful expressions that are new. In the next section a detailed 
numerical analysis of these formulae will be presented. 
Unless explicitely stated all the formulae below are given for 
$\lambda=0.2248$. The dependence on $\lambda$ can easily be found from the 
formulae of the previous section. When it is introduced, it is often 
useful to replace $\lambda^2 A$ by $\vcb$ to avoid high powers of 
$\lambda$. On the whole, the issue of the error in $\lambda$ in
$K\to\pi\nu\bar\nu$ decays is really not 
an issue if changes are made consistently in all places as emphasized 
before. 
\boldmath
\subsection{Unitarity Triangle and $\kpn$}
\unboldmath
\subsubsection{Basic Formulae}
Using (\ref{IMRE}) in (\ref{bkpnn}) we obtain \cite{Buras:2004ub} 
\begin{equation}\label{bkpnn1}
Br(\kpn)=\kappa_+\left[\tilde r^2 A^4 R_t^2 X^2(x_t)
+2 \tilde r \bar P_c(X) A^2 R_t X(x_t)\cos\beta_{\rm eff}
+ \bar P_c(X)^2  \right]
\end{equation}
with $\beta_{\rm eff}=\beta-\beta_s$, 
$\tilde r$  given in (\ref{LAMT}) and 
\be\label{Pbar}
\bar P_c(X)=\left(1-\frac{\lambda^2}{2}\right) P_c(X).
\ee

In the context of the unitarity triangle also the
expression following from (\ref{bkpnn}) and  (\ref{IMREBLO}) 
is useful \cite{Buras:1994ec}
\begin{equation}\label{108}
Br(K^{+} \to \pi^{+} \nu \bar\nu) = \bar\kappa_+ \vcb^4 X^2(x_t)
\frac{1}{\sigma} \left[ (\sigma \bar\eta)^2 +
\left(\varrho_c - \bar\varrho \right)^2 \right]\,,
\end{equation}
where
\begin{equation}\label{109}
\sigma = \left( \frac{1}{1- \frac{\lambda^2}{2}} \right)^2\,.
\end{equation}

The measured value of $Br(K^{+} \to \pi^{+} \nu \bar\nu)$ then
determines  an ellipse in the $(\bar\varrho,\bar\eta)$ plane  centered at
$(\varrho_c,0)$ (see Fig.~\ref{fig:KPKL}) with 
\begin{equation}\label{110}
\varrho_c = 1 + \frac{\lambda^4 P_c(X)}{\vcb^2 X(x_t)}
\end{equation}
and having the squared axes
\begin{equation}\label{110a}
\bar\varrho_1^2 = r^2_0, \qquad \bar\eta_1^2 = \left( \frac{r_0}{\sigma}
\right)^2\,,
\end{equation}
where
\begin{equation}\label{111}
r^2_0 = \left[
\frac{\sigma \cdot Br(K^{+} \to \pi^{+} \nu \bar\nu)}
{\bar\kappa_+ \vcb^4 X^2(x_t)} \right]\,.
\end{equation}
Note that $r_0$ depends only on the top contribution.
The departure of $\varrho_c$ from unity measures the relative importance
of the internal charm contributions. $\varrho_c\approx 1.37$.

Imposing then the constraint from $\vub$
allows to determine
$\bar\varrho$ and $\bar\eta$  with 
\begin{equation}\label{113}
\bar\varrho = \frac{1}{1-\sigma^2} \left( \varrho_c - \sqrt{\sigma^2
\varrho_c^2 +(1-\sigma^2)(r_0^2-\sigma^2 R_b^2)} \right), \qquad
\bar\eta = \sqrt{R_b^2 -\bar\varrho^2}
\end{equation}
where $\bar\eta$ is assumed to be positive.
Consequently
\begin{equation}\label{113aa}
R_t^2 = 1+R_b^2 - 2 \bar\varrho, \qquad
V_{td}=A \lambda^3(1-\bar\varrho-i\bar\eta),\qquad
|V_{td}|=A \lambda^3 R_t.
\end{equation}

The determination of $|V_{td}|$ and of the unitarity triangle in this way
 requires
the knowledge of $\vcb$ (or $A$) and of $|V_{ub}/V_{cb}|$. Both
values are subject to theoretical uncertainties present in the existing
analyses of tree level decays \cite{Battaglia:2003in}. Whereas the dependence on
$|V_{ub}/V_{cb}|$ is rather weak, the very strong dependence of
$Br(\kpn)$ on $A$ or $\vcb$, as seen in (\ref{bkpnn1}) and (\ref{108}),
 made in the past a precise prediction for this
branching ratio and the construction of the UT difficult.
With the more accurate value of $\vcb$ obtained recently \cite{Battaglia:2003in} 
and given in 
(\ref{CKM1}), the situation 
improved significantly.
 We will return to this in Section \ref{sec:num}.
The dependence of $Br(\kpn)$ on $\mt$ is also strong. However, $\mt$
is known already  within $\pm 1\%$ and
consequently the related uncertainty in 
$Br(\kpn)$ is substantially smaller than the corresponding uncertainty 
due to $\vcb$. 

As $\vub$ is subject to theoretical uncertainties, a cleaner strategy 
is to use $Br(K^{+} \to \pi^{+} \nu \bar\nu)$ in conjunction with $\beta$ 
determined through the mixing induced CP asymmetry $a_{\psi K_S}$. We will
investigate this strategy in the next section.

\boldmath
\subsubsection{{$Br(\kpn)$}, {$\beta$}, 
{$\Delta M_d/\Delta M_s$} or 
 {$\gamma$}.}
\unboldmath
In \cite{Buchalla:1998ba} an upper bound on $Br(K^+ \rightarrow \pi^+
\nu \bar{\nu})$ 
has been derived within the SM. This bound depends only on $\vcb$, $X$, 
$\xi$ and $\Delta M_d/\Delta M_s$. With the precise value for the angle 
$\beta$ now available this bound can be turned into a useful formula for 
$Br(K^+ \rightarrow \pi^+ \nu \bar{\nu})$ \cite{D'Ambrosio:2001zh}
that expresses 
this branching ratio in terms of theoretically clean observables. 
In the SM and any MFV model this formula reads:
\be \label{AIACD}
Br(K^+ \rightarrow \pi^+ \nu \bar{\nu})=
\bar\kappa_+~\vcb^4 X^2
\Bigg[ \sigma   R^2_t\sin^2\beta+
\frac{1}{\sigma}\left(R_t\cos\beta +
\frac{\lambda^4P_c(X)}{\vcb^2X}\right)^2\Bigg],
\ee
with $\sigma$ defined in (\ref{109})
 and $\bar\kappa_+$ given in (\ref{kapbar}).
It can be considered as the fundamental formula for a correlation between 
$Br(\kpn)$, $\beta$ and any observable used to determine $R_t$.
This formula is theoretically very clean with
the uncertainties residing only 
in $\vcb$, $P_c(X)$ and $\bar\kappa_+$. However, when one relates $R_t$ to some observable 
new uncertainties could enter. 
In \cite{Buchalla:1998ba} and \cite{D'Ambrosio:2001zh} it has been proposed to express
$R_t$ through $\Delta M_d/\Delta M_s$ by means of (\ref{Rt}).
This implies an additional uncertainty due to
the value of $\xi$ in (\ref{paramet}).

Here we would like to point out that
if the strategy $(\beta,\gamma)$ is used to determine 
$R_t$ by means of (\ref{S3}), the resulting formula that relates 
$Br(\kpn)$, $\beta$ and $\gamma$ is even cleaner than the one 
that relates $Br(\kpn)$, $\beta$ and $\Delta M_d/\Delta M_s$.
We have then

\be \label{AJBNEW}
Br(K^+ \rightarrow \pi^+ \nu \bar{\nu})=
\bar\kappa_+~\vcb^4 X^2
\Bigg[ \sigma   T_1^2+
\frac{1}{\sigma}\left(T_2 +
\frac{\lambda^4P_c(X)}{\vcb^2X}\right)^2\Bigg],
\ee
where
\be\label{T1T2}
T_1=\frac{\sin\beta\sin\gamma}{\sin(\beta+\gamma)}, \qquad
T_2=\frac{\cos\beta\sin\gamma}{\sin(\beta+\gamma)}.
\ee 
Similarly, the following formulae for $R_t$ could be used in conjunction 
with (\ref{AIACD})
\begin{equation}\label{bxnn}
R_t=\frac{\tilde r}{\lambda}
\sqrt{\frac{Br(B\to X_d\nu\bar\nu)}{Br(B\to X_s\nu\bar\nu)}},
\end{equation}
\begin{equation}\label{bmumu}
R_t=\frac{\tilde r}{\lambda}
\sqrt{\frac{\tau({B_s})}{\tau({B_d})}\frac{m_{B_s}}{m_{B_d}}}
\left[\frac{F_{B_s}}{F_{B_d}}\right]
\sqrt{\frac{Br(B_d\to\mu^+\mu^-)}{Br(B_s\to\mu^+\mu^-)}},
\end{equation}
with $\tilde r$ given in (\ref{VTDVTS}).
In particular, (\ref{bxnn}) is essentially free
of hadronic uncertainties \cite{Buchalla:1998ux} and (\ref{bmumu}), not involving
$\hat B_{B_s}/\hat B_{B_d}$, is a bit cleaner than (\ref{Rt}). 
\boldmath
\subsection{$\klpn$, $\bar\eta$, $\imlt$ and the $(\beta,\gamma)$ Strategy}
\subsubsection{$\bar\eta$ and $\imlt$}
\unboldmath
Using (\ref{bklpn}) and (\ref{IMRE}) we find
\begin{equation}\label{bklpn1}
Br(K_{\rm L}\to\pi^0\nu\bar\nu)= \kappa_L
\tilde r^2 A^4 R_t^2 X^2(x_t)\sin^2\beta_{\rm eff}.
\end{equation}

In the context of the unitarity triangle the
expression following from (\ref{bklpn}) and  (\ref{IMREBLO}) is useful:
\begin{equation}\label{BKL}
Br(\klpn)=
\bar\kappa_L \eta^2 \vcb^4 X^2(x_t), \qquad
\bar\kappa_L=\frac{\kappa_L}{\lambda^8}={ (3.39\pm0.03)\cdot 10^{-5} }
\end{equation}
from which { $\bar\eta=\eta (1-\frac{\lambda^2}{2})$} can be determined
\be\label{bareta}
\bar\eta=0.351 
\sqrt{\frac{3.34\cdot 10^{-5}}{\bar\kappa_L}}
\left[\frac{1.53}{X(x_t)}\right]
\left[{\frac{0.0415}{\vcb}}\right]^2
\sqrt{\frac{Br(\klpn)}{3\cdot 10^{-11}}}.
\ee
The determination of $\bar\eta$ in this manner requires the knowledge
of $\vcb$ and $\mt$. With the improved determination of these two
parameters a useful determination of $\bar\eta$ should be possible.

On the other hand, the uncertainty due to $\vcb$ is not present in the 
determination of $\imlt$ as \cite{Buchalla:1996fp}:
\begin{equation}\label{imlta}
\IM\lambda_t=1.39\cdot 10^{-4} 
\left[\frac{\lambda}{0.2248}\right]
\sqrt{\frac{3.34\cdot 10^{-5}}{\bar\kappa_L}}
\left[\frac{1.53}{X(x_t)}\right]
\sqrt{\frac{Br(\klpn)}{3\cdot 10^{-11}}}.
\end{equation}
This formula  offers
 the cleanest method to measure $\IM\lambda_t$ in the SM and all MFV models
in which the function $X$ takes generally different values than $X(x_t)$.
This determination is even better than the one with the help of 
the CP asymmetries
in $B$ decays that require the knowledge
of $\vcb$ to determine $\IM\lambda_t$. Measuring $Br(\klpn)$
with $10\%$ accuracy allows to determine $\IM\lambda_t$
with an error of $5\%$ \cite{Buchalla:1995vs,Buras:1998ra,Buchalla:1996fp}.

The importance of the precise measurement of $\imlt$ is clear: 
the areas $A_{\Delta}$ of all unitarity triangles are equal and related 
to the measure of CP violation $J_{\rm CP}$ \cite{Jarlskog:1985ht,Jarlskog:1985cw}:
\begin{equation}
\left|J_{\rm CP}\right| = 2 A_{\Delta}=
\lambda \left(1-\frac{\lambda^2}{2}\right)|\imlt|.
\end{equation}

\boldmath
\subsubsection{A New ``Golden Relation"}
\unboldmath
Next, in the spirit of the analysis in \cite{Buras:1994rj} we can use 
the clean CP asymmetries in $B$ decays and determine $\bar\eta$ through 
the $(\beta,\gamma)$ strategy. Using (\ref{S1}) and (\ref{S3}) in 
(\ref{bareta})
we obtain a new ``golden relation" 
\be\label{newrel}
\frac{\sin\beta\sin\gamma}{\sin(\beta+\gamma)}
=0.351 
\sqrt{\frac{3.34\cdot 10^{-5}}{\bar\kappa_L}}
\left[\frac{1.53}{X(x_t)}\right]
\left[{\frac{0.0415}{\vcb}}\right]^2
\sqrt{\frac{Br(\klpn)}{3\cdot 10^{-11}}}.
\ee

This relation between $\beta$, $\gamma$ and $Br(\klpn)$, 
is very clean and offers an excellent test of the SM and of its 
extensions. Similarly to the ``golden relation" in (\ref{R7}) it connects 
the observables in $B$ decays with those in $K$ decays. Moreover, it has the
following two important virtues:
\begin{itemize}
\item
It allows to determine $|X|$;
\be\label{modX}
|X|=F_1(\beta,\gamma,\vcb,Br(K_L))
\ee
with $Br(K_L)=Br(\klpn)$. The analytic expression for the function $F_1$ 
can easily be extracted from (\ref{newrel}).
\item
As $X(x_t)$ should be known with high precision once the error on $m_t$ has
been decreased, the relation (\ref{newrel}) allows to determine $\vcb$ with 
a remarkable precision \cite{Buras:1994rj}
\be\label{vcbdet}
\vcb=F_2(\beta,\gamma,X,Br(K_L)).
\ee
The analytic formula for $F_2$ can easily be obtained from 
(\ref{newrel}).
\end{itemize}
At first sight one could question the usefulness of the 
determination of $\vcb$ in this manner, since it is usually determined
from tree level $B$ decays. On the other hand, one should realize
that one determines here actually the parameter $A$ in the Wolfenstein
parametrization that enters the elements $V_{ub}$, $V_{cb}$, $V_{ts}$ and 
$V_{td}$ of the CKM matrix. Moreover this determination of $A$ benefits from 
the very weak dependence on $Br(\klpn)$, which is only with a power of
$0.25$. The weak point of this determination of $\vcb$ is the pollution 
from new physics that could enter through the function $X$, whereas the 
standard determination of $\vcb$ through tree level $B$ decays is free 
from this dependence. Still, a determination of $\vcb$ that in precision 
can almost compete with the usual tree diagrams determinations and 
is theoretically
cleaner, is clearly of interest 
within the SM.

\boldmath
\subsection{Unitarity Triangle from $\kpn$ and $\klpn$}
\label{sec:Kpnn:Triangle}
\unboldmath

The measurement of $Br(\kpn)$ and $Br(\klpn)$ can determine the
unitarity triangle completely (see Fig.~\ref{fig:KPKL}), 
provided $\mt$ and $\vcb$ are known \cite{Buchalla:1994tr}.
Using these two branching ratios simultaneously allows to eliminate
$|V_{ub}/V_{cb}|$ from the analysis which removes a considerable
uncertainty in the determination of the UT, even if it is less important 
for $\vtd$. Indeed it is evident from (\ref{bkpnn}) and
(\ref{bklpn}) that, given $Br(\kpn)$ and $Br(\klpn)$, one can extract
both $\imlt$ and $\relt$. One finds \cite{Buchalla:1994tr,Buchalla:1995vs,Buras:1998ra}
\begin{equation}\label{imre}
\imlt=\lambda^5{\sqrt{B_2}\over X(x_t)},\qquad
\relt=-\lambda^5{{\relc\over\lambda}P_c(X)+\sqrt{B_1-B_2}\over X(x_t)}\,,
\end{equation}
where we have defined the ``reduced'' branching ratios
\begin{equation}\label{b1b2}
B_1={Br(\kpn)\over \kappa_+},\qquad
B_2={Br(\klpn)\over \kappa_L}\,.
\end{equation}
Using next the expressions for $\imlt$, $\relt$ and $\relc$ given
in (\ref{IMREBLO}) and (\ref{REC}) one finds
\begin{equation}\label{rhetb}
\bar\varrho=1+{P_c(X)-\sqrt{\sigma(B_1-B_2)}\over A^2 X(x_t)}\,,\qquad
\bar\eta={\sqrt{B_2}\over\sqrt{\sigma} A^2 X(x_t)}
\end{equation}
with $\sigma$ defined in (\ref{109}). An exact treatment of the CKM
matrix shows that the formulae (\ref{rhetb}), in particular the one 
for $\bar\eta$, are rather precise \cite{Buchalla:1994tr}.

\begin{figure}[hbt]
\vspace{0.10in}
\centerline{
\epsfysize=2.3in
\epsffile{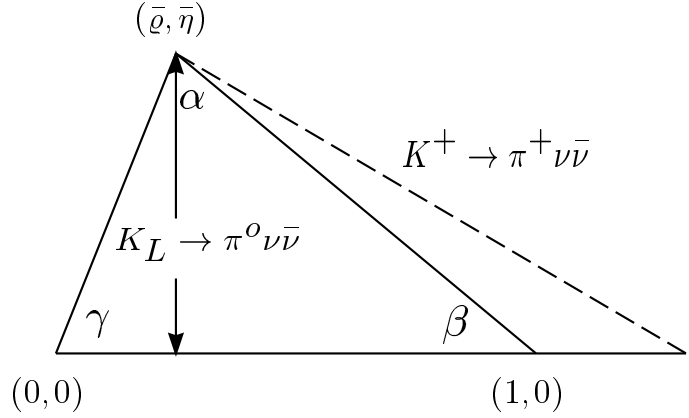}
}
\vspace{0.08in}
\caption{Unitarity triangle from $K\to\pi\nu\bar\nu$.}\label{fig:KPKL}
\end{figure}

\boldmath
\subsection{$\sin 2\beta$ from $K\to \pi\nu\bar\nu$}
\unboldmath
Using (\ref{rhetb}) one finds subsequently \cite{Buchalla:1994tr}
\begin{equation}\label{sin}
\sin 2\beta=\frac{2 r_s}{1+r^2_s}, \qquad
r_s=\sqrt{\sigma}{\sqrt{\sigma(B_1-B_2)}-P_c(X)\over\sqrt{B_2}}
=\cot\beta.
\end{equation}
Thus, within the approximation of (\ref{rhetb}), $\sin 2\beta$ is
independent of $V_{cb}$ (or $A$) and $m_t$ and as we will see in
 Section \ref{sec:num} these dependences are fully negligible.

It should be stressed that $\sin 2\beta$ determined this way depends
only on two measurable branching ratios and on the parameter
$P_c(X)$ which is dominantely calculable in perturbation theory as discussed 
in the previous section. { $P_c(X)$ contains a small non-perturbative
contribution, $\delta P_{c,u}$.}
Consequently this determination is almost free from any hadronic
uncertainties and its accuracy can be estimated with a high degree
of confidence.  The recent calculation of 
NNLO QCD corrections to $P_c(X)$ 
 improved significantly the accuracy of the determination of $\sin 2
\beta$ from the $K\to\pi\nu\bar\nu$ complex.  

Alternatively, combining (\ref{bkpnn1}) and (\ref{bklpn1}), one finds
\cite{Buras:2004ub}
\be\label{sin2bnunu}
\sin 2\beta_{\rm eff}= \frac{2 \bar r_s}{1+\bar r_s^2}, \qquad 
\bar r_s=\frac{\sqrt{B_1-B_2}-\bar P_c(X)}
{\sqrt{B_2}}=\cot\beta_{\rm eff}
\ee
where $\beta_{\rm eff}=\beta-\beta_s$.
As $\beta_s=\ord(\lambda^2)$, we have 
\be
\cot\beta=\sigma\cot\beta_{\rm eff}+ {\ord(\lambda^2)}
\ee 
and consequently one can verify that (\ref{sin2bnunu}), while being 
slightly more accurate, is numerically very close to (\ref{sin}). 
This formula turns out to be more useful than (\ref{sin}) when SM extensions
with new complex phases in $X$ are considered. 
We will return to it in Section \ref{sec:newphys}.

Finally, as in the SM and more generally in all MFV models there are no 
phases beyond the CKM phase, the MFV relation (\ref{R7}) should be satisfied. 
The confirmation of this relation would be a very important test for the 
MFV idea. Indeed, in $K\to\pi\nu\bar\nu$ the phase $\beta$ originates in 
the $Z^0$ penguin diagram, whereas in the case of $a_{\psi K_S}$ in 
the $B^0_d-\bar B^0_d$ box diagram. 
We will discuss the violation of this relation in particular new physics
scenarios in Sections \ref{sec:newphys} and \ref{sec:models}.
\boldmath
\subsection{The Angle $\gamma$ from $K\to\pi\nu\bar\nu$}
\unboldmath
We have seen that a precise value of $\beta$ can be obtained both from 
the CP asymmetry $a_{\psi K_S}$ and from the $K\to\pi\nu\bar\nu$ complex 
in a theoretically clean manner. The determination of the 
angle $\gamma$ is much harder. As briefly discussed in Section \ref{sec:decays} and in great detail in 
\cite{Fleischer:2002ys,Fleischer:2004xw,Ali:2003te,Hurth:2003vb,Nir:2001ge,Buchalla:2003ux}, there are several strategies for $\gamma$ in $B$ decays but 
only few of them can be considered as theoretically clean. They all are 
experimentally very challenging and a determination of $\gamma$ with 
a precision of better than $\pm 5^\circ$ from these strategies alone 
will only be possible at LHCB and after a few years of running 
\cite{Ball:2000ba,Anikeev:2001rk}.  A determination of $\gamma$ with precision of
$\pm (1-2)^\circ$ should be
 possible at Super-B \cite{Super-B}.

Here, we would like to point out that the $K\to\pi\nu\bar\nu$ decays 
offer a clean determination of 
$\gamma$ that in accuracy can compete with the strategies in $B$ decays, 
provided the uncertainties present in $\vcb$, in $m_t$ and in particular in 
$m_c$ present in $P_c$ can 
be further reduced and the two branching ratios measured with an accuracy 
of $5\%$. 

The relevant formula, that has not been presented in the literature 
so far, can be directly obtained from (\ref{rhetb}). It reads
\begin{equation}\label{bgamma}
\cot\gamma=\sqrt{\frac{\sigma}{B_2}}
  \left(A^2 X(x_t)-\sqrt{\sigma(B_1-B_2)}+P_c(X)\right).
\end{equation}
We will investigate 
it numerically in Section \ref{sec:num}.

\boldmath
\subsection{A Second Route to UT from $K\to\pi\nu\bar\nu$}
\unboldmath
Instead of using the formulae for $\imlt$ and $\relt$ in (\ref{imre}), 
it is instructive to construct the UT by using (\ref{sin2bnunu}) to find 
$\beta$ and subsequently determine $R_t$ from (\ref{bkpnn1}) with the result 
\be\label{RtNEW}
R_t=\frac{\sqrt{B_1-\bar P_c^2\sin^2\beta_{\rm eff}}
-\bar P_c \cos\beta_{\rm eff}}{\tilde r A^2X(x_t)}.
\ee 
This $(R_t,\beta)$ strategy by means of $K\to\pi\nu\bar\nu$ decays gives 
then $(\bar\varrho,\bar\eta)$ as given in (\ref{S1}) and in particular 
\be\label{gammaKP}
\cot\gamma=\frac{1-R_t\cos\beta}{R_t\sin\beta}.
\ee

\section{Numerical Analysis in the SM}\label{sec:num}
\setcounter{equation}{0}
\subsection{Introducing Scenarios}
In our numerical analysis we will consider various scenarios for the CKM
elements and the values of the branching ratios $Br(\kpn)$ and $Br(\klpn)$ 
that should be measured in the future. In choosing the values of these 
branching ratios we will be guided in this section by their values predicted 
in the SM. We will consider then 
\begin{itemize}
\item
Scenario A for the present elements of the CKM matrix and a future Scenario 
B with improved elements of the CKM matrix and the improved 
value of $P_c$ through the reduction in the error of $m_c$ and $\alpha_s$. 
They are summarized in table~\ref{SAB}.
The accuracy on $\beta$ in table~\ref{SAB} corresponds to the error in 
$\sin 2\beta$ of $\pm 0.023$ for Scenario A and $\pm 0.013$ for Scenario B. It should be achieved respectively at $B$ factories,
and LHCB. As discussed in \cite{Boos:2004xp}, even at this level of 
experimental precision, theoretical uncertainties in the determination of
$\beta$ through $a_{\psi K_S}$ can be neglected.
The accuracy on $\gamma$ given in table~\ref{SAB} in the Scenarios A and B 
 can presumably  be
achieved through the clean tree diagrams strategies in $B$ decays that will 
only become effective at LHC and Super-B. We will briefly discuss them in 
Section \ref{sec:decays}.
\item
Scenarios I and II for the measurements of $Br(\kpn)$ and $Br(\klpn)$ 
that together with future values of $\vcb$, $m_t$ and $P_c$ should allow the
determination of the UT, that is of the angles $\beta$ and $\gamma$ and of
the sides $R_b$ and $R_t$, from $K\to\pi\nu\bar\nu$ alone. These scenarios 
are summarized in table~\ref{tabin}.
Scenario I corresponds to the first half of the next decade, while Scenario II is more
futuristic. 
\end{itemize}

In the rest of the review we will frequently refer to tables~\ref{SAB} and 
\ref{tabin} indicating which observables listed there are used at a given
time in our numerical calculations.

\begin{center}
\begin{table}
\centering
\caption[]{Input for  the determination of the branching ratios
$Br(\kpn)$ and $Br(\klpn)$ in three scenarios. The corresponding 
$(\bar\varrho,\bar\eta)$ are given too. \label{tab:scenarios}
\label{SAB}}
\begin{tabular}{|c|c|c|}
\hline
& {\rm Scenario A} & {\rm Scenario B} \\ 
\hline
\hline
$\beta$  & $ (22.2\pm 0.9)^\circ$ & $(22.2\pm 0.5)^\circ$\\ 
\hline
$\gamma$  & $ (64.6\pm 4.2)^\circ $ &$ (64.6\pm 2.0)^\circ $\\
\hline
$\vcb/10^{-3}$ &  $ 41.6\pm 0.6$ &$ 41.6\pm 0.3$\\
\hline
$R_b$  & $ 0.381\pm0.014$ &$ 0.381\pm 0.007$\\
\hline 
$m_t[{\rm GeV}]$     & $161\pm 1.7$  & $ 161\pm 1.0$ \\
\hline 
$P_c(X)$  & $0.41\pm 0.05 $ &$ 0.41\pm 0.02$\\
\hline
\hline
$\bar\eta$  & $0.344\pm 0.016$ & $0.344\pm 0.008$ \\
\hline
$\bar\varrho$  & $0.163\pm 0.028$ & $0.163\pm 0.014$ \\
\hline
\end{tabular}
\end{table}
\end{center}

\begin{center}
\begin{table}
\centering
\caption[]{Input for  the determination of CKM
parameters from $K\to\pi\nu\bar\nu$ in two scenarios.
\label{tabin}}
\begin{tabular}{|c|c|c|}
\hline
& {\rm Scenario I} & {\rm Scenario II}\\ 
\hline
\hline
$Br(\kpn)/10^{-11}$ & $ 8.0\pm 0.8$&$ 8.0\pm 0.4$\\ 
\hline
$Br(\klpn)/10^{-11}$ & $ 3.0\pm 0.3$ &$ 3.0\pm 0.2$\\
\hline 
$m_t[{\rm GeV}]$     & $161\pm 1.7$  & $ 161\pm 1.0$ \\
\hline 
$P_c(X)$  & $0.41\pm 0.05 $ &$ 0.41\pm 0.02$\\
\hline
$\vcb/10^{-3}$ & $ 41.6\pm 0.6$ &$ 41.6\pm 0.3$\\
\hline
\end{tabular}
\end{table}
\end{center}

\subsection{Branching Ratios in the SM}

With the CKM parameters of Scenario A given in table
\ref{tab:scenarios} we find
using (\ref{bkpnn}) and (\ref{bklpn}) 
{ \begin{equation}\label{SMkp1}
Br(\kpn)_{\rm SM}=(8.1 \pm 0.6_{P_c} \pm 0.5 )\cdot 10^{-11}=
(8.1 \pm 1.1 )\cdot 10^{-11},
\ee
\be\label{SMkl1}
 Br(\klpn)_{\rm SM}=
(2.6 \pm 0.3)\cdot 10^{-11} .
\end{equation}}

The parametric errors come from the CKM parameters and the value of 
$m_t$ and have been added in quadrature. In the case of $Br(\klpn)$ only parametric
uncertainties matter. 
For $Br(\kpn)$ in the SM (\ref{SMkp1}) we additionally have the error due to $P_c(X)$ which was added
linearly.\\

The central value of $Br(\kpn)$ in (\ref{SMkp1}) is below the central 
experimental value
in (\ref{EXP1}), but within theoretical, parametric and experimental 
uncertainties, the SM result is fully consistent with the data.
We also observe that the error in $P_c(X)$ constitutes still a significant 
portion of the full error. 

 \begin{table}[t]
\caption[]{ Values of $Br(\kpn)$ and $Br(\klpn)$ in the
SM in units of $10^{-11}$ obtained through various strategies 
described in the text.
\label{SCENA}}
\begin{center}
\begin{tabular}{|c|c|c|}\hline
 Strategy & $Br(\kpn)$ $[10^{-11}]$ & $Br(\klpn)$ $[10^{-11}]$ 
 \\ \hline

{Scenario A} & $8.10\pm 1.11$                & $2.64\pm 0.30$ \\
            & $8.10\pm 0.62_{P_c}\pm 0.49$ & \\
{Scenario B} & $8.10\pm  0.52 $ & $2.64\pm 0.15$ \\
           & $8.10\pm 0.25_{P_c}\pm 0.27$ &      \\
\hline
 \end{tabular}
\end{center}
\end{table}

One of the main origins of the parametric uncertainties in both branching 
ratios 
is the value of $\vcb$. As pointed out in  \cite{Kettell:2002ep}
with the help of $\varepsilon_K$ the dependence on $\vcb$ can be eliminated.
Indeed, from the expression for $\varepsilon_K$ in 
(\ref{eq:epsformula}) and the relation
\be
\frac{\IM\lambda_t}{\RE\lambda_t}=-\tan\beta_{\rm eff}, \qquad 
\beta_{\rm eff}=\beta-\beta_s,
\ee
that follows from (\ref{IMRE}), $\IM\lambda_t$ and $\RE\lambda_t$ can 
be determined subject mainly to the uncertainty in $\hat B_K$ that 
should be decreased through lattice simulations in the future.
Note that $\beta$ will soon be  determined with high precision from
the $a_{\psi K_S}$ asymmetry.

{ \begin{table}[t]
\caption[]{ The anatomy of parametric uncertainties in 
$Br(\kpn)$ and $Br(\klpn)$ corresponding to the results of table~\ref{SCENA}.
\label{ANATOMY}}
\begin{center}
\begin{tabular}{|c|c|c|}\hline
Strategy & ${\bf \sigma}Br(\kpn)$ $[10^{-11}]$ & ${\bf \sigma}Br(\klpn)$ $[10^{-11}]$ 
 \\ \hline
{ Scenario A} & $\pm 0.33_{\bar\varrho}\pm 0.06_{\bar\eta}\pm 0.33_{\vcb} 
\pm 0.13_{m_t} $   & $\pm 0.25_{\bar\eta}\pm 0.16_{\vcb}\pm 0.06_{m_t}$ \\
{ Scenario B}& $\pm 0.17_{\bar\varrho}\pm 0.03_{\bar\eta}\pm 0.16_{\vcb} 
\pm 0.08_{m_t} $   & $\pm 0.12_{\bar\eta}\pm 0.08_{\vcb}\pm 0.04_{m_t}$ \\
\hline
 \end{tabular}
\end{center}
\end{table}}

We can next investigate what kind of predictions one will get in a few years 
when $\beta$ and $\gamma$ will be  measured with high precision 
through theoretically clean strategies at LHCB \cite{Ball:2000ba} and 
BTeV \cite{Anikeev:2001rk}.
As pointed out in \cite{Buras:2002yj},
the use of $\beta$ and $\gamma$  is the most powerful
strategy to get $(\bar\varrho,\bar\eta)$.
With the input of Scenario B of table~\ref{SAB}, 
we find 
{ \be
 \imlt= (1.38\pm 0.04)\cdot 10^{-4}, 
\qquad \relt=-(3.19\pm 0.07)\cdot 10^{-4}\quad  {\rm (Scenario~~B)}
\ee}

The results for the branching ratios in this scenario 
is given in table~\ref{SCENA}, where we have separated the error due to $P_c$ 
from the parametric uncertainties.

In table~\ref{ANATOMY} we present the anatomy of parametric uncertainties 
given in table~\ref{SCENA}. Adding these uncertainties in quadrature gives 
the values in table~\ref{SCENA}. We observe that $\vcb$ plays a prominent
role in these uncertainties.

\begin{figure}[hbt]
\vspace{0.10in}
\centerline{
\epsfysize=2.7in
\epsffile{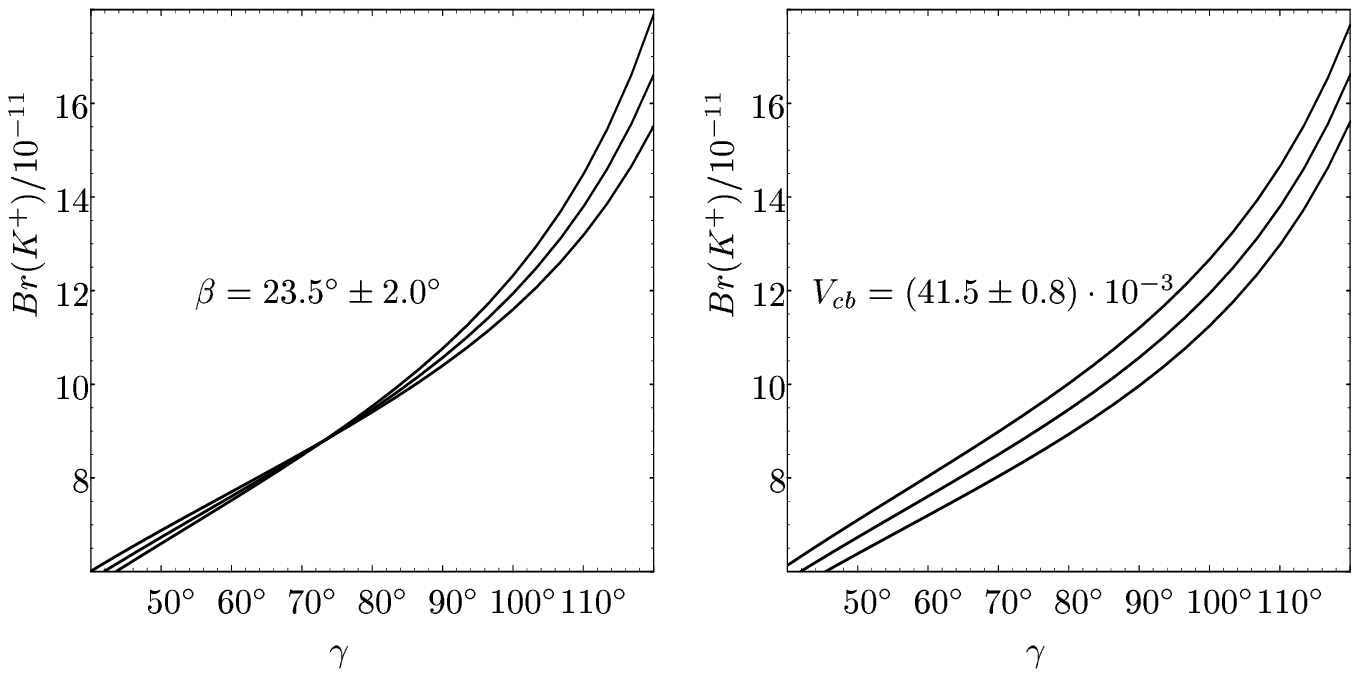}
}
\vspace{0.08in}
\caption{$Br(\kpn)$ as a function of $\gamma$ for different values of
$\beta$ and $\vcb$.}\label{KPGAMMA}
\end{figure}

Finally in Fig.~\ref{KPGAMMA} we show $Br(\kpn)$ as a function of 
$\gamma$ for different 
values of $\beta$ and $\vcb$. We observe that 
the dependence on $\beta$ is rather weak, while the dependence on $\gamma$ 
is very strong. Also the dependence on $\vcb$ is significant.
This implies that a precise measurement of $\gamma$ one day will also 
have a large impact on the prediction for $Br(\kpn)$.

\boldmath
\subsection{Impact of $Br(\kpn)$ on the UT} 
\unboldmath
\subsubsection{Preliminaries}
Let us then reverse the analysis and investigate the impact of present
and future
measurements of $Br(\kpn)$ on $\vtd$ and on the UT. To this end one can take 
as additional inputs the values of $\vcb$ and $\beta$. 
One finds immediately that now a precise value of $\vcb$ is required in
order to obtain a satisfactory result for $(\bar\varrho,\bar\eta)$. Indeed 
$K\to \pi\nu\bar\nu$ decays are excellent means to determine $\imlt$ and
$\relt$ or equivalently the ``$sd$" unitarity triangle and in this respect have
no competition from any $B$ decay, but in order to construct the standard
``$bd$" triangle of Fig.~\ref{fig:utriangle} from these decays, $\vcb$ is
required. Here the CP-asymmetries in $B$ decays measuring directly angles of
the UT are superior as the value of $\vcb$ is not required. Consequently the
precise value of $\vcb$ is of utmost importance if we want to make useful
comparisons between various observables in $K$ and $B$ decays. 
 On the other
hand, in some relations such as (\ref{R7}), the $\vcb$ dependence is absent 
to an excellent accuracy.

\boldmath
\subsubsection{$\vtd$ from $K^+\to\pi^+\nu\bar\nu$}
\unboldmath
Taking  the present experimental value of 
$Br(\kpn)$ in (\ref{EXP1}),
we determine first the UT side $R_t$ and next
the CKM element $|V_{td}|$. 
Using then the accurate expression for $Br(\kpn)$ 
in (\ref{AIACD}) and the  values of $\vcb$ and $\beta$ 
in the present Scenario A of table~\ref{SAB},
 we find 
\be\label{RT1}
 R_t= 1.35\pm 0.70, \qquad   \vtd= (12.6\pm 6.6)\cdot 10^{-3}~,
\ee 
where the dominant error arises due to the error in the branching ratio. 
The central values obtained here are large compared to the SM ones, but in view
of the large errors one cannot say anything conclusive yet.

We consider then Scenarios I and II of table~\ref{tabin} but do not take yet 
the values for $Br(\klpn)$ into account. As an additional variable we take 
$\beta$ or $R_b$ in the Scenario B  of table~\ref{SAB}. In table~\ref{RTVTD}
we give the values of $R_t$ and $\vtd$ resulting from this exercise. 
The precise value of $\beta$ or $R_b$ does not matter much 
in the determination of $R_t$ and $\vtd$, which is evident from 
the inspection of the $(\bar\varrho,\bar\eta)$ plot. This is 
also the reason why  with the assumed errors on $\beta$ and $R_b$ 
the two exercises in table~\ref{RTVTD} give essentially the same results.

\begin{center}
\begin{table}
\centering
\caption[]{The values for $R_t$ and $\vtd/10^{-3}$ (in parentheses)  
from $\kpn$ for various cases considered in the text.
\label{RTVTD}}
\begin{tabular}{|c|c|c|}
\hline
& {\rm Scenario I} & {\rm Scenario II}\\ 
\hline
\hline
{ Scenario B ($\beta$) }& $ 0.897\pm 0.086~(8.42\pm 0.80)$&  
$ 0.897\pm 0.056~(8.42\pm 0.51)$  \\ 
\hline
{ Scenario B ($R_b$)} & $ 0.897\pm 0.086~(8.42\pm 0.80)$&  
$ 0.897\pm 0.056~(8.42\pm 0.51)$  \\ 
\hline
\end{tabular}
\end{table}
\end{center}

In order to judge the precision achievable in the future, 
it is instructive to 
show the separate contributions of the uncertainties involved. 
In general, $\vtd$ is subject to 
various uncertainties of which the dominant ones are given below 
\be\label{vtda}
\frac{\sigma(\vtd)}{\vtd}=
\pm 0.39 \frac{\sigma(P_c)}{P_c}
\pm 0.70 \frac{\sigma( Br(K^+))}{Br(K^+)} \pm \frac{\sigma(\vcb)}{\vcb}~.
\ee

We find then
\be\label{vtdI}
 \frac{\sigma(\vtd)}{\vtd}=
\pm 5.0\%_{P_c} \pm 7.0\%_{Br(K^+)} \pm 1.4\%_{\vcb},\qquad {\rm (Scenario~I)} 
\ee
and
\be\label{vtdII}
\frac{\sigma(\vtd)}{\vtd}=
\pm 2.0\%_{P_c} \pm 3.5\%_{Br(K^+)} \pm 1.0\%_{\vcb}. \qquad {\rm
(Scenario~II)}
\ee
Adding the errors in quadrature, we find that
$\vtd$ can be determined with an accuracy of $\pm 8.7\%$ and 
$\pm 4.2\%$, respectively.
These numbers are increased to $\pm 9.2\%$  and $\pm 4.3\%$ 
once the uncertainties due to $\mt$,
$\alpha_s$ and $\beta$ (or $\vub$) are taken into account. As a 
measurement of $Br(\kpn)$ with a precision of $5\%$ is 
very challenging, the 
determination of $\vtd$ with
an accuracy better than $\pm 5\%$ from $Br(\kpn)$ seems very difficult from 
the present perspective.

\subsubsection{Impact on UT}
The impact of $\kpn$ on the UT is illustrated in Fig.~\ref{fig:figmfv}, 
where  we show the lines corresponding to
several selected values of $Br(\kpn)$.
The construction of the UT from both decays shown there is described below.

\begin{figure}[htb!]
\begin{center}
\vspace{0.5cm}
{\epsfig{figure=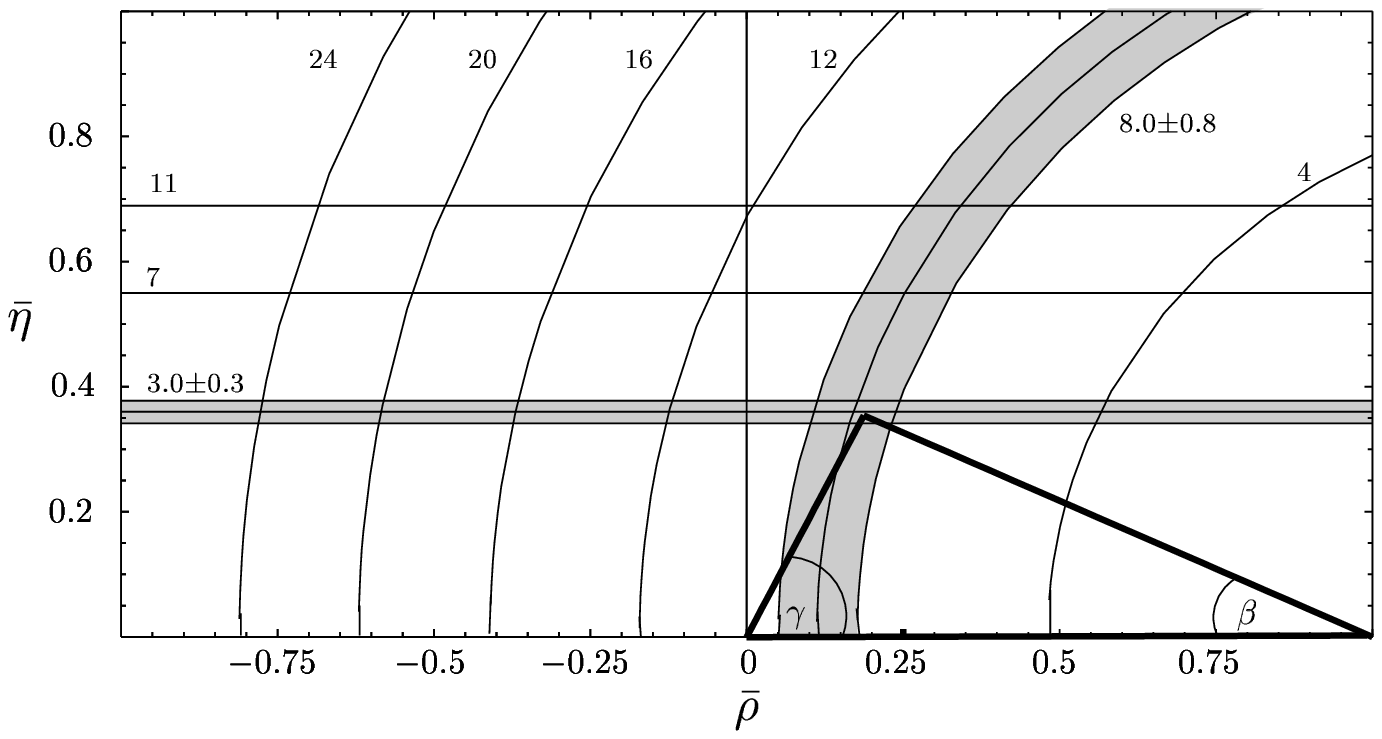,height=6.5cm}}
\caption[]{
The UT from $K\to\pi\nu\bar\nu$ in Scenario I of 
table~\ref{tabin}. Also lines corresponding to several values of $Br(\kpn)$ 
and $Br(\klpn)$ (in units of $10^{-11}$) are shown.
}
\label{fig:figmfv}
\end{center}
\end{figure}

\boldmath
\subsection{Impact of $Br(\klpn)$ on the UT } 
\unboldmath
\boldmath
\subsubsection{$\bar\eta$ and $\imlt$}
\unboldmath
We consider next the impact of a future measurement of $Br(\klpn)$ 
on the UT. As already discussed in the previous section, this measurement 
will offer a theoretically clean determinations of $\bar\eta$ and in
particular of $\imlt$. The relevant formulae are given in
(\ref{bareta}) and (\ref{imlta}), respectively. 
Using Scenarios I and II of table~\ref{tabin}
 we find
\be\label{KLR1}
 \bar\eta= 0.367\pm 0.019, \qquad \imlt=(1.47\pm 0.07)\cdot 10^{-4} 
\qquad {\rm (Scenario~I)}.
\ee
\be\label{KLR2}
\bar\eta= 0.367\pm 0.013, \qquad \imlt=(1.47\pm 0.05)\cdot 10^{-4}.\qquad
{\rm (Scenario~II)}
\ee
The obtained precision in the case of Scenario II is truely impressive.
We stress the very clean character of these 
determinations.

\subsubsection{Completing the Determination of the UT}
In order to construct the UT we need still another input. It could be 
$\beta$, $\gamma$, $R_b$ or $R_t$. It turns out that the most effective 
in this determination is $\gamma$, as in the classification of \cite{Buras:2002yj} 
the $(\bar\eta,\gamma)$ strategy belongs to the top class together with 
the $(\beta,\gamma)$ pair. The angle $\gamma$ should be known with high
precision in five years.
Still it is of interest to see what one finds when $\beta$ instead of
$\gamma$ is used. $R_b$ is not useful here as it generally gives two 
solutions for the UT.

In analogy to table~\ref{RTVTD} we show in table~\ref{RHOVTD} the 
values of $\bar\varrho$ and $\vtd$ resulting from Scenarios I and II 
without using $Br(\kpn)$. As an additional variable we use $\beta$ or 
$\gamma$. We observe that, with the assumed errors on $\beta$ and $\gamma$, 
the use of $\gamma$ is more effective than the use of $\beta$. Moreover, while 
going from Scenario I to Scenario II for $Br(\klpn)$ has a significant 
impact when $\beta$ is used, the impact is rather small when $\gamma$ 
is used instead. Both features are consistent with the observations 
made in \cite{Buras:2002yj} in the context of $(\beta,\bar\eta)$ and  
$(\gamma,\bar\eta)$ strategies. In particular, the last feature 
is directly related to the fact that $\gamma$ is by 
a factor of three larger than $\beta$.

The main message from table~\ref{RHOVTD} is that,
 using a rather precise value of $\gamma$, a
   very precise determination of $\vtd$ becomes possible,
   where the branching fraction of $\klpn$ needs to be known only to
   about $10 \%$ accuracy.

\begin{center}
\begin{table}
\centering
\caption[]{The values for $\bar\varrho$ and $\vtd/10^{-3}$ (in parentheses)   
from $\klpn$ for various cases considered in the text.
\label{RHOVTD}}
\begin{tabular}{|c|c|c|}
\hline
& {\rm Scenario I} & {\rm Scenario II}\\ 
\hline
\hline
 Scenario B ($\beta$) & $ 0.101\pm 0.052~(9.12\pm 0.51)$&  
$ 0.101\pm 0.040~(9.12\pm 0.37)$  \\ 
\hline
Scenario B ($\gamma$) & $ 0.174\pm 0.018~(8.49\pm 0.16)$&  
$ 0.174\pm 0.017~(8.49\pm 0.16)$  \\
\hline
\end{tabular}
\end{table}
\end{center}

\boldmath
\subsubsection{A Clean and Accurate Determination of $\vcb$ and $\vtd$} 
\unboldmath
Next, combining $\beta$ and $\gamma$ with the values of $Br(\klpn)$ and 
$m_t$, a clean determination of $\vcb$ 
by means of (\ref{vcbdet}) is possible. In turn also $\vtd$ can be
determined. In table~\ref{VCBVTD} we show the values of $\vcb$ and $\vtd$ 
obtained using Scenarios I and II for $Br(\klpn)$ in table~\ref{tabin} 
with $\beta$ and $\gamma$ in Scenario B of table~\ref{SAB}. 

We observe that the errors on $\vcb$ 
are larger than presently obtained from semi-leptonic $B$ decays.
But one should emphasize that this determination is essentially without
any theoretical uncertainties. The high precision on $\vtd$ is a result 
of a very precise measurement of $R_t$ by means of the $(\beta,\gamma)$
strategy and a rather accurate value of $\vcb$ obtained with the 
help of $Br(\klpn)$. Again also in this case the determination is 
theoretically very clean.

\begin{center}
\begin{table}[thb]
\centering
\caption[]{The values for $\vcb$ and $\vtd$ (in parentheses) in units 
of $10^{-3}$  
from $\klpn$, $\beta$ and $\gamma$ for various cases considered in the text.
\label{VCBVTD}}
\begin{tabular}{|c|c|c|}
\hline
& {\rm Scenario I} & {\rm Scenario II}\\ 
\hline
\hline
{ Scenario B } & $ 43.1\pm 1.2~(8.23\pm 0.76)$&  
$ 43.1\pm 0.9~(8.23\pm 0.48)$  \\ 
\hline
\end{tabular}
\end{table}
\end{center}

\boldmath
\subsection{Impact of $Br(\kpn)$ and $Br(\klpn)$ on UT} 
\unboldmath
In \cite{Buchalla:1996fp} the 
determination of the UT from both decays has been discussed 
in explicit terms. The relevant formulae have been given in Section
\ref{sec:phen}. Here we confine our discussion to the determination of $\imlt$,
$\sin 2\beta$ and $\gamma$.
We consider again two scenarios for which the input parameters are collected 
in table~\ref{tabin}. 
This time no other parameters beside those given in this table are required 
for the construction of the UT and the determination of these three quantities
in question.

\boldmath
\subsection{$\imlt$ from $\klpn$}
\unboldmath
As oposed to $\sin 2\beta$ and $\gamma$ only $\klpn$ is relevant here. Using
(\ref{imlta}) we find that the error from $m_t$ is roughly $1 \% $ and will soon
be decreased even below that. Neglecting it, we find
\be
\frac{\sigma (\imlt)}{\imlt}= \pm 0.5 \frac{\sigma (Br(K_L))}{Br(K_L)}=\left\{\begin{array}{ll}
5.0 \% &  {\rm Scenario \,I}\\
3.3 \% &  {\rm Scenario \,II}
\end{array}\right.
\ee
which already in the case of Scenario II is an impressive accuracy.

\boldmath
\subsection{The Angle $\beta$ from $K\to\pi\nu\bar\nu$}
\unboldmath
Let us next investigate the separate uncertainties in the 
determination of $\sin 2\beta$ coming from $P_c$, 
$Br(\kpn)\equiv Br(K^+)$ and $Br(\klpn)\equiv Br(K_L)$. 
We find first 
\be\label{sinerr}
\frac{\sigma(\sin 2\beta)}{\sin 2\beta}=\pm 0.31\frac{\sigma(P_c)}{P_c}
\pm 0.55 \frac{\sigma( Br(K^+))}{Br(K^+)}
\pm 0.39 \frac{\sigma( Br(K_L))}{Br(K_L)}~.
\ee
This leads to
\be\label{sinI}
 \sigma(\sin 2\beta)=0.030_{P_c}+0.041_{Br(K^+)}+0.029_{Br(K_L)}
=0.080 \qquad ({\rm Scenario~I})
\ee
and 
\be\label{sinII}
\sigma(\sin 2\beta)=0.011_{P_c}+0.020_{Br(K^+)}+0.018_{Br(K_L)}
=0.038, \qquad ({\rm Scenario~II})
\ee
where the
errors have been added in quadrature  apart from the one in $P_c$ which has been
added linearly. The uncertainties due to 
$\vcb$ and $m_t$ are fully negligible.

We observe that 
\begin{itemize}
\item
The uncertainty in $\sin 2\beta$ due to $P_c$ alone amounted to
$0.04$ at NLO, implying that a NNLO calculation of $P_c$ was very desirable. 
On the other hand, now, at NNLO, the pure perturbative uncertainty in $\sin 2 \beta$
 amounts to $\pm 0.006 \%$ \cite{Buras:2006gb}, to be compared with $\pm 0.025 \%$ at NLO.
\item
The accuracy of the determination of $\sin 2\beta$, 
 after the NNLO result became available, 
 depends dominantly on the accuracy with which both branching
ratios will be
measured.  In order to decrease $\sigma(\sin 2\beta)$ down to 
$0.02$ they have to be measured with an accuracy 
 better than $5\%$. Also, the reduction of the error in $m_c$ 
relevant for $P_c$ would be desirable.
\end{itemize}

\boldmath
\subsection{The Angle $\gamma$ from $K\to\pi\nu\bar\nu$}
\unboldmath
Let us next investigate, in analogy to (\ref{sinerr}), 
the separate uncertainties 
in the 
determination of $\gamma$ coming from $P_c$, 
$Br(\kpn)$, $Br(\klpn)$ and $\vcb$. 
The relevant expression for $\gamma$ in terms of these quantities 
is given in (\ref{bgamma}). We find then 
\be\label{gamerr}
\frac{\sigma(\gamma)}{\gamma}=\pm 0.75\frac{\sigma(P_c)}{P_c}
\pm 1.32 \frac{\sigma( Br(K^+))}{Br(K^+)}
\pm 0.07 \frac{\sigma( Br(K_L))}{Br(K_L)}\pm 4.11\frac{\sigma(\vcb)}{\vcb}
\pm 2.34\frac{\sigma(m_t)}{m_t}.
\ee

This gives
\be\label{gamI}
\sigma(\gamma)=5.7^\circ_{P_c}+8.2^\circ_{Br(K^+)}+0.4^\circ_{Br(K_L)}
+3.7_{\vcb}^\circ +1.5_{m_t}^\circ= 19.6^\circ
\ee
and 
\be\label{gamII}
 \sigma(\gamma)=2.3^\circ_{P_c}+4.1^\circ_{Br(K^+)}+0.3^\circ_{Br(K_L)}
+1.9_{\vcb}^\circ + 0.9_{m_t}^\circ= 9.4^\circ
\ee
for Scenario I and II, respectively, where the
errors have been added in quadrature.

We observe that 
\begin{itemize}
\item
The uncertainty in $\gamma$ due to $P_c$ alone amounted to
 $8.6^\circ$ at the NLO level, implying that a NNLO calculation of $P_c$ was very desirable. The pure
perturbative uncertainty in $\gamma$ amounts to $\pm 1.2 \%$ at NNLO, compared
to $\pm 4.9 \%$ at NLO. Again, the reduction of the error in $m_c$ 
relevant for $P_c$ would be desirable.
\item
The dominant uncertainty in the determination of $\gamma$ in Scenarios I and
II besides the one of $P_c$ resides in $Br(\kpn)$. In order to lower $\sigma(\gamma)$ below $5^\circ$,
a measurement of this branching ratio with an accuracy of better than $5\%$ 
is required. The measurement of $Br(\klpn)$ has only a small impact on this 
determination.
\end{itemize}

\

\subsection{Summary}
In this section we have presented a very detailed numerical analysis of the
formulae of Section \ref{sec:phen}. First working in two scenarios, A and B, for the
input parameters that should be measured precisely through $B$ physics
observables in this decade, we have shown how the accuracy on the predictions
of the branching ratios will improve with time. 

In the case of $Br(\klpn)$ there are essentially no theoretical uncertainties
and the future of the accuracy of the prediction on this branching ratio
within the SM depends fully on the accuracy with which $\imlt$ and $m_t$ can be
determined from other processes. We learn from table~\ref{SCENA} that the
present error of roughly $12\%$ will be decreased to  $6\%$ when the
Scenario B will be realized.
 As seen in table~\ref{ANATOMY}, 
the progress on the error on $Br(\klpn)$ will depend importantly on the
progress on $\vcb$.

The case of $\kpn$ is a bit different as now also the uncertainty in $P_c$
enters. As discussed in Section \ref{sec:basic}, 
this uncertainty comes on the one hand from
the scale uncertainty and on the other  hand from the error in $m_c$. 
The scale uncertainty dominated at NLO
 while the error on $m_c$ is mainly responsible for the present error in $P_c$
 after NNLO has been completed. 
Formula (\ref{FS})
quantifies this explicitely.
The anatomy of parametric uncertainties in $Br(\kpn)$ is presented in
table~\ref{ANATOMY}. As in the case of $Br(\klpn)$ also here the reduction of
the error in $\vcb$ will be important.

As seen in table~\ref{SCENA} the present error in $Br(\kpn)$ due to $P_c$
amounts roughly to $\pm 8\%$, which is  roughly by a factor of  1.5 smaller than 
before the NNLO results for $P_c$ where available. 
It is also clearly seen in this 
table that in order to benefit from the improved values of the CKM parameters
and of $\mt$, also the uncertainty in $P_c$ has to be reduced through the improvement of $m_c$. It appears to us that the present
error of $8\%$ due to $P_c$ could be decreased to $3\%$ 
one day with the present total error of $14\%$ reduced to $7\%$.

In the main part of this section we have investigated the impact of the
future measurements of $Br(\kpn)$ and $Br(\klpn)$
on the determination of the CKM matrix. 
The results are self-explanatory and demonstrate very clearly
that the $K\to\pi\nu\bar\nu$ decays offer powerful means in the
determination of the UT and of the CKM matrix.

Clearly, the future determination of various observables by means of 
$K\to\pi\nu\bar\nu$ will crucially depend on the accuracy with which 
$Br(\kpn)$ and $Br(\klpn)$ can be measured.
Our discussion shows that it is certainly desirable to measure both
branching ratios with an accuracy of at least $5\%$.

On the other hand the uncertainties due to $P_c$, $\vcb$ and to a lesser
extent $m_t$ are also  important ingredients of these investigations.

\boldmath
\section{A Guide to Sections \ref{sec:MFV}-\ref{sec:models}}\label{sec:guide}
\unboldmath

Until now our discussion was confined to the SM. In the next three sections we will 
discuss the decays $K\to\pi\nu\bar\nu$ in various extensions of the SM.

In the case of most $K$ and $B$ meson decays the effective Hamiltonian in the 
extensions of the SM becomes generally much more complicated than in the SM in that
new operators, new complex phases, and new one-loop short distance functions and generally
new flavour violating couplings can be present. A classification of various
possible extensions of the SM from the point of view of an effective Hamiltonian
and valid for all decays can be found in \cite{Buras:2005xt}.

As we already emphasized at the beginning of this review in the case of 
$K\to\pi\nu\bar\nu$, the effective Hamiltonian in essentially all extensions of
the SM is found simply from ${\cal H}^{\rm SM}_{\rm eff}$ in (\ref{hkpn}) by replacing $X(x_t)$ as
follows \cite{Buras:1997ij}
\begin{equation}\label{repl}
X(x_t) \rightarrow X = \left| X \right| e^{i \theta_X}.
\end{equation}
Thus, the only effect of new physics is to modify the magnitude of the SM function 
$X(x_t)$ and/or introduce a new complex phase $\theta_X$ that vanishes in the SM.

Clearly, the simplest class of extensions are models with minimal flavour
violatio
in which $\theta_X=0,\pi$ and $ \left| X \right|$ is only modified by loop diagrams
with new particles exchanges but the driving mechanism of flavour and CP violation 
remains to be the CKM matrix. As in this class of models the basic structure of 
effective Hamiltonians in other decays is unchanged relative to the SM and only the 
modifications in the one-loop functions, analogous to $X$, are allowed, the correlations 
between $K\to\pi\nu\bar\nu$ and other $K-$ and, in particular, $B-$decays, valid in the 
SM remain true. A detailed review of these correlations has been given in 
{\cite{Buras:2003jf}}.

In the following section, we will summarize the present status of $K\to\pi\nu\bar\nu$ in
the models with MFV. As we will see, the recently improved bounds on rare $B$ decays, 
combined with the correlations in question, do not allow for a large departure of 
$K\to\pi\nu\bar\nu$ from the SM within this simplest class of new physics.

Much more spectacular effects in $K\to\pi\nu\bar\nu$ are still possible in models in 
which the phase $\theta_X$ is large. We discuss this in Section \ref{sec:newphys} in a model independent manner. We also discuss there 
situations in which simultaneously to $\theta_X \not= 0$, also new complex phases in 
$B_d^0- \bar B_d^0$ mixing are present, and illustrate how these new phases, including 
$\theta_X$, could be extracted from future data. 

While Section \ref{sec:MFV} and \ref{sec:newphys} have a more model independent
character and 
basically analyze the implications of the replacement (\ref{repl}) with arbitrary 
$\left| X \right|$ and $\theta_X$, Section \ref{sec:models} can be considered as a 
guide to the rich literature on the new physics effects in $K\to\pi\nu\bar\nu$. In
particular, we discuss the Littlest Higgs Model with
T-parity,  $Z^\prime$ models, the MSSM with MFV, general supersymmetric models, models with
universal extra dimensions and models with lepton flavor mixing. Finally, we briefly 
comment on essentially all new physics analyses done until the summer of 2007.

\boldmath
\section{\boldmath{$K\to\pi\nu\bar\nu$} and MFV}\label{sec:MFV}
\unboldmath
\setcounter{equation}{0}
\subsection{Preliminaries}
A general discussion of the decays $\kpn$ and $\klpn$ in the framework of
minimal flavour violation (MFV) has been presented in \cite{Buras:2001af}. 
Earlier papers in specific MFV scenarios like two Higgs doublet 
can be found in \cite{Belanger:1992ti,Cho:1998bf}, where additional references are given.
We would like to recall that in almost all extensions of the SM the effective
Hamiltonian for $K \to \pi \nu \bar \nu$ decays involves only the $(V-A)\otimes
(V-A)$ operator of (\ref{hkpn}) and consequently for these decays there is no
distinction between the constrained MFV (CMFV) \cite{Buras:2000dm,Blanke:2006ig} and more general
formulation of MFV \cite{D'Ambrosio:2002ex} in which additional non-SM operators are present in
certain decays. Consequently in MFV or CMFV, all formulae of Section
\ref{sec:basic} and \ref{sec:phen} for $K \to \pi \nu \bar \nu$  remain valid except that
\begin{itemize}
\item
the function $X(x_t)$ is replaced by the real valued 
master function 
\cite{Buras:2005xt,Buras:2003jf,Buras:2004sc} $X(v)$ with $v$ denoting collectively the parameters
of a given MFV model,
\item
if the function $X(v)$ is allowed to take also negative values, the 
following replacements should effectively be made in all formulae of 
Sections \ref{sec:basic} and \ref{sec:phen} \cite{Buras:2001af}
\be
X \to |X|, \qquad P_c(X) \to {\rm sgn}(X) P_c(X).
\ee
\end{itemize}
 Here we will also assume that the $B^0-\bar B^0$ function $S(v)>0$, as in the SM. In fact 
as found recently in \cite{Blanke:2006yh,Altmannshofer:2007cs} in all models
with CMFV $S(v)>S(v)_{\rm SM}$.
On the other hand, we will allow first for negative 
values of the function $X(v)$. The values of $X(v)$ and $S(v)$ can be calculated 
 in any MFV
model.

\boldmath
\subsection{$\kpn$ versus $\klpn$}
\unboldmath
An important consequence of (\ref{sin}) and (\ref{R7}) is the following
MFV relation \cite{Buras:2001af}
\be\label{B1B2}
B_1=B_2+\left[\frac{\cot\beta\sqrt{B_2}+
{\rm sgn}(X)\sqrt{\sigma}P_c(X)}
{\sigma}\right]^2,
\ee
that, for a given $\sin 2\beta$ extracted from $a_{\psi K_S}$ and $Br(\kpn)$, 
allows to predict $Br(\klpn)$. We observe that 
in the full class of MFV models, independent of any new parameters 
present in these models, only two values for $Br(\klpn)$, 
corresponding to two signs of $X$, are possible. 
Consequently, measuring $Br(\klpn)$ will 
either select one of these two possible values or rule out all MFV models.  In fact the recent
analysis \cite{Haisch:2007ia} shows that $X<0$ is basically ruled out and these are good news as $X>0$ gives larger branching
ratios for the same $|X|$. We will therefore not consider $X<0$ any further.

In \cite{Buras:2001af} a detailed numerical analysis of the relation (\ref{B1B2}) 
has been presented. In view of the improved data on $\sin 2\beta$ and 
$Br(\kpn)$ we update and extend this analysis.
This is shown in Fig.~\ref{fig:Xpos},  
where we show $Br(\kpn)$ as a function 
of $Br(\klpn)$ for several values of $a_{\psi K_S}$. 
These plots  
are universal for all MFV models.

\begin{figure}
\begin{center}
\includegraphics[angle=0,width=6.8cm]{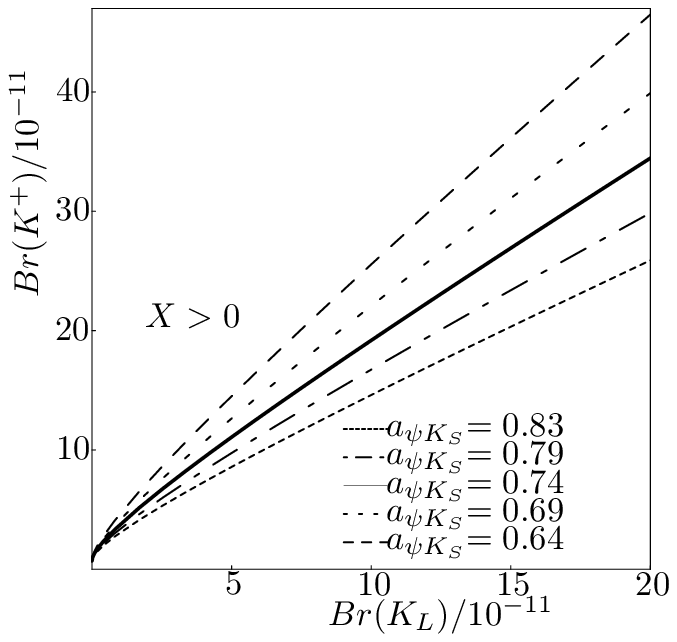}
\caption{$Br(\kpn)$ as a function of $Br(\klpn)$ for several 
values of $a_{\psi K_S}$ in the case of ${\rm sgn}(X)=1$.}
\label{fig:Xpos}
\end{center}
\end{figure}

We also observe, as in \cite{Buras:2001af},
that the upper bound on $Br(\klpn)$ following from the data 
on $Br(\kpn)$ and $\sin 2\beta\le 0.719$
is substantially stronger than the 
model independent bound following from isospin symmetry \cite{Grossman:1997sk}
\begin{equation}
\label{iso}
Br(\klpn) < 4.4 \cdot Br(\kpn).
\end{equation}
With the data in (\ref{EXP1}), that imply 
\be
Br(\kpn) < 3.8 \cdot 10^{-10}~(90\%~\mbox{C.L.}),
\ee
one finds from (\ref{iso})
\be\label{absbound}
Br(\klpn) < 1.7 \cdot 10^{-9}~(90\%~\mbox{C.L.}),
\ee
that is still two orders of magnitude lower than
the upper bound from the KTeV experiment at Fermilab \cite{E391KL}, yielding
$Br(\klpn)<2.9 \cdot 10^{-7}$ and the bound from KEK, $Br(\klpn)<2.1 \cdot
10^{-7}$ \cite{Ahn:2006uf}. 

On the other hand,
taking the experimental bound  
$Br(\kpn)$ in (\ref{EXP1}) and $a_{\psi K_S}\le 0.719$, we find
from (\ref{B1B2})
\begin{equation}\label{KL-bound}
Br(\klpn)_{\rm MFV}\le
2.0 \cdot 10^{-10}, \qquad  {\rm sgn}(X)=+1.
\end{equation}

In \cite{Bobeth:2005ck} a  detailed analysis of several branching ratios
for rare $K$ and $B$ decays in MFV models has been performed. Using the presently available information on the UUT, 
summarized in \cite{Bona:2005eu}, and from the measurements of $Br(B\to
X_s\gamma)$, $Br(B\to X_sl^+l^-)$ and $Br(\kpn)$ the upper bounds on various
branching ratios within the CMFV scenario have been found. Very recently this analysis has been updated
and generalized to include the constraints from the observables in $Z\to b \bar
b$ decay \cite{Haisch:2007ia}. The results of this analysis are collected 
in Table~\ref{brMFV} together with the results within the SM.

Finally, anticipating that the leading role in constraining this kind of
physics will eventually be taken over by $\kpn$, $\klpn$ and
$B_{s,d}\to\mu^+\mu^-$, that are dominated by the function $C(v)$, references 
\cite{Bobeth:2005ck,Haisch:2007ia} provide plots for several branching ratios as functions of 
$C(v)$.

\begin{table*}[t]
\small{
\begin{center}
\begin{tabular}{|c|c|c|c|c|}
\hline
Observable & CMFV ($95 \% \, {\rm CL}$) & \hspace{0mm} SM ($68 \% \,
{\rm CL}$) & \hspace{0mm} SM ($95 \% \, {\rm CL}$) & \hspace{0mm}
Experiment \\[0.5mm] 
\hline 
$\BRKp \times 10^{11}$ & $[4.29, 10.72]$ & \hspace{0mm} $7.15 \pm 1.28$ &  
\hspace{0mm} $[5.40, 9.11]$ & \hspace{0mm} $\left ( 14.7^{+13.0}_{-8.9}
\right )$ \cite{Anisimovsky:2004hr}\\
$\BRKL \times 10^{11}$ & $[1.55, 4.38]$ & \hspace{0mm} $2.79 \pm 0.31$
& \hspace{0mm} $[2.21, 3.45]$ & \hspace{0mm} $< 2.1 \times 10^4 \;\;
(90 \% \, \text{CL})$ \cite{Ahn:2006uf} \\
$\BRKm \times 10^9$ & $[0.30, 1.22]$ & \hspace{0mm} $0.70 \pm 0.11$ &
\hspace{0mm} $[0.54, 0.88]$ & \hspace{0mm} -- \\
$\BRXd \times 10^6$ & $[0.77, 2.00]$ & \hspace{0mm} $1.34 \pm 0.05$ &
\hspace{0mm} $[1.24, 1.45]$ & \hspace{0mm} -- \\
$\BRXs \times 10^5$ & $[1.88, 4.86]$ & \hspace{0mm} $3.27 \pm 0.11$ &
\hspace{0mm} $[3.06, 3.48]$ & \hspace{0mm} $< 64 \;\; (90 \% \,
\text{CL})$ \cite{Barate:2000rc} \\ 
$\BRBd \times 10^{10}$ & $[0.36, 2.03]$ & \hspace{0mm} $1.06 \pm 0.16$ &
\hspace{0mm} $[0.87, 1.27]$ & \hspace{0mm} $< 3.0 \times 10^2 \;\; (95
\% \, \text{CL})$ \cite{Bernhard:2006fa} \\  
$\BRBs \times 10^9$ & $[1.17, 6.67]$ & \hspace{0mm} $3.51 \pm 0.50$ &
\hspace{0mm} $[2.92, 4.13]$ & \hspace{0mm} $< 5.8 \times 10^1
\;\; (95 \% \, \text{CL})$ \cite{bsmumu} \\[1mm] 
\hline 
\end{tabular}
\caption[]{ Bounds for various rare decays in CMFV models at $95 \%$
  probability, the corresponding values in the SM at $68 \%$ and $95
  \% \, {\rm CL}$, and the available experimental information \cite{Haisch:2007ia}. See
  text for details.}
\label{brMFV}
\end{center}
}
\end{table*}

{The main messages from \cite{Bobeth:2005ck, Haisch:2007ia} are the following ones:}

The existing constraints coming from $K^+\to\pi^+\nu\bar\nu$, $B\to
X_s\gamma$, $B\to X_s l^+l^-$ and $Z\to b \bar b$ do not allow within the
CMFV scenario
of \cite{Buras:2000dm} for substantial departures of the branching ratios for
all rare $K$ and $B$ decays from the SM estimates. This is evident
{}from Table~\ref{brMFV}.

This could be at first sight a rather pessimistic message. On the other 
hand it implies that finding practically any branching ratio enhanced 
by more than a factor of two with respect to the SM will automatically 
signal either the presence of new CP-violating phases or new operators, 
strongly suppressed within the SM, at work.
 In particular, recalling that in most extensions of the SM the
  decays $K\to\pi\nu\bar\nu$ are governed by the single $(V-A)\otimes
  (V-A)$ operator, the violation of the upper bounds on at least one
  of the $K\to\pi\nu\bar\nu$ branching ratios, will either signal the
  presence of new complex weak phases at work or new contributions
  that violate the correlations between the $B$ decays and $K$ decays.

As $a_{\psi K_S}$ in MFV models determines the true value of $\beta$ and the
true value of $\gamma$ can be determined in tree level strategies in $B$
 decays one day, the true value of $\bar\eta$ can also be determined in a
clean manner.
Consequently, using 
(\ref{modX}) offers probably the cleanest measurement of $|X|$ in the field 
of weak decays.

\boldmath
\section{Scenarios with New Complex Phases in
$\Delta F=1$ and $\Delta F=2$ Transitions}\label{sec:newphys}\unboldmath
\setcounter{equation}{0}
\subsection{Preliminaries}
In this section we will consider three simple scenarios 
beyond the framework of MFV, in 
which $X$ becomes a complex quantity as given in 
(\ref{NX}),
and the universal box function $S(v)$ entering
$\varepsilon_K$ and $\Delta M_{d,s}$ not only becomes complex but generally
becomes non-universal with  
\be\label{SCOMP}
S_K(v)=|S_K(v)|e^{i2\varphi_K}, \quad 
S_d(v)=|S_d(v)|e^{i2\varphi_d}, \quad
S_s(v)=|S_s(v)|e^{i2\varphi_s},
\ee
for $K^0-\bar K^0$,  $B_d^0-\bar B_d^0$ and $B_s^0-\bar B_s^0$ mixing,
respectively. If these three functions are different from each other, 
some universal properties found in the SM and MFV models, that have been
reviewed in \cite{Buras:2005xt,Buras:2003jf,Buras:2004sc}, are lost. 
In addition, the mixing
induced CP asymmetries in $B$ decays do not measure the angles of the UT
but only sums of these angles and of $\varphi_i$. 
In particular
\be
S_{\psi K_S}=\sin (2 \beta+2 \varphi_{B_d}).
\ee
Equally importantly the rare $K$ and $B$ decays, governed in models with MFV by
the real universal functions $X$, $Y$ and $Z$, are described now by nine complex
functions ($i=K,d,s$) \cite{Blanke:2006eb}
\be\label{MFVfunctions}
X_i=|X_i|e^{i \theta_K^i}, \quad Y_i=|Y_i|e^{i \theta_Y^i}, \quad Z_i=|Z_i|e^{i
\theta_Z^i}
\ee  
that result from th SM box and penguin diagrams and analogous diagrams with new
particle exchanges. In the SM and in CMFV models the independence of the
functions in (\ref{MFVfunctions}) of $i$ implies very strong correleations between
various branching ratios in $K$, $B_d$ and $B_s$ system and consequently strong
upper bounds as shown in table \ref{brMFV}. In models with new complex phases
this universality is generally broken and consequently as we will see in the
next section the bounds in Table \ref{brMFV} can be strongly violated.\\

As in the $K\to \pi \nu \bar \nu$ system only one function is present, we will
drop the index $i$ and denote it by
\be
X=|X| e^{i\theta_X}.
\ee

In order to simplify the presentation we will assume here that 
$S_s=S_0(x_t)$ as in the SM but we will take $S_d(v)$ to be complex with
$S_d(v)\not=S_0(x_t)$. This will allow to change the relation between $R_t$
and $\Delta M_d/\Delta M_s$ in (\ref{Rt}).
We will leave open whether $S_K(v)$ receives new physics contributions. We will
relax these assumptions in concrete models in the next chapter.

An example of general scenarios with new complex phases is the scenario in which  new physics enters dominantly through 
enhanced $Z^0$ penguins involving a new CP-violating weak phase.  It
was first considered in 
\cite{Buras:1997ij,Colangelo:1998pm,Buras:1998ed,Buras:1999da} in the context of rare 
$K$ decays and the ratio $\epe$ measuring direct CP violation
in the neutral kaon system, and was generalized to rare $B$ decays in 
\cite{Buchalla:2000sk,Atwood:2003tg}. Subsequently this particular extension of the SM has been revived 
in \cite{Buras:2003dj,Buras:2004ub}, where it has been pointed out that the 
anomalous behaviour in $B\to\pi K$ decays 
observed by CLEO, BABAR and  Belle \cite{Bornheim:2003bv,Aubert:2002jb,Aubert:2003sg,Aubert:2003qj,Chao:2003ue}
 could be due to 
the presence 
of enhanced $Z^0$ penguins carrying a large new CP-violating phase around 
$-90^\circ$. 

The possibility of important electroweak penguin contributions behind
the anomalous behaviour of the $B\to\pi K$ data has been pointed out 
already in \cite{Buras:2000gc}, but only in 2005 has this behaviour been 
independently 
observed by the three collaborations in question. Recent discussions 
related to electroweak penguins can be also found in \cite{Yoshikawa:2003hb,Beneke:2003zv}. 
Other conjectures in connection with these data can be found in \cite{Gronau:2003kj,Gronau:2003br,Chiang:2004nm}.

The implications of the large CP-violating phase in electroweak penguins  
for rare $K$ and $B$ decays and $B\to X_s l^+l^-$
have been analyzed in detail in 
\cite{Buras:2003dj,Buras:2004ub} and subsequently the analyses of $B\to X_s l^+l^-$
and $K_L\to\pi^0 l^+l^-$ have been extended in \cite{RaiChoudhury:2004pw} and
\cite{Isidori:2004rb}, respectively. 
It turns out that in this scenario 
several predictions differ significantly from the SM 
expectations with most spectacular effects found precisely in the 
$K\to\pi\nu\bar\nu$ system. 

Meanwhile the data on $B\to \pi K$ decays have changed considerably and the case
for large electroweak penguin contributions in these decays is much less
convincing \cite{Fleischer:2007mq,Fleischer:2007hj,Baek:2007yy,Silvestrini:2007yf,Gronau:2006xu,Jain:2007dy}. Still the general formalism developed for the $K\to \pi \nu \bar
\nu$ system in the presence of new complex phases \cite{Buras:1998ra} and
\cite{Buras:2003dj,Buras:2004ub} remains valid and we will present it below. Moreover, in
the next section we will discuss three explicit models, Littlest Higgs model with
T-Parity (LHT), a $Z^{\prime}$ model and the
MSSM in which the functions $X$ becomes a complex quantity and the departures of
the $K\to \pi \nu\bar\nu$ rates from the SM ones can be spectacular.

The scenarios with complex phases in  
$B^0_d-\bar B^0_d$ mixing
have been considered in many papers with the subset of references given
in \cite{D'Ambrosio:2001zh,Bergmann:2001pm,Bergmann:2000ak,Bertolini:1987cw,Nir:1990hj,Nir:1990yq,Laplace:2002ik,Laplace:2002mb},
\cite{Fleischer:2003xx,Bona:2005eu}. 

Very recently this scenario has been revived through the possible inconsistencies between UUT and the RUT
signalled by the discrepancy between the value of $\sin 2 \beta$ from $S_{\psi K_S}$ and its value obtained from
tree-level measurements. We will return to this issue below.

In what follows, we will first briefly review the formulae
for $\kpn$ and $\klpn$ decays obtained in \cite{Buras:2003dj,Buras:2004ub}, for the case of a complex $X$. 
Subsequently, we will discuss the implications of this general scenario for 
the relevant branching ratios. 

Next we will consider scenarios with new physics present only in 
$B^0_d-\bar B^0_d$ mixing and the function $X$ as in the SM. Here the impact 
on $Br(\kpn)$ and $Br(\klpn)$ comes only through modified values of the CKM
parameters but, as we will see below, this impact is rather interesting.

Finally we will consider a hybrid scenario with new physics entering both
$K\to\pi\nu\bar\nu$ decays and $B^0_d-\bar B^0_d$ mixing. In this discussion 
the $(R_b,\gamma)$ strategy (RUT) for the determination of the UT will play a very 
important role.

\boldmath
\subsection{A Large New CP-Violating Phase $\theta_X$}\label{ssec:newphase}
\unboldmath
In this general scenario the function $X$ becomes a complex quantity \cite{Buras:1997ij}, 
as given in (\ref{NX}), 
with $\theta_X$ being a new complex phase that originates 
from new physics contributions to the relevant Feynman diagrams.
Explicit realizations of such extension of the SM will be discussed in 
Section \ref{sec:models}.
In what follows it will be useful to  define the following combination of 
weak phases,
\be\label{betas}
\beta_X\equiv \beta-\beta_s-\theta_X=\beta_{\rm eff}-\theta_X.
\ee

Following \cite{Buras:2004ub},
the branching ratios for $\kpn$ and $\klpn$ are now given as follows:
\begin{equation}\label{bkpnZ}
{Br}(\kpn)=\kappa_+\left[\tilde r^2 A^4 R_t^2 |X|^2
+2 \tilde r \bar P_c(X) A^2 R_t |X|\cos\beta_X
+ \bar P_c(X)^2  \right]
\end{equation}
\begin{equation}\label{bklpnZ}
Br(\klpn)=\kappa_L
\tilde r^2 A^4 R_t^2 |X|^2\sin^2\beta_X,
\end{equation}
with $\kappa_+$ given in (\ref{kapp}), 
$\kappa_L$ given in (\ref{kapl}), 
$\bar P_c(X)$ defined in (\ref{Pbar}),
$\beta_X$ in (\ref{betas}) and 
$\tilde r$ in (\ref{LAMT}). 

Once $Br(\kpn)$ and $Br(\klpn)$ have been measured, the
parameters $|X|$ and $\beta_X$ can be determined, subject to ambiguities that
can be resolved by considering other processes, such as the non-leptonic 
$B$ decays and the rare decays discussed in \cite{Buras:2004ub}. 
Combining (\ref{bkpnZ}) and (\ref{bklpnZ}), the generalization of 
(\ref{sin2bnunu}) to 
the scenario considered can be found \cite{Buras:2004ub,Buras:1997ij}
\be\label{sin2bnunuZ}
\sin 2\beta_X= \frac{2 \bar r_s}{1+\bar r_s^2}, \qquad 
\bar r_s=\frac{\varepsilon_1\sqrt{B_1-B_2}-\bar P_c(X)}
{\varepsilon_2\sqrt{B_2}}=\cot\beta_X,
\ee
where $\varepsilon_i=\pm 1$.
Moreover,
\be\label{modXg}
|X|=\frac{\varepsilon_2\sqrt{B_2}}{\tilde r A^2R_t\sin\beta_X},\quad
\varepsilon_2\sin\beta_X >0.
\ee
 The ``reduced'' branching ratios
$B_i$ are given in (\ref{b1b2}).

These formulae are valid for arbitrary 
$\beta_X\not=0^\circ$. 
For $\theta_X=0^\circ$ and $\varepsilon_1=\varepsilon_2=1$, one obtains from 
(\ref{sin2bnunu}) the SM result in (\ref{sin2bnunu}). 
On the other hand for 
 $99^\circ\le \beta_X\le 125^\circ$ one has
$\varepsilon_1=-1$ and $\varepsilon_2=1$.

As in this scenario it is assumed that there are no significant contributions to
$B_{s,d}^0-\bar B_{s,d}^0$ mixings 
and $\varepsilon_K$, in particular no complex phases, 
the determination of the CKM parameters through the standard analysis
of the unitarity triangle proceeds as in the SM with the input parameters
given in Section \ref{ssec:CKM}.
Consequently, $\beta$ and $\beta_s$ are already known from the usual 
analysis of the UT and the measurement 
of $\bar r_s$ in $K\to\pi\nu\bar\nu$ decays 
will provide  a theoretically clean determination of 
$\theta_X$ and $\beta_X$. Similarly, 
a clean determination of $|X|$ can be obtained from (\ref{modXg}),  with $R_t$ determined 
by means of (\ref{S3}).

It has been pointed out in \cite{Buras:2003dj} that in the case of $\beta_X \approx 90^\circ$, in spite of the enhanced value of $|X|$, 
$Br(\kpn)$ does not significantly
differ from the SM estimate because the enhancement of the first term in 
(\ref{bkpnZ}) can be to a large extent compensated by the suppression of the
second term ($\cos \beta_X\ll\cos(\beta-\beta_s)$).
Consequently, $Br(\kpn)$ in this case is very 
strongly
dominated by the ``top" contribution given by the function $X$ and 
charm-top interference is either small or even destructive.

On the other hand, $\beta_X \approx 90^\circ$ implies a spectacular enhancement of $Br(\klpn)$
by one order of magnitude.
Consequently,  while $Br(\klpn)\approx (1/3)Br(\kpn)$ 
in the SM, it is substantially larger than 
$Br(\kpn)$ in such a scenario.
The huge enhancement of $Br(\klpn)$ seen here
is mainly due to 
the large weak phase $\beta_X$, as 
\be
\frac{Br(\klpn)}{Br(\klpn)_{\rm SM}}=
\left|\frac{X}{X_{\rm SM}}\right|^2
\left[\frac{\sin\beta_X}{\sin(\beta-\beta_s)}\right]^2 
\ee
and to a lesser extent due to the enhanced value of $|X|$, which generally could be bounded by other processes.

Inspecting (\ref{bkpnZ}) and (\ref{bklpnZ}), one observes \cite{Buras:2004ub} 
that the very strong 
dominance of the ``top" contribution in these expressions implies a simple 
approximate expression:
\be
\frac{Br(\klpn)}{Br(\kpn)}\approx 4.4\times (\sin\beta_X)^2
\approx 4.2\pm 0.2.  
\ee
We note that $Br(\klpn)$ is then rather close to its model-independent
upper bound~\cite{Grossman:1997sk} given in (\ref{iso}).
It is evident from (\ref{sin2bnunuZ}) that this bound
is reached when the reduced branching ratios $B_1$ and $B_2$ in
(\ref{b1b2}) are equal to each other.

A spectacular 
implication of such a scenario is a strong violation of the MFV relation 
\cite{Buchalla:1994tr} in (\ref{R7}).
Indeed, with $\beta_X\approx \pm 90^\circ$
\be
(\sin 2 \beta)_{\pi \nu\bar\nu}=\sin 2\beta_X \neq (\sin 2 \beta)_{\psi K_{\rm S}}= 
0.675\pm0.026.
\ee

\begin{figure}
\vspace*{0.3truecm}
\begin{center}
\includegraphics[width=12.5cm]{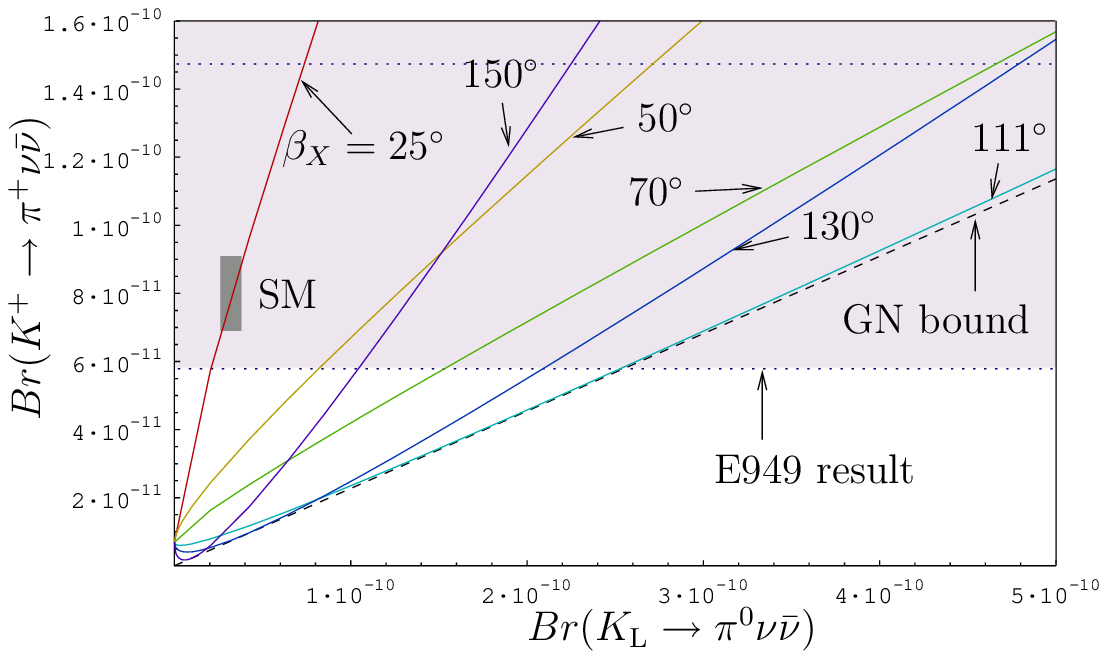}
\end{center}
\caption{$Br(\kpn)$ as a function of $Br(\klpn)$
for various values of $\beta_X$ \cite{Buras:2004ub}. 
The dotted horizontal lines indicate 
the lower part of the experimental range
(\ref{EXP0}) and the grey area the SM prediction. We also show the 
bound in (\ref{iso}).  \label{fig:KpKl}}
\end{figure}

In the next section, we will investigate this violation in two specific models.
In Fig.~\ref{fig:KpKl}, we show -- in the spirit of the plot in 
Fig. \ref{fig:Xpos}  -- 
$Br(\kpn)$ as a function of $Br(\klpn)$ for fixed values of
$\beta_X$ that has been presented in \cite{Buras:2004ub}. 
As this plot is independent of $|X|$, it offers a direct
measurement of the phase $\beta_X$. 
The first line on the left represents the MFV models with
$\beta_X=\beta_{\rm eff}=\beta-\beta_s$, already discussed in Section \ref {sec:MFV}, 
whereas the first line on the right corresponds 
to the model-independent Grossman--Nir bound \cite{Grossman:1997sk} given in 
(\ref{iso}). Note 
that the value of $\beta_X$ corresponding to this bound depends on 
the actual value of $Br(\kpn)$ and $Br(\klpn)$ as at this 
bound ($B_1=B_2$) we have \cite{Buras:2004ub} 
\be\label{BX-bound}
(\cot\beta_X)_{\rm Bound}=-\frac{\bar P_c(X)}
{\varepsilon_2\sqrt{B_2}}.
\ee
For the central values of $\bar P_c(X)$ and $B_2$ found in the latter paper 
the bound corresponds 
to $\beta_X=107.3^\circ$. As only $\cot\beta_X$ and not $\beta_X$ is 
directly determined by the values of the branching ratios in question, 
the angle $\beta_X$ is determined only up to discrete ambiguities,  
seen already in Fig.~\ref{fig:KpKl}. These ambiguities can be resolved by
considering simultaneously other quantities discussed in \cite{Buras:2004ub}.

\boldmath
\subsection{General Discussion of $\theta_X$ and $|X|$}
\unboldmath

In Fig.~\ref{KLKPR}, we show the ratio of the two branching ratios in question as a
function of $\beta_X$ for three values of $|X|=1.25,~1.5,~2.0$. We observe 
that for $\beta_X$ in the ballpark of $110^\circ$ this ratio is very close 
to the bound in (\ref{iso}). However, even for $\beta_X=50^\circ$ the 
ratio is close to unity and by a factor of 3 higher than in the SM.

\begin{figure}[hbt]
\vspace{0.10in}
\centerline{
\epsfysize=2.7in
\epsffile{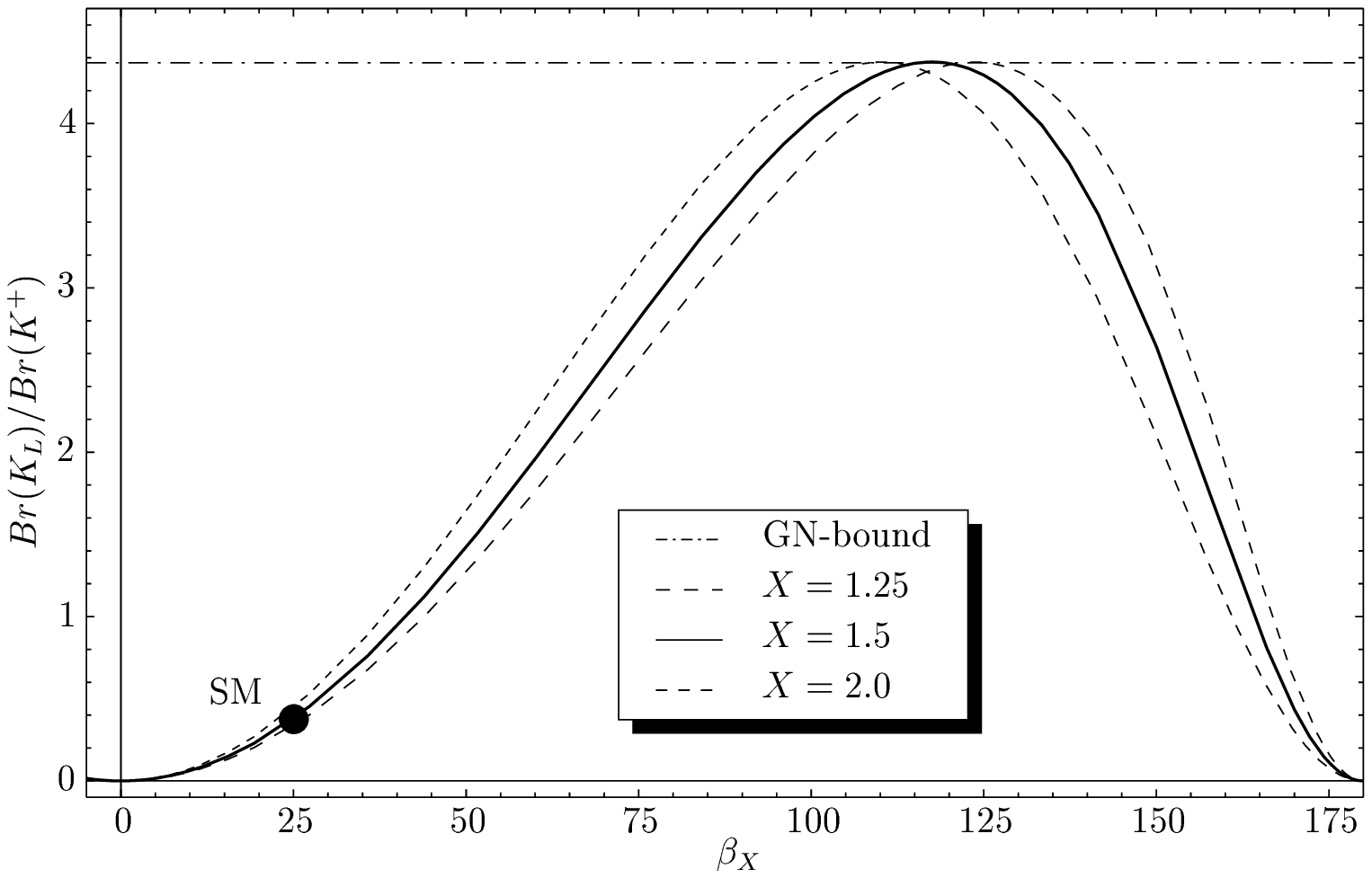}
}
\vspace{0.08in}
\caption{{ The ratio of the  $K\to\pi\nu\bar\nu$ branching ratios as a
function of $\beta_X$ for $|X|=1.25,~1.5,~2.0$. The horizontal line is 
the bound in (\ref{iso}). }}\label{KLKPR}
\end{figure}

Finally, in table~\ref{tab:KPKL}, we give the values of $Br(\kpn)$ and 
$Br(\klpn)$ for different values of $|X|$ and $\theta_X$, $\beta=22.2^\circ$ 
and $\vcb=41.6\cdot 10^{-3}$. In this context we would like to refer to 
scaling laws for FCNC processes pointed out in \cite{Buras:1992uf}, from which 
it follows that the dependence of $K\to\pi\nu\bar\nu$ branching ratios 
on $\vcb$ and $|X|$ is encoded in a single variable
\be\label{scaling}
Z=A^2 |X|.
\ee 
This observation allows to make the following replacement in 
table~\ref{tab:KPKL}
\be\label{res}
|X| \to  |X|_{\rm eff}=\left[\frac{\vcb}{41.5\cdot 10^{-3}}\right]^2 |X|,
\ee
so that for $\vcb\not=41.6\cdot 10^{-3}$ the results in this table correspond
to different values of $|X|$ obtained by rescaling the values for $|X|$ there
by means of (\ref{res}).

As beyond the SM the uncertainties in the value of $|X|$ are substantially 
larger than the ones in $\vcb$, the error in $\vcb$ can be absorbed into 
the one of $|X|_{\rm eff}$.

\begin{table}[thb]
\caption[]{ Values of $Br(\kpn)$ and of $Br(\klpn)$ (in parentheses) 
in units of $10^{-11}$ for different values of $\theta_X$ and $|X|$ with 
$\beta=22.2^\circ$ 
and $\vcb=41.6\cdot 10^{-3}$. \label{tab:KPKL}}
\begin{center}
\begin{tabular}{|c||c|c|c|c|c|}\hline
$\theta_X/|X| $& $ 1.25 $ & $ 1.50$ & $ 1.75$ & $ 2.00$ & $2.25$ 
\\ \hline
$-90^\circ$      & $ 2.3 $ & $ 3.3 $ & $ 4.5 $  & $ 6.0 $ & $ 7.6$ \\
                 & $(10.1)$ & $(14.5)$ & $(19.8)$ & $(25.8)$ & $(32.7)$\\

$-60^\circ$      & $ 3.8 $ & $ 5.0$  & $ 6.5 $  & $ 8.3 $ & $ 10.2$ \\
                 & $(12.1)$ & $(17.4)$ & $(23.6)$ & $(30.9)$ & $(39.1)$\\
$-30^\circ$      & $ 5.1 $ & $ 6.7 $ & $ 8.4 $  & $ 10.4$ & $ 12.6$ \\
                 & $(8.1)$ & $(11.6)$ & $(15.8)$ & $(20.7)$ & $(26.1)$\\
$0^\circ$      & $ 6.0 $ & $ 7.8$ & $ 9.7 $  & $ 11.9 $ & $ 14.3$ \\
                 & $(2.1)$ & $(3.0)$ & $(4.1)$ & $(5.4)$ & $(6.8)$\\
$30^\circ$      & $ 6.3 $ & $ 8.0 $ & $ 10.0 $  & $ 12.3 $ & $ 14.7$ \\
                 & $(0.11)$ & $(0.16)$ & $(0.22)$ & $(0.29)$ & $(0.36)$\\
$60^\circ$      & $ 5.8 $ & $ 7.4$ & $ 9.3 $  & $ 11.5$ & $ 13.8$ \\
                 & $(4.1)$ & $(5.9)$ & $(8.0)$ & $(10.5)$ & $(13.3)$\\
$90^\circ$      & $ 4.6 $ & $ 6.1 $ & $ 7.8 $  & $ 9.7 $ & $ 11.8 $ \\
                 & $(10.1)$ & $(14.5)$ & $(19.8)$ & $(25.8)$ & $(32.7)$\\
\hline
 \end{tabular}
\end{center}
\end{table}

\boldmath
\subsection{New Complex Phases in the $B^0_d-\bar B^0_d$ Mixing}
\unboldmath
We next move to the scenario in which $X=X_{\rm SM}$ but there are new
contributions to $B^0_d-\bar B^0_d$ mixing. This scenario has been 
considered in detail in many papers
\cite{D'Ambrosio:2001zh,Bergmann:2001pm,Bergmann:2000ak,Bertolini:1987cw,Nir:1990hj,Nir:1990yq,Laplace:2002ik,Laplace:2002mb,Fleischer:2003xx}. 
As summarized in the latter paper, this scenario can be realized 
in supersymmetric
models with a) a heavy scale for the soft-breaking terms, b) new sources of
flavour symmetry breaking only in the soft-breaking terms which do not
involve the Higgs fields and c) Yukawa interactions very similar to the SM
case. However, as emphasized in \cite{Fleischer:2003xx} and discussed briefly in
Section \ref{sec:models}, this scenario is not representative for all supersymmetric
scenarios, in particular those with important mass insertions of the
left-right type and Higgs mediated FCNC amplitudes with large $\tan\beta$.
Non-supersymmetric examples like Littlest Higgs with T-Parity and $Z^\prime$-models can also provide new phase
effects in $B^0_d-\bar B^0_d$ mixing but generally such effects are
simultaneously present in $B^0_s-\bar B^0_s$ mixing and $K\to \pi \nu \bar \nu$.

Let us recall that, in the presence of a complex function $S_d$, the
off-diagonal term $M^d_{12}$ in the neutral $B^0_d$ meson mass matrix has
the phase structure
\be
M^d_{12}=
\frac{\langle B^0_d|H_{eff}^{\Delta B=2}|\bar B^0_d\rangle}{2 m_{B_d}}
         \propto e^{i2\beta} e^{i2\varphi_d} |S_d|
\ee
with
$|S_d|$ generally differing from $S_0(x_t)$. If 
$S_s$ remains unchanged, then
\begin{itemize}
\item
The asymmetry $a_{\psi K_S}$ does not measure $\beta$ but $\beta+\varphi_d$
\item
The expression for $R_t$ in (\ref{Rt}) becomes
\be\label{Rtnew}
r_d R_t=0.920~\tilde r\left[\frac{\xi}{1.24}\right] 
\left[\frac{0.2248}{\lambda}\right]
\sqrt{\frac{18.4/ps}{\Delta M_s}} 
\sqrt{\frac{\Delta M_d}{0.50/ps}},
\qquad
r^2_d\equiv\left|\frac{S_d}{S_0(x_t)}\right|.
\ee
\end{itemize}
As a consequence of these changes, the true angle $\beta$ differs from the
one extracted from $a_{\psi K_S}$ and also $R_t$ and $\vtd$ will be modified 
if $r_d\not= 1$.

As $X$ is not modified with respect to the SM, the impact on
$K\to\pi\nu\bar\nu$ amounts exclusively to 
the change of the true $\beta_{\rm eff}$ and $R_t$ in 
the formulae (\ref{bkpnn1}) and (\ref{bklpn1}). 
A particular pattern of a possible impact on $K\to\pi\nu\bar\nu$ in the 
scenario in question has been presented in
\cite{Fleischer:2003xx}. 

In the meantime the data on the CP asymmetry $S_{\psi K_S}$ and the observables in $B_{s,d}^0-\bar B_{s,d}^0$
systems have so much improved that the allowed values for $r_d$ and $\varphi_{B_d}$ are strongly constrained.
Also, there is now a slight tension between the values of $|V_{ub}|$ and $\sin 2 \beta$ as inputed into the fits, potentially
hinting towards some non-vanishing (negative) phase $\varphi_{B_d}$
\cite{Bona:2006sa,Blanke:2006ig}. However, since there are some
open questions concerning the value of $|V_{ub}|$, it remains to
be seen how this situation develops further. The implication of this for the $K\to \pi \nu \bar \nu$ decays is that, due to the higher value of $\bar \eta$ 
obtained from the RUT fit, the values for both branching ratios are larger than found using CKM
values from an overall fit of the unitarity triangle.

\boldmath
\subsection{A Hybrid Scenario}\label{ssec:hybrid}
\unboldmath
The situation is more involved if new physics effects enter both $X$ and $S$.
Similarly to previous two scenarios, the golden relation in (\ref{R7}) 
is violated, but now the structure of a possible violation is more involved
\be
\left[\sin 2(\beta-\theta_X)\right]_{\pi\nu\bar\nu}\not= 
\left[\sin 2(\beta+\varphi_d)\right]_{\psi K_S}.
\ee
Since $\theta_X$ originates in new contributions to the decay amplitude
$K\to\pi\nu\bar\nu$ and $\theta_d$ in new contributions to the 
$B^0_d-\bar B^0_d$ mixing, it is very likely that $\theta_X\not=\varphi_d$.

The most straightforward strategy to
disentangle new physics contributions in $K\to\pi\nu\bar\nu$ and the
$B^0_d-\bar B^0_d$ mixing in this scenario is to use the reference unitarity triangle that
results from the $(R_b,\gamma)$ strategy. Having the true CKM parameters at
hand, one can determine $\theta_X$ and $|X|$ from
$K\to\pi\nu\bar\nu$ and $\varphi_d$ and $|S_d|$  from the $B^0_d-\bar B^0_d$
mixing and $a_{\psi K_S}$.

In order to illustrate these ideas in explicit terms let us investigate,
in the rest of this section,
how the presence of new contributions in $K\to\pi\nu\bar\nu$ and 
the $B^0_d-\bar B^0_d$ mixing could be signaled in the
$(\bar\varrho,\bar\eta)$ plane.

Beginning with $K\to\pi\nu\bar\nu$, let us write
\be\label{W1}
X=r_X X_{\rm SM} e^{i\theta_X}.
\ee 
Then formulae (\ref{bkpnZ}) and (\ref{bklpnZ}) apply with
\be
|X| \to X_{\rm SM}, \qquad R_t \to r_X R_t.
\ee
We proceed then as follows:
\begin{itemize}
\item
>From the measured $Br(\kpn)$ and $Br(\klpn)$ we determine the ``fake" 
angle $\beta$ in the unitarity triangle with the help of (\ref{sin2bnunuZ}).
We denote this angle by $\beta_X$, that we defined in (\ref{betas}). 
In what follows we neglect $\beta_s$ 
but it can be taken straightforwardly into account if necessary.
\item
The height of the fake UT from $K\to\pi\nu\bar\nu$ is then given by
\be\label{W2} 
\bar\eta_{\pi\nu\bar\nu}=r_X R_t \sin\beta_X=\frac{\sqrt{B_2}}{\tilde r A^2
X_{\rm SM}},
\ee
where we set $\varepsilon_2=+1$ in order to be concrete. As seen this height
can be found from $Br(\klpn)$ and $X_{\rm SM}$.
\end{itemize}

Now let us go to the $B^0_d-\bar B^0_d$ mixing where we introduced the 
parameter 
$r_d$ defined in (\ref{Rtnew}). We proceed then as follows:
\begin{itemize}
\item
The asymmetry $a_{\psi K_S}$ determines the fake angle $\beta$, that we denote
by $\beta_d=\beta+\theta_d$.
\item
The fake side $R_t$, to be denoted by $(R_t)_d$, is now given as follows
\be
(R_t)_d=r_d R_t. 
\ee
It can be calculated from (\ref{Rtnew}) subject to uncertainties in $\xi$.
\end{itemize}

Clearly, generally the fake UT's resulting from $K\to\pi\nu\bar\nu$ and the
$(\Delta M_d/\Delta M_s,\beta)$ strategy, discussed above,
 will differ from each other, 
from the true
reference triangle and also from the UT obtained from the $(\gamma,\beta)$
and $(\bar\eta,\gamma)$ strategies, if the determinations of
$\bar\eta$ and $\beta$ are polluted by new physics.

We show these five different triangles in Fig.~\ref{bsuplot}. 
Comparing the fake
triangles with the reference triangle, all new physics parameters in
$K\to\pi\nu\bar\nu$ and  $B^0_d-\bar B^0_d$ mixing can be easily
extracted. Fig. \ref{bsuplot} has only illustrative character. We know already
from the recent analyses og the UT \cite{Blanke:2006ig, Bona:2006sa} that the phase
$\varphi_{B_d}$ is constrained to be much smaller than depicted in this figure. Moreover, a
negative value seems to be favoured.

\begin{figure}[hbt]
\begin{center}
\epsfig{file=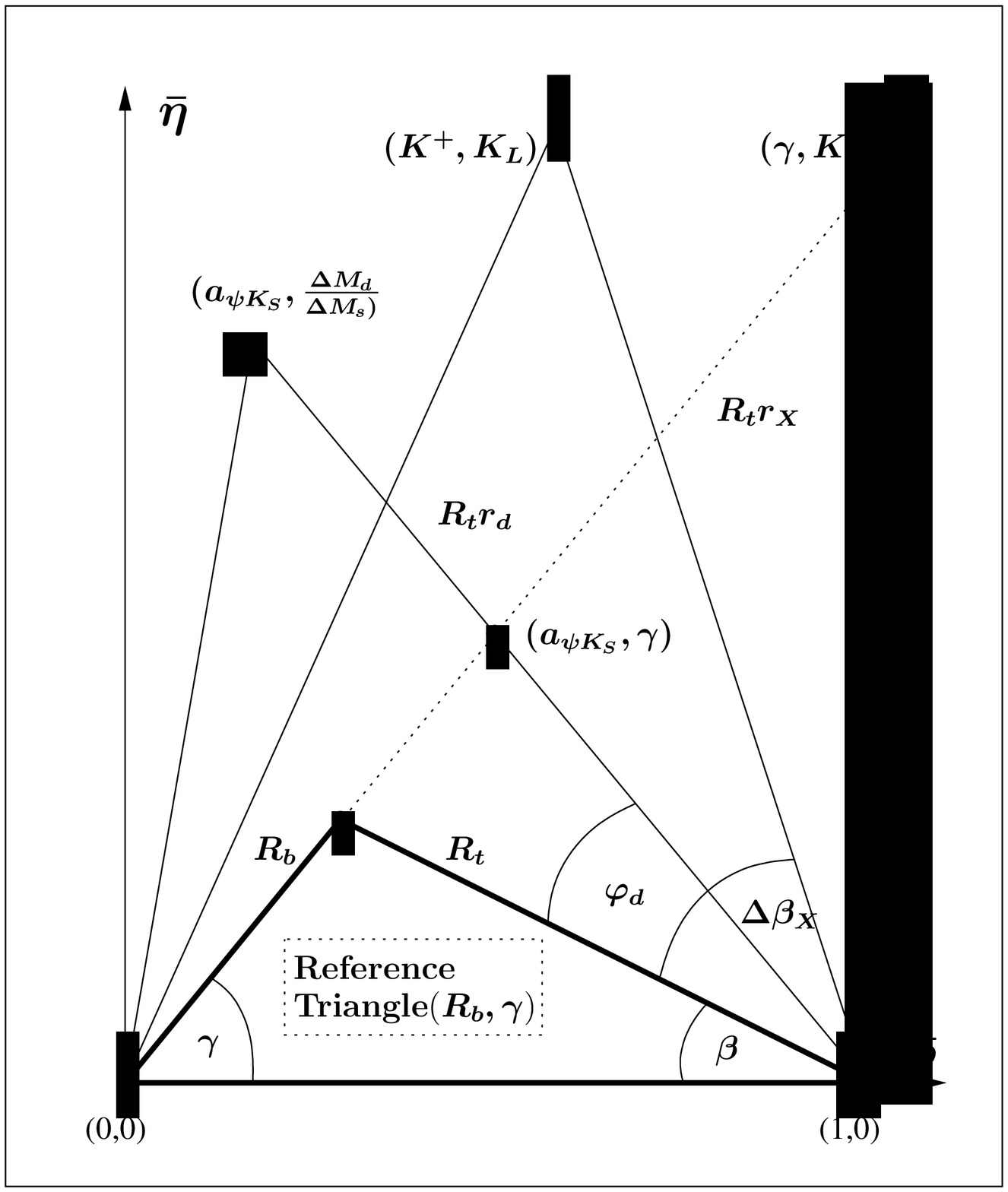,height=10cm,angle=0,clip=}
\end{center}
\caption{Fake unitarity triangles as discussed in the text compared to the 
reference triangle.
$\Delta\beta_X=-\theta_X$.
\label{bsuplot}}
\end{figure}

\boldmath
\subsection{Correlation between {$Br(\klpn)$} and
{$Br(B\to X_{s,d}\nu\overline{\nu})$}}
\unboldmath
The branching ratios for the inclusive rare decays $B\to X_{s,d}\nu\overline{\nu}$ 
can be written in the models with a new complex phase in $X$ as follows 
\cite{Buras:2004ub} $(q=d,s)$
\be\label{BXS}
Br(B\to X_q\nu\overline{\nu})=1.58\cdot 10^{-5}
\left[\frac{Br(B\to X_c e\overline{\nu})}{0.104}\right]
\left|\frac{V_{tq}}{V_{cb}}\right|^2
\left[\frac{0.54}{f(z)}\right]~|X|^2,
\ee
where  $f(z)=0.54\pm0.04$ is the phase-space factor for 
$B\to X_c e\overline{\nu}$ with $z=m_c^2/m_b^2$, and 
$Br(B\to X_c e\overline{\nu})=0.104\pm 0.004$. 

Formulae (\ref{bklpnZ}) and (\ref{BXS}) imply  interesting relations 
between the decays $\klpn$ and $B\to X_{s,d}\nu\overline{\nu}$ that are
generalizations of similar relations within the MFV models 
\cite{Buras:2001af,Bergmann:2001pm,Bergmann:2000ak} 
to the scenario considered here
\be\label{KLBXS}
\frac{Br(\klpn)}{Br(B\to X_s\nu\bar\nu)}
=\frac{\kappa_L}{1.58\cdot 10^{-5}}
\left[\frac{0.104}{Br(B\to X_c e\bar\nu)}\right]
\left[\frac{f(z)}{0.54}\right] A^4 R_t^2 \sin^2\beta_X,
\ee

\be\label{KLBXD}
\frac{Br(\klpn)}{Br(B\to X_d\nu\bar\nu)}
=\frac{\kappa_L}{1.58\cdot 10^{-5}}
\left[\frac{0.104}{Br(B\to X_c e\bar\nu)}\right]
\left[\frac{f(z)}{0.54}\right] \frac{A^4 \tilde r^2}{\lambda^2} \sin^2\beta_X.
\ee

The experimental upper bound on $Br(B\to X_s\nu\overline{\nu})$ reads 
\cite{Barate:2000rc}
\be\label{BBOUND}
Br(B\to X_s\nu\overline{\nu})<6.4 \cdot 10^{-4}\quad (90\%~\mbox{C.L.}).
\ee
Using this bound and setting $R_t=0.95$, 
$f(z)=0.58$ and $Br(B\to X_c e\overline{\nu})=0.10$, we 
find from (\ref{KLBXS}) the upper bound
\begin{equation}\label{KLXS-bound}
Br(\klpn)\le 4.4\cdot 10^{-9} (\sin\beta_X)^2=\left\{\begin{array}{ll}
6.3 \cdot 10^{-10} &  \beta_X=22.2^\circ\\
3.9 \cdot 10^{-9}  &  \beta_X=111^\circ
\end{array}\right.
\end{equation}
at $90\%~\mbox{C.L.}$ for the MFV models and a scenario with a large new phase,
respectively.
In the case of the MFV models this bound is  weaker than the bound 
in (\ref{KL-bound}) but, 
as the bound in (\ref{BBOUND}) should be improved in the $B$-factory era, 
the situation could change in the next years.
Concerning the scenario with a complex phase $\theta_X$ of Section \ref{ssec:newphase}, 
no useful bound on $Br(\klpn)$ from (\ref{BBOUND}) results at present as 
the bound in (\ref{KLXS-bound}) is weaker than the model independent bound in 
(\ref{absbound}).

\boldmath
\section{\boldmath{$K\to\pi\nu\bar\nu$} in Selected New Physics Scenarios}\label{sec:models}
\unboldmath
\subsection{Preliminaries}
In this section we will briefly review  the results for decays 
$\kpn$ and $\klpn$ in selected new physics scenarios. Our goal is 
mainly to indicate the size of new physics contributions in the branching 
ratios in question. 
Due to several free parameters present in some of these extensions 
the actual predictions for the branching ratios are not very precise 
and often depend sensitively on some of the parameters involved.
The latter could then be determined or bounded efficiently once 
precise data on $K\to\pi\nu\bar\nu$ and other rare decays 
will be available. While we will  
only present the results for $Br(\kpn)$ and $Br(\klpn)$, most of 
the analyses discussed below used all available constraints from other 
observables known at the time of a given analysis. 
A detailed analysis of these constraints is clearly beyond the scope of this
review. 
A general discussion of $K\to\pi\nu\bar\nu$ beyond the SM can be found in 
\cite{Grossman:1997sk}. In writing this section we also benefited from
\cite{Isidori:2003ij,D'Ambrosio:2001zh,Bryman:2005xp}. 

\boldmath
\subsection{Littlest Higgs Models}
\unboldmath
One of the most attractive solutions to the so-called {\it little hierarchy
  problem} that affects the Standard Model (SM) is provided by Little Higgs models.
They are perturbatively computable up to $\sim 10$ TeV and have a rather
  small number of parameters, although their predictivity can be weakened by a
  certain sensitivity to the unknown ultraviolet (UV) completion of the
  theory.
In these models, in contrast to supersymmetry, the problematic quadratic
divergences to the Higgs mass are cancelled by loop contributions of new
particles with the same spin-statistics of the SM ones and with masses around
1 TeV.\\
The basic idea of Little Higgs models \cite{Arkani-Hamed:2001nc} is that the Higgs is 
naturally light as it is identified with a Nambu-Goldstone 
boson of a spontaneously broken global symmetry.

The most economical, in matter content, Little Higgs model is the Littlest
Higgs (LH) model \cite{Arkani-Hamed:2002qy}, where the global group $SU(5)$ is spontaneously broken
into $SO(5)$ at the scale $f \approx \mathcal{O}(1 {\rm TeV})$ and
the electroweak sector of the SM is embedded in an $SU(5)/SO(5)$ non-linear
sigma model. 
Gauge and Yukawa Higgs interactions are introduced by gauging the subgroup of
$SU(5)$: $[SU(2) \times U(1)]_1 \times [SU(2) \times U(1)]_2$. 
In the LH model, the new particles appearing at the TeV scales are the heavy
gauge bosons ($W^\pm_H, Z_H, A_H$), the heavy top ($T$) and the scalar triplet 
$\Phi$.

In the original Littlest Higgs model (LH) \cite{Arkani-Hamed:2002qy}, the
custodial $SU(2)$ symmetry, of fundamental importance for electroweak precision
studies, is unfortunately broken already at tree level, implying
that the relevant scale of new physics, $f$, must be at least 2-3 TeV
 in order to be consistent with electroweak precision data
\cite{Han:2003wu, Csaki:2002qg, Hewett:2002px, Chen:2003fm, Chen:2004ig, Yue:2004xt,
Kilian:2003xt, Han:2003gf}. As a consequence, the contributions of the new
particles to FCNC processes turn out to be at most $10-20\%$
\cite{Buras:2004kq, Choudhury:2004bh, Buras:2005iv, Huo:2003vd, Buras:2006wk}, which will not be easy to distinguish from the
SM due to experimental and theoretical uncertainties.  In particular, 
a detailed analysis of particle-antiparticle mixing in the LH model has been
published in \cite{Buras:2004kq} and the corresponding analysis of rare $K$
and $B$ decays has recently been presented in \cite{Buras:2006wk}.\\

More promising and more interesting from the point of view of FCNC
processes is the Littlest Higgs model with a discrete symmetry
(T-parity)~\cite{Cheng:2003ju, Cheng:2004yc} under which all new particles listed
above, except $T_+$, are odd and do not contribute to processes
with external SM quarks (T-even) at tree level. As a
consequence, the new physics scale $f$ can be lowered down to 1 TeV and even below it, without violating electroweak precision
constraints \cite{Hubisz:2005tx}.\\

A consistent and phenomenologically viable Littlest Higgs model
with T-parity (LHT) requires the introduction of three doublets of
``mirror quarks'' and three doublets of ``mirror leptons'' which
are odd under T-parity, transform vectorially under $SU(2)_L$ and
can be given a large mass. Moreover, there is an additional
heavy $T_-$ quark that is odd under T-parity \cite{Low:2004xc}.\\

Mirror fermions are characterized by new flavour interactions with SM fermions
and heavy gauge bosons, which involve in the quark sector two new unitary 
mixing
matrices analogous to the CKM matrix  \cite{Chau:1984fp,Hagiwara:2002fs}.
They are $V_{Hd}$ and
$V_{Hu}$, respectively involved when the SM quark is of down- or up-type,
and satisfying $V_{Hu}^\dagger V_{Hd}=V_{\rm CKM}$ \cite{Kobayashi:1973fv}.
$V_{Hd}$ contains $3$ angles, like $V_{\rm CKM}$, but $3$
  (non-Majorana) phases \cite{Blanke:2006xr}, i.e. 
  two additional phases relative to the SM matrices, that cannot be rotated
  away in this case.

Because of these new mixing matrices, the LHT model does not belong to the
Minimal Flavour Violation (MFV) class of models~\cite{Buras:2000dm, Buras:2003jf, D'Ambrosio:2002ex} and 
significant effects in flavour observables are possible, without adding new 
operators to the  SM ones.
Finally, it is important to recall that Little Higgs models are low
energy non-linear sigma models, whose unknown UV-completion introduces a
theoretical uncertainty, as discussed in detail in~\cite{Buras:2006wk, Blanke:2006eb}.\\

The flavour physics analysis in the LHT model can be found in the case of quark sector in
\cite{Hubisz:2005bd, Blanke:2006eb, Blanke:2006sb} and in the lepton sector in
\cite{Choudhury:2006sq, Blanke:2007db}. Here we summarize the results obtained for $K \to \pi
\nu \bar \nu$ decays obtained in \cite{Blanke:2006eb}.\\

The presence of new flavour violating interactions between ordinary quarks and mirror quarks
described by the matrix 
\[ V_{Hd}=
\left( 
\begin{array}{ccc}
 c_{12}^d c_{13}^d & s_{12}^d c_{13}^d e^{-i\delta^d_{12}}& s_{13}^d e^{-i\delta^d_{13}}\\
-s_{12}^d c_{23}^d e^{i\delta^d_{12}}-c_{12}^d s_{23}^ds_{13}^d e^{i(\delta^d_{13}-\delta^d_{23})} &
c_{12}^d c_{23}^d-s_{12}^d s_{23}^ds_{13}^d e^{i(\delta^d_{13}-\delta^d_{12}-\delta^d_{23})} &
s_{23}^dc_{13}^d e^{-i\delta^d_{23}}\\
s_{12}^d s_{23}^d e^{i(\delta^d_{12}+\delta^d_{23})}-c_{12}^d c_{23}^ds_{13}^d e^{i\delta^d_{13}} &
-c_{12}^d s_{23}^d e^{i\delta^d_{23}}-s_{12}^d c_{23}^d s_{13}^d
e^{i(\delta^d_{13}-\delta^d_{12})} & c_{23}^d c_{13}^d\\
\end{array}
\right)
\]
introduces complex phases in the short distance functions $X_i$, $Y_i$ and $Z_i$ and breaks the
universality and correlations between $K$, $B_d$ and $B_s$ systems characteristic for the MFV
models. Spectacular results are found in particular for $\kpn$ and $\klpn$ decays. First one finds
\be
0.7 \leq |X| \leq 4.7, \qquad -130^\circ \leq \theta_X \leq 55^\circ
\ee 
to be compared with { $|X|=1.44$} and $\theta_X=0$ in the SM.
As already advertised in Section \ref{ssec:newphase}, a large phase $\theta_X$ can change totaly the pattern of
branching ratios in the $K\to \pi \nu \bar \nu$ system. This is clearly seen in Fig. \ref{fig:KLKp},
where we show the correlation between $Br(\kpn)$ and $Br(\klpn)$ in the LHT model. The experimental
$1\sigma$-range for $Br(\kpn)$ \cite{Adler:2001xv,Anisimovsky:2004hr} and the
model-independent Grossman-Nir (GN) bound \cite{Grossman:1997sk} are also
shown. The different colours in the figure correspond to different scenarios for the $V_{Hd}$
matrix whose detailed discussion is beyond the scope of this review.\\

We observe that there are two branches of possible points. The first one is parallel to the GN-bound
and leads to possible huge enhancements in $Br(\klpn)$ so that values
as high as $5\cdot 10^{-10}$ are possible, being at the same time
consistent with the measured value for $Br(\kpn)$. The second branch corresponds
to values for $Br(\klpn)$ being rather close to its SM prediction,
while  $Br(\kpn)$ is allowed to vary in the range $[1\cdot
10^{-11},5\cdot 10^{-10}]$, however, values above $4\cdot 10^{-10}$
are experimentally not favored. We note also that for certain values of the parameters of the
model $Br(\kpn)$ can be significantly suppressed.

\begin{figure}
\center{\epsfig{file=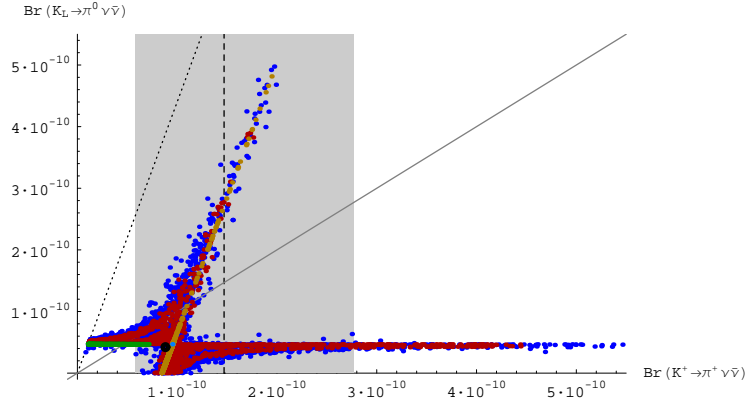,scale=0.6}}
\caption{\it $Br(\klpn)$ as a function of $Br(\kpn)$ in the LHT model. The shaded
    area represents the experimental $1\sigma$-range for $Br(\kpn)$. The
    GN-bound is displayed by the dotted line, while the solid line
    separates the two areas where $Br(\klpn)$ is larger or smaller than
    $Br(\kpn)$.}
\label{fig:KLKp}
\end{figure}

\begin{figure}
\center{\epsfig{file=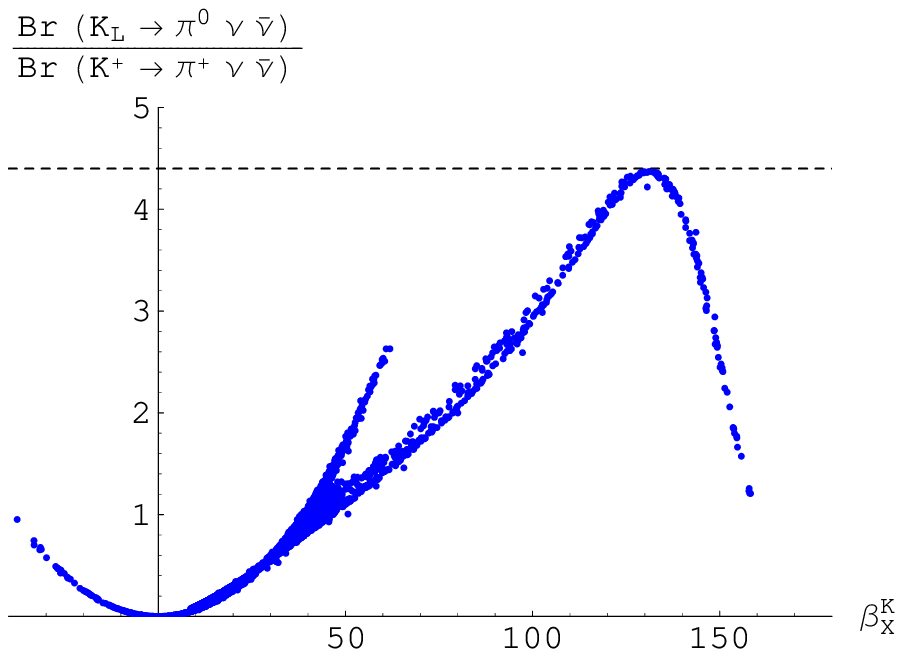,scale=0.7}}
\caption{\it $Br(\klpn)/Br(\kpn)$ in the LHT model as a function of $\beta_X^K$. 
  The dashed line represents the GN-bound.}
\label{fig:KrTK}
\end{figure}

In Fig.~\ref{fig:KrTK} we show the ratio $Br(\klpn)/Br(\kpn)$ as a function of
the phase $\beta_X^K$, displaying again the GN-bound.
We observe that the ratio can be significantly different from the SM
prediction, with a possible enhancement of an order of magnitude.\\

The most interesting implications of this analysis are:
\begin{itemize}
\item
If $Br(\kpn)$ is found sufficiently above the SM prediction but below 
$2.3\cdot 10^{-10}$, basically only two values for $Br(\klpn)$ are 
possible within the LHT model. One of these values is very close to
the SM value in (\ref{SMkp0}) and the second much larger.
\item
If $Br(\kpn)$ is found above $2.3\cdot 10^{-10}$, then only 
$Br(\klpn)$ with a value close to the SM one in (\ref{SMkl0})
is possible.
\item 
The violation of the MFV relation (\ref{R7}). We show this in Fig.  \ref{fig:rbetad13d}, where the
ratio of $\sin2\beta_X^K$ over $\sin(2\beta+2\varphi_{B_d})$ is plotted versus $\delta^d_{13}$.
As $\varphi_{B_d}$ is constrained by the measured $S_{\psi K_S}$ asymmetry 
to be at most a few degrees \cite{Blanke:2006ig, Bona:2006sa}, large violations of the relation in question 
can only follow from the $K\to\pi\nu\bar\nu$ decays. As seen in Fig.~\ref{fig:rbetad13d}, they can be spectacular.
\end{itemize}

\begin{figure}
\center{\epsfig{file=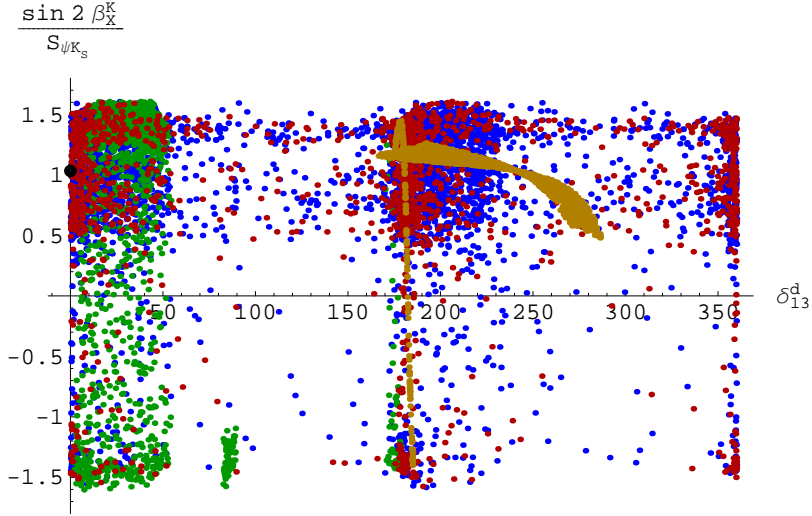,scale=0.9}}
\caption{\it $\sin2\beta_X^K/\sin(2\beta+2\varphi_{B_d})$ as a
  function of $\delta^d_{13}$ in the LHT model.}
\label{fig:rbetad13d}
\end{figure}

\begin{figure}
\center{\epsfig{file=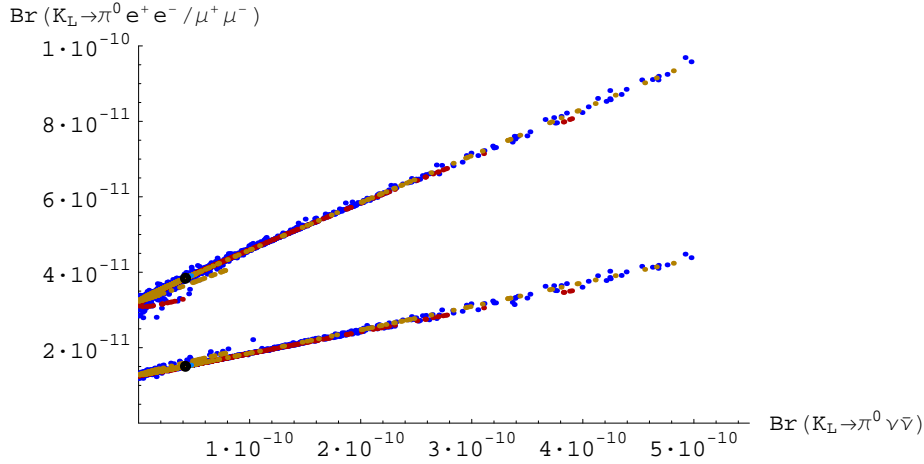,scale=0.90}}
\caption{\it $Br(K_L\to \pi^0 e^+e^-)$ (upper curve) and  $Br(K_L \to \pi^0
  \mu^+\mu^-)$ (lower curve) as functions of $Br(\klpn)$ in the LHT model. The corresponding SM
  predictions are represented by dark points.}
\label{fig:KemuKp}
\end{figure}

Finally in Fig.~\ref{fig:KemuKp}  we show
 $Br(K_L\to\pi^0 e^+e^-)$ and 
$Br(K_L\to\pi^0 \mu^+\mu^-)$ versus $Br(\klpn)$. 
We observe a strong correlation between $K_L\to \pi^0\ell^+\ell^-$ and 
$\klpn$ decays that we expect to be valid beyond the LHT model, at least
in models with the same operators present as in the SM. We note that
a large enhancement of $Br(\klpn)$ automatically implies significant 
enhancements of $Br(K_L\to \pi^0\ell^+\ell^-)$ and that
different models and their parameter sets can than be distinguished
by the position on the correlation curve.
Moreover, measuring $Br(K_L\to\pi^0\ell^+\ell^-)$ should 
allow a rather precise prediction of $Br(\klpn)$ at least in models
with the same operators as the SM. This should distinguish the LHT model from
models with more complicated operator structure in $K_L \to \pi^0 l^+ l^-$
\cite{Mescia:2006jd}, and consequently different correlations between $\klpn$ and
$K_L \to \pi^0 l^+ l^-$.

As emphasized in \cite{Buras:1998ed,Buras:1999da}, there exist correlations 
between $K\to\pi\nu\bar\nu$ 
decays, $K_L\to\mu^+\mu^-$ and $\epe$, that could bound the size of the
enhancement of  $Br(\kpn)$ and $Br(\klpn)$. Unfortunately, the 
hadronic uncertainties in $K_L\to\mu^+\mu^-$ and in particular in $\epe$
lower the usefulness of these correlations at present. More promising,  
in the context of supersymmetric models and also generally, appear the
correlations between $K\to\pi\nu\bar\nu$ and rare FCNC semileptonic decays
like $B\to X_{s,d} l^+l^-$, $B_{s,d}\to l^+l^-$ and in particular 
$B\to X_{s,d} \nu\bar\nu$, because also in these decays the main deviations
from the SM can be encoded in an effective $Z\bar b q$ ($q=s,d$) vertex 
\cite{Buchalla:2000sk,Atwood:2003tg}. We have discussed the correlation with $B\to X_{s,d} \nu\bar\nu$
in the previous section. \\

Recently the correlation between $\epsilon'/\epsilon$ and the decays $K \to \pi
\nu \bar \nu$ has been investigated in the context of the LHT model for specific
values of the relevant hadronic matrix elements entering $\epsilon'/\epsilon$
\cite{Blanke:2007wr}.
The resulting correlation between $\klpn$ and $\epsilon'/\epsilon$ is very
strong but less pronounced in the case of $\kpn$. With the hadronic matrix
elements evaluated in the large-N limit, $(\epsilon'/\epsilon)_{\rm SM}$ turns
out to be close to the experimental data and significant departures of
$Br(\klpn)$ and $Br(K_L \to \pi^0  l^+ l^-)$ from the SM expectations are
unlikely, while $Br(\kpn)$ can be enhanced by a factor of 5. On the other hand,
modest departures of the relevant hadronic matrix
elements from their large-N values allow for a consistent description of
$\epsilon'/\epsilon$ within the LHT model accompanied by large enhancements of
$Br(\klpn)$ and $Br(K_L \to \pi^0 l^+ l^-)$, but only modest
enhancements of $Br(\kpn)$. 
This analysis demonstrates very clearly that without a significant progress in
the evaluation of the hadronic parameters in $\epsilon'/\epsilon$, the role of
this ratio in constraining physics beyond the SM will remain limited.

\boldmath
\subsection{$Z^{\prime}$ Models}
\unboldmath

An additional neutral gauge boson can appear in several extensions of the standard model, such as Left-Right Symmetric Models, SUSY models with an additional $U(1)$ Factor, 
often arising in the breaking process of several GUT models, such as the breaking chain $SO(10) \to SU(5) \times U(1)$ or $E_6 \to SO(10) \times U(1)$, or in 331 models,
where the $SU(2)_L$ of the SM is extended to an $SU(3)_L$. In general, direct collider searches have already placed some lower bounds on a general $Z'$ mass, but FCNC 
processes can also provide valuable information on these particles, since additional contributions appear at tree level, if the $Z'$ transmits flavour changes. General,
model independent, analyses of $B$ decays as well as the mass differences $\Delta M_s$ can be found in \cite{Langacker:2000ju,Grossman:1999av,Barger:2003hg,Barger:2004hn}.
Additional interest in these contributions with respect to the $B$ meson system has arisen in the context of the CP asymmetries in $B^0_d \to \phi K_S$. In general, one
finds that sizeable contributions are still well possible but are rather unpredictive in this model independent context. On the other hand, predictive power
increases if the analysis is performed in a specific model.

As an example for this situation, we discuss the recent analysis \cite{Promberger:2007py} performed in the minimal 331 model \cite{Frampton:1992wt, Pisano:1991ee}. Here,
one has an $SU(3)_c \times SU(3)_L \times U(1)$ which is broken down to the electromagnetic $U(1)$ in two steps. In this process, the additional $Z'$ boson appears, along
with several charged gauge bosons, that play no role in low energy processes involving quarks, since they always couple to the additional heavy quarks that fill the 
left-handed triplet. Finally, the third generation of quarks is treated differently from the first two, transforming as an anti-triplet. In this setup, taking into account 
also the asymptotic freedom of QCD, one finds that anomalies are canceled precisely in the case of the three generations, thereby explaining this feature of the
SM, where the numbers of generations is fixed from observation. Apart from FCNC processes, constraints on the $Z'$ mass come also from electroweak precision observables,
but the new contributions here appear at the one loop level, so that the constraints from FCNCs are actually more interesting. Also, the model develops a Landau pole
at a scale of several $\mathrm{TeV}$, which constrains the new energy scale from above, thereby complementing the bounds from direct searches and FCNC observables.

The flavour non-universality reflected in the different transformation property of the third generation leads to the flavour changing $Z'$ vertices. The FCNC processes under
investigation then depend on the $Z'$ mass as well as the weak mixing matrix required to diagonalize the Yukawa coupling of the down quark sector (as long as one is studying
$B$ or $K$ meson processes). Also, one finds that the different processes decouple from each other, i.e. that $sd$, $bd$ and $bs$ transitions are constrained independently, so that only
the constraints from $\Delta M_K$ and $\varepsilon_K$ are used to constrain the branching fractions of $\klpn$ and $\kpn$. This leads to an allowed region 
in the $\klpn-\kpn$ plane shown in Fig. \ref{KLKPscat}. We show the corresponding areas for $M_{Z'}=1~\mathrm{TeV}$ as well as $M_{Z'}=5~\mathrm{TeV}$, where one observes
that the allowed region shrinks with  increasing $M_{Z'}$. The pattern is similar to the one shown in the LHT model, in that there exist two possible branches, where 
$Br(\klpn)$ is close to the SM on one of them, while $Br(\kpn)$ is on the other. This is due to the different strength of the $\varepsilon_K$ and $\Delta M_K$ bounds, 
respectively, and the large modifications arise in those areas, where the phase of the new contribution is such, that it does not modify $\varepsilon_K$ strongly.
Therefore a similar structure should appear whenever the $K \to \pi^0 \nu \bar \nu$ decays are constrained mainly by these two quantities.  
On the other hand, the minimal 331 model has a somewhat leptophobic nature, so that the effects are
not expected to be as large as, for example, in the LHT model, but the current experimental central value can be reached, in particular for $M_{Z'}<2 \mathrm{TeV}$.

Additionally, a measurement of both branching fractions fixes both the absolute value and phase of the new contributions (this is true in all $Z'$ models) and
allows predictions for the observables $\Delta M_K$ and $\varepsilon_K$ (this is of course only true if the model is explicitely fixed). 
Another interesting feature of this model is that there are significant differences 
between the vector and axial vector coupling, which cancel each other out in the $V-A$ difference, to which $Br(\klpn)$ is sensitive, so that, in comparison, one finds 
stronger modifications in the $K_L \to \pi^0 l^+ l^-$ branching fraction than in $\klpn$ \cite{Promberger:2007py}. Finally, significant modifications can also be found in
the angle $\beta|_{K\pi \nu \nu}$, which may be as large as $45^{\circ}$ for small values of $M_{Z'}$.

\begin{figure}
\begin{center}
\includegraphics[height=7cm]{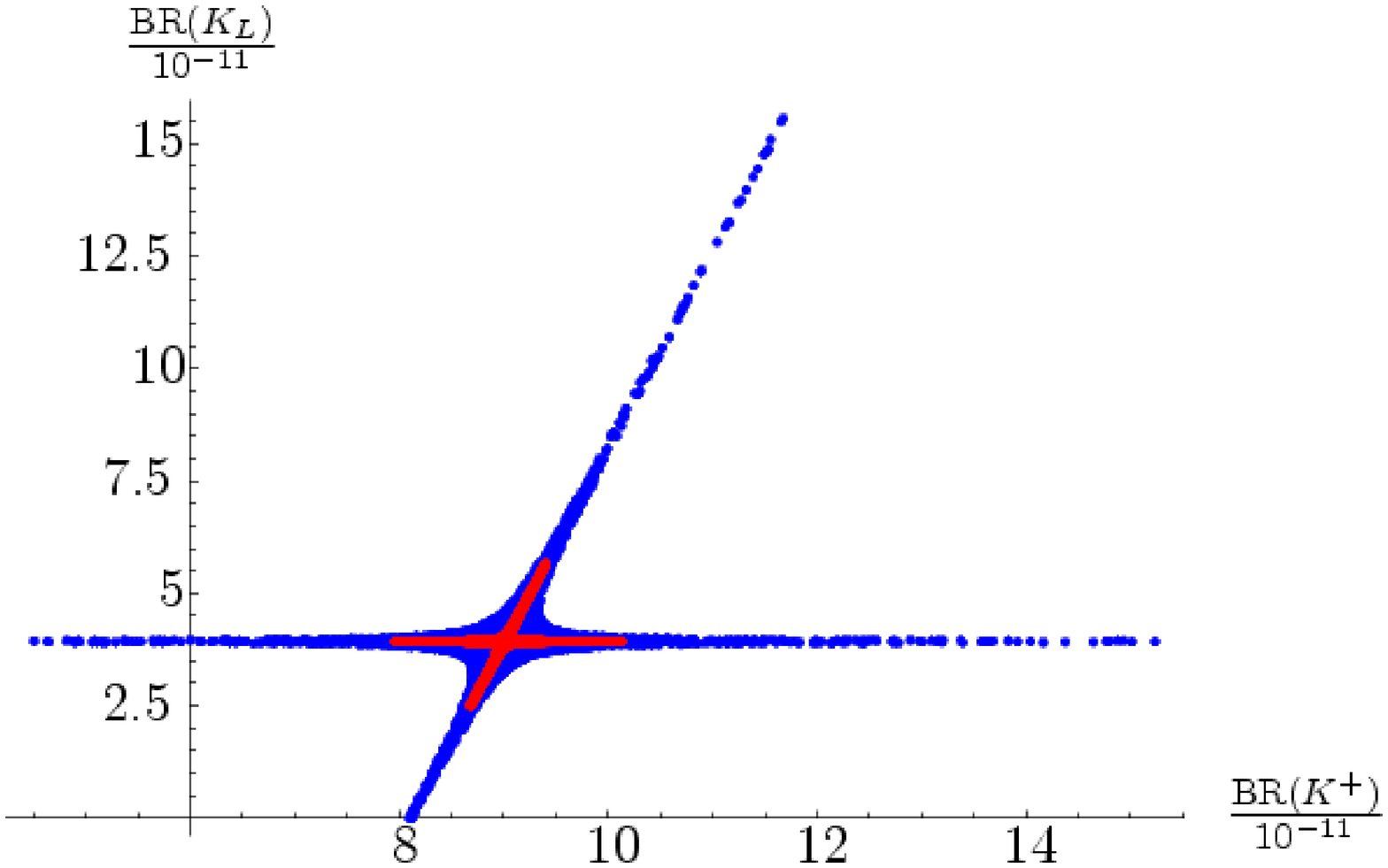}
\end{center}
\caption{\label{KLKPscat}A projection onto the $\klpn$-$\kpn$ plane including the upper bounds from $\Delta M_K$ and $\epsilon_K$ for $M_{Z'}=5~\mathrm{TeV}$ (red)
and  $M_{Z'}=1~\mathrm{TeV}$ (blue).}
\end{figure}

On the other hand, recently \cite{He:2004it,He:2006bk} the decays $K\to\pi\nu\bar\nu$ have been analyzed in
models that are variations of left-right symmetric models in which
right-handed interactions, involving in particular a heavy $Z^\prime$ boson, single out the third generation \cite{He:2002ha,He:2003qv}. 
The contributions of these new non-universal FCNC interactions appear both 
at the tree and one-loop level. The tree level contributions involving $Z^\prime$ of the type 
$(\bar s d)_{V+A}(\bar\nu_\tau\nu_{\tau})_{V+A}$ can be severely constrained by other rare decays, $\varepsilon_K$  and in particular $B^0_s-\bar B^0_s$ 
mixing. Before the measurement of $\Delta M_s$, these could enhance $Br(\kpn)$ to the central experimental value in (\ref{EXP1}) and 
$Br(\klpn)$ could be as high as $1.4\cdot 10^{-10}$. These enhancements where accompanied by an enhancement of $\Delta M_s$ and finding $\Delta M_s$ 
in the ball park of the SM expectations has significantly limited these possibilities \cite{He:2006bk}.
On the other hand new one loop contributions involving $Z^\prime$ boson may be important, because of the particularly large $\tau$ neutrino coupling.
They are not constrained by $B^0_s-\bar B^0_s$ mixing and can give significant enhancements of both branching ratios even if $\Delta M_s\approx (\Delta M_s)_{SM}$. 
Unfortunately the presence of many free parameters in these new one-loop contributions does
not allow to make definite predictions, but an enhancement by a factor of two still seems possible \cite{He:2006bk}. 

Finally, FCNC processes at tree level arise also if there is an additional vector-like quark generation, or, if there is only one additional isosinglet 
down-type or up-type quark, as one can encounter in certain $E_6$ GUT theories, or some models with extra dimensions.
In this case, the SM $Z$ itself boson can transmit flavor changes, since the mixing matrix of the respective quark sector is no longer unitary
and therefore does not cancel out in the neutral $Z$-current, causing FCNCs in the respective sector where the additional quark appears. 
The most recent analysis of the $K \to \pi \nu \bar \nu$ decays in this model has been presented in \cite{Deshpande:2004xc},
while a very complete analysis of FCNC processes in this type of scenario can be found 
in \cite{Barenboim:2001fd}. Here, the authors obtain constraints on the matrix element $U_{sd}$ (here $U= V^\dagger V$, with $V$ being
the mixing matrix that diagonalizes the down quark sector) from $\kpn$, $\varepsilon_K$, $\varepsilon'/\varepsilon_K$. 
Additionally, they emphasize that the $K \to \pi \nu \bar \nu$ decays can be very valuable for constraining this element further, if the decays are precisely measured.
In fact, one finds there a figure somewhat similar in spirit to the one shown in Fig. \ref{KLKPscat}, which shows an analogous interplay of constraints in the $K$ physics sector. 
We have included this figure as Fig. \ref{fig:constraints}.
\begin{figure}
\begin{center}
\epsfxsize = 0.5 \linewidth
\epsffile{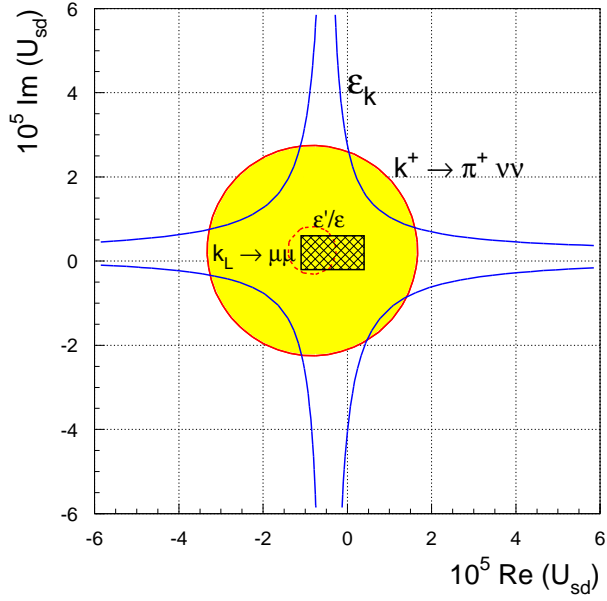}
\leavevmode
\end{center}
\caption{Effect of the constraints from $\varepsilon_K$, 
$K^+ \rightarrow \pi^+ \nu \bar{\nu}$, 
$K_L \rightarrow \mu^+ \mu^-$ and 
$\varepsilon^\prime/\varepsilon$ on the $U_{sd}$ FCNC coupling in case of an extra isosinglet down quark \cite{Barenboim:2001fd}.}
\label{fig:constraints}
\end{figure}

\boldmath
\subsection{MSSM with MFV}
\unboldmath
There are many new contributions in MSSM
such as charged Higgs, chargino, neutralino and gluino contributions.
However, in the case of
$K \to \pi \nu \bar \nu$ and MFV it is a good approximation to keep only
charged Higgs and chargino contributions.

To our knowledge the first analyses of $K\to\pi\nu\bar\nu$  in this scenario
can be found in \cite{Bertolini:1986bs,Giudice:1986jx,Mukhopadhyaya:1986dm,Bigi:1991qt}, subsequently in \cite{Couture:1995wd,Goto:1998qv} and 
 in \cite{Buras:2000qz}. 
In the latter analysis constraints on the
supersymmetric parameters from $\varepsilon_K$, $\Delta M_{d,s}$,
$B\to X_s\gamma$, $\Delta\varrho$ in the electroweak precision
studies and from the lower bound on the neutral Higgs mass
have been imposed. Supersymmetric contributions affect both the loop
functions like $X(v)$ present in the SM and the values of the extracted
CKM parameters like $\vtd$ and $\IM\lambda_t$. As the supersymmetric
contributions to the function $S(v)$ relevant for the analysis of 
the UT are always positive (see also \cite{Altmannshofer:2007cs}),
the extracted values of $\vtd$ and $\IM\lambda_t$ are always smaller
than in the SM. Consequently, $Br(\kpn)$ and 
$Br(\klpn)$, that are sensitive to $\vtd$ and 
$\IM\lambda_t$, respectively, are generally suppressed relative to the 
SM expectations.
The supersymmetric contributions to the loop function $X(v)$
can compensate the suppression of $\vtd$ and $\IM\lambda_t$
only for special values of supersymmetric parameters,
so that in these cases the results are very close to the SM expectations.

Setting $\lambda$, $|V_{ub}|$ and $\vcb$, 
all unaffected by SUSY contributions,
at their central values one finds \cite{Buras:2000qz}
\be\label{RR1}
0.65\le \frac{Br(\kpn)}{Br(\kpn)_{\rm SM}}\le 1.02,\qquad
0.41\le \frac{Br(\klpn)}{Br(\klpn)_{\rm SM}} \le 1.03.
\ee
We observe that significant suppressions of the branching ratios
 relative to the SM expectations are still possible.
More importantly, finding experimentally at least 
one of these branching ratios above the SM value would
exclude this scenario, indicating new flavour violating 
sources beyond the CKM matrix. Similarly, in the MSSM based on 
supergravity a reduction of both $K\to\pi\nu\bar\nu$ rates up to $10\%$ 
is possible \cite{Goto:1998qv}.

Reference \cite{Buras:2000qz} provides a compendium of phenomenologically
relevant formulae in the MSSM, that should turn out to be useful
once the relevant branching ratios have been accurately measured
and the supersymmetric particles have been discovered at Tevatron, LHC and
the $e^+e^-$ linear collider.
The study of the unitarity triangle can be found in \cite{Ali:1999we,Ali:1999xi,Ali:1999xm,Ali:2000hy}.
The inclusion of NLO QCD corrections to the processes discussed 
in \cite{Buras:2000qz} has been performed in \cite{Bobeth:2001jm}. These corrections 
reduce mainly the renormalization scale uncertainties present in
the analysis of \cite{Buras:2000qz}, without modifying the results in 
(\ref{RR1}) significantly.

\subsection{General Supersymmetric Models}

In general supersymmetric models the effects of supersymmetric 
contributions to rare branching ratios can be
larger than discussed above. In these models new CP-violating
phases and new operators are present. Moreover the structure
of flavour violating interactions is much richer than in the
MFV models. \\

The new flavour violating interactions are present
because generally the sfermion mass matrices $\tilde M_q^2$ can be non-diagonal
in the basis in which all neutral quark-squark-gaugino vertices and
quark and lepton mass matrices are flavour diagonal.
Instead
of diagonalizing sfermion mass matrices it is convenient
to consider their off-diagonal terms as new flavour violating
interactions. 
This  so--called mass--insertion
approximation \cite{Hall:1985dx} has been reviewed in the classic
papers \cite{Gabbiani:1996hi,Misiak:1997ei}, where further references can be found.

Within the MSSM with $R$-parity conservation, sizable non-standard contributions 
to $K \to \pi \nu\overline{\nu}$ decays can be generated 
if the soft-breaking terms have a non-MFV structure. 
The leading amplitudes giving rise to large effects are induced by:
i) chargino/up-squark 
loops~\cite{Nir:1997tf,Buras:1997ij,Colangelo:1998pm,Buras:1999da}
ii) charged Higgs/top quark loops~\cite{Isidori:2006jh}.
In the  first case, large effects are generated if the left-right mixing
($A$ term) of the  up-squarks has a non-MFV structure \cite{D'Ambrosio:2002ex}.
In the second case, deviations from the SM are induced by non-MFV terms 
in the right-right down sector, provided the ratio of the two Higgs  vacuum expectation values
($\tan \beta = v_u/v_d$) is large ($\tan \beta \sim 30-50$).

The effective Hamiltonian encoding SD contributions in the 
general MSSM has the following structure:
\be
\label{Ht}
{\mathcal H}_{\rm eff}^{( \rm SD)} \propto 
\sum_{l=e,\mu,\tau} V^{\ast}_{ts}V_{td} 
\left[X_L (\bar s_L \gamma^\mu d_L)(\bar\nu_{l L} \gamma_\mu \nu_{l L}) + 
X_R (\bar s_R \gamma^\mu d_R)(\bar\nu_{l L} \gamma_\mu \nu_{l L})\right]~,
\ee
where the SM case is recovered for ${X_R=0}$ and $X_L=X_{\rm SM}$. 
In general, both $X_R$ and  $X_L$ are  non-vanishing, 
and the misalignment between quark and squark flavour 
structures implies that they are both complex quantities. 
Since the $K\to\pi$ matrix elements of  $(\bar s_L \gamma^\mu d_L)$ and
$(\bar s_R \gamma^\mu  d_R)$ are equal, the combination $X_L+X_R$
allows us to describe all the  SD
contributions to $K \to \pi \nu\overline{\nu}$ decays.
More precisely, we can simply use the SM expressions 
for the branching ratios 
with the following replacement 
\be
X_{\rm SM} \to  X_{\rm SM} + X^{\rm SUSY}_L + X^{\rm SUSY}_R~,
\ee
with $X_{L,R}^{\rm SUSY}$ being complex quantities.
In the limit of almost degenerate superpartners, the leading chargino/up-squarks
contribution is~\cite{Colangelo:1998pm}:
\be
X^{\chi^{\pm}}_L
\approx \frac{1}{96}
\left[\frac{(\delta^{u}_{LR})_{23} (\delta^{u}_{RL})_{31}}{\lambda_t}\right] 
~=~ \frac{1}{96 \lambda_t }
\left[\frac{(\tilde M^{2}_{u})_{2_L 3_R}}{(\tilde M^{2}_{u})_{LL} 
(\tilde M^{2}_{u})_{RR}} \right]
\left[\frac{(\tilde M^{2}_{u})_{3_R 1_L}}{(\tilde M^{2}_{u})_{LL}
(\tilde M^{2}_{u})_{RR}} \right]
\,.
\label{eq:XL_eff}
\ee

Here $(\delta_{AB}^q)_{ij}$ result from a convenient parametrization
\cite{Gabbiani:1996hi,Misiak:1997ei} of the non-diagonal terms $(\tilde
M_u^2)_{iAjB}$ in squark mass matrices with $A,B=L,R$ and $i,j=1,2,3$ standing
for quark generation indices.
As pointed out  in \cite{Colangelo:1998pm}, a remarkable feature of 
the above result is that no extra $\mathcal O(M_{W}/M_{\rm SUSY})$ suppression 
and no explicit CKM suppression is present (as it happens in
the chargino/up-squark contributions to other processes). 
 Furthermore, the $(\delta^{u}_{LR})$-type mass insertions 
are not strongly constrained by other  $B$- and $K$-observables.
This implies that large departures from the SM 
expectations in  $K \to \pi \nu\overline{\nu}$ decays are allowed,
as confirmed by the complete analyses  in \cite{Buras:2004qb,Isidori:2006qy}. 
In particular in \cite{Buras:2004qb} one finds that both branching ratios can be as large as few times $10^{-10}$ with
$Br(\klpn)$ often larger than $Br(\kpn)$ and close to the GN bound. One also finds \cite{Isidori:2006qy} that 
$K \to \pi \nu\overline{\nu}$ are the best observables 
to determine/constrain from experimental data the size of the off-diagonal 
 $(\delta^{u}_{LR})$ mass insertions or, equivalently, the
up-type trilinear terms $A_{i3}$ 
[$(\tilde M^{2}_{u})_{i_L 3_R} \approx m_t A_{i3}$].
Their measurement is therefore extremely interesting 
also in the LHC era. 

In the large $\tan\beta$ limit, 
the charged Higgs/ top quark exchange leads to \cite{Isidori:2006jh}:
\beqa
X^{H^{\pm}}_R \approx
\left[\left(\frac{m_{s}m_{d}\,t^{2}_{\beta}}{2 M_W^2}\right)+
\frac{(\delta^d_{RR})_{31} (\delta^d_{RR})_{32}}{\lambda_t}
\left(\frac{m^{2}_{b}\,t^{2}_{\beta}}{2 M_W^2}\right)
\frac{\epsilon^{2}_{RR}t^{2}_{\beta}}{(1+\epsilon_{i}t_{\beta})^4}
\right]f_H(y_{tH})\,,
\label{eq:XR_eff}
\eeqa
where $y_{tH}=m_t^2/M_H^2$, $f_H (x)=x/4(1-x)+x\log x/4(x-1)^2$
and $\epsilon_{i, RR} t_{\beta} = \mathcal O(1)$ for 
$t_{\beta}=\tan\beta\sim 50$.
The first term of Eq.~(\ref{eq:XR_eff}) arises from MFV effects and
its potential $\tan\beta$ enhancement is more than compensated
by the smallness of $m_{d,s}$.
The second term on the r.h.s.~of Eq.~(\ref{eq:XR_eff}),
which would appear only at the three-loop level in a standard loop
expansion can be largely enhanced by the $\tan^4\beta$ factor
and does not contain any suppression due to light quark masses.
Similarly to the double mass-insertion mechanism of Eq.~(\ref{eq:XL_eff}),
also in this case the potentially leading effect is the one generated
when two off-diagonal squark mixing terms replace the two CKM
factors $V_{ts}$ and $V_{td}$.

The coupling of the $ (\bar s_R \gamma^\mu d_R) (\bar\nu_L \gamma_\mu \nu_L)$
effective FCNC operator, generated by charged-Higgs/top quark loops
is phenomenologically relevant only at large $\tan\beta$ and with non-MFV 
right-right soft-breaking terms: a specific but well-motivated scenario
within grand-unified theories (see  e.g. \cite{Moroi:2000tk,Chang:2002mq}).
These non-standard effects do not vanish in the limit of heavy squarks and 
gauginos, and have a slow decoupling with respect to the charged-Higgs boson mass.
As shown  in \cite{Isidori:2006jh}
the $B$-physics constraints still allow 
a large room of non-standard effects in $K \to \pi \nu\overline{\nu}$
even for flavour-mixing terms of CKM size (see Fig.~\ref{fig_SUSY}).

\begin{figure}[t]
\begin{center}
\includegraphics[width=7cm]{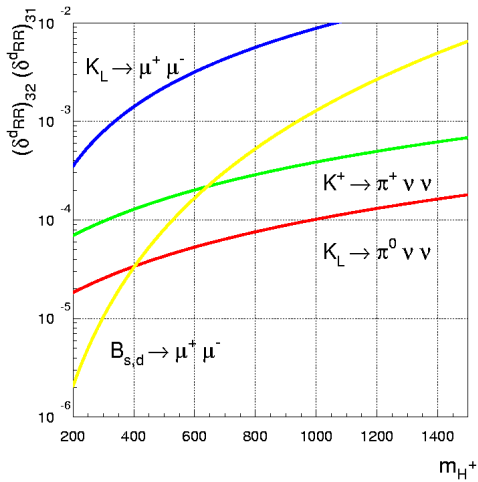}
\end{center}
\caption{\label{fig_SUSY} Supersymmetric contributions to 
$K \to \pi \nu\bar\nu$.  
Sensitivity to $(\delta^{d}_{RR})_{23}(\delta^{d}_{RR})_{31}$
of various rare  $K$- and $B$-decays as a function 
of $M_{H^+}$,  setting $\tan\beta\!=\!50$, $\mu\!<\!0$
and assuming almost degenerate superparteners
(the bounds from the two $K\rightarrow \pi\nu\bar{\nu}$ modes are 
obtained assuming a $10\%$ measurement of their 
branching ratios while the $B_{s,d} \rightarrow \mu^+\mu^-$ bounds 
refer to the present experimental limits \cite{Isidori:2006jh}). }
\begin{picture}(0,0)(0,0)
\put(0,300){a)} 
\put(270,300){b)} 
\end{picture}
\end{figure}

A systematic study of $K\to\pi\nu\bar\nu$ decays in flavour supersymmetric 
models has been performed in \cite{Nir:1997tf,Nir:2002ah}. These particular models 
are designed to solve naturally the CP and flavour problems characteristic
for supersymmetric theories.\footnote{See the review in \cite{Grossman:1997pa}.}
 They are more constrained than the general
supersymmetric models just discussed, in which parameters are tuned to
satisfy the experimental constraints.

Models with exact universality of squark masses at a high energy scale 
with the $A$ terms proportional  to the corresponding Yukawa couplings, 
models with approximate CP, quark and squark alignment, approximate
universality and heavy squarks have been analyzed in  
\cite{Nir:1997tf,Nir:2002ah} in general terms. It has been concluded that in most of
these models the impact of new physics on $K\to\pi\nu\bar\nu$ is sufficiently
small so that in these scenarios one can get information on the CKM matrix 
from these decays even in the presence of supersymmetry.
On the other hand, supersymmetric contributions to $B^0_d-\bar B^0_d$ 
mixing in models with alignment, with approximate universality and heavy
squarks can significantly affect the asymmetry $a_{\psi K_S}$, so that 
in these models the golden relation (\ref{R7}) can be violated.  However
such scenarios have been put under large pressure in
view of the recent data on $D^0-\bar D^0$ mixing
\cite{Nir:2007ac,Ciuchini:2007cw}.

Finally, in
supersymmetric models with non-universal $A$ terms, enhancements of $Br(\kpn)$
and $Br(\klpn)$ up to $1.5\cdot 10^{-10}$ and $2.5\cdot 10^{-10}$ are
possible, respectively \cite{Chen:2002eh}. Significant departures from the SM
expectations have also been found in supersymmetric models with R-parity 
breaking \cite{Deandrea:2004ae}, { but all these analyses should be reconsidered in view of experimental
constraints.}

\subsection{Models with Universal Extra Dimensions}
The decays $\kpn$ and $\klpn$ have been studied in
the SM model with one extra 
universal dimension in \cite{Buras:2002ej}. 
In this model (ACD) \cite{Appelquist:2000nn}  all the SM fields 
are allowed to propagate in all available dimensions  
and the relevant 
penguin and box diagrams receive additional contributions from 
Kaluza-Klein (KK) modes.  
This model belongs to the class of CMFV models and
the only additional free parameter relative to the SM is the
compactification scale $1/R$. 
Extensive analyses of the precision electroweak data, the analyses of the 
anomalous magnetic moment
of the muon and of the $Z\to b\bar b$ vertex have shown the consistency of the 
ACD model with the data for $1/R\ge 300\gev$.
We refer to \cite{Buras:2002ej,Buras:2003mk} for the list of relevant papers.

For $1/R=300\gev$ and $1/R=400\gev$ the function $X$ is found with 
$m_t=167\gev$ to be $X=1.67$ and $X=1.61$, respectively. This should 
be compared with $X=1.53$ in the SM. In contrast to the analysis in 
the MSSM discussed in \cite{Buras:2000qz} and above, this $5-10\%$ enhancement 
of the function $X$ is only insignificantly compensated by the change 
in the values of the CKM parameters. Consequently, the clear prediction 
of the model are the enhanced branching ratios $Br(\kpn)$ and $Br(\klpn)$, 
albeit by at most $15\%$ relative to the SM expectation.
These enhancements allow to distinguish this scenario from the MSSM with 
MFV. 

The enhancement of $Br(\kpn)$ in the ACD model is interesting in view
of the experimental results in (\ref{EXP1})
with the central value  by a factor of $1.8$ higher than the central 
value in the SM. Even if 
the errors are substantial and this result is compatible with the SM, 
the ACD model with a low compactification scale is  closer to the 
data. In  table~\ref{Bound} we show 
the upper bound on $Br(\kpn)$ in the ACD model 
obtained in \cite{Buras:2002ej} by means of the formula (\ref{AIACD}), with $X$ 
replaced by its enhanced value in the model in question. 
To this end 
$\vcb\le 0.0422$, $P_c(X)<0.47$, $\mt(m_t)<172~\gev$
and $\sin 2 \beta = 0.734$ have been used.  
Table~\ref{Bound} illustrates the dependence of the bound on the 
nonperturbative parameter $\xi$, $1/R$ and $\Delta M_s$.
We observe that for $1/R=300~\gev$ and $\xi=1.30$ the maximal value
for $Br(\kpn)$ in the ACD model is rather close to the central value in
(\ref{EXP1}). 

\begin{table}[hbt]
\vspace{0.4cm}
\begin{center}
\caption[]{\small Upper bound on $Br(\kpn)$ in units of $10^{-11}$ for 
different 
values of $\xi$, $1/R$ and $\Delta M_s=18/{\rm ps}~(21/{\rm ps})$ from 
\cite{Buras:2002ej}.  
\label{Bound}}
\begin{tabular}{|c||c|c|c|}\hline
{$\xi$ } & {$1/R=300~\gev$} 
& {$1/R=400~\gev$} & SM
 \\ \hline
$1.30$ &  $ 12.0~(10.7)$ & 
$ 11.3~(10.1)$ &  $10.8~(9.3)$ \\ \hline
$1.25$ &  $ 11.4~(10.2) $ & 
$ 10.7~(9.6)$ & $10.3~(8.8)$ \\ \hline
$1.20$ &  $ 10.7~(9.6) $ & 
$ 10.1~(9.1)$ & $9.7~(8.4)$\\ \hline
$1.15$ &  $ 10.1~(9.0) $ & $ 9.5~(8.5)$ &
$9.1~(7.9)$ \\ \hline
\end{tabular}
\end{center}
\end{table}

Clearly, in order to distinguish these results and the ACD model from the SM,
other quantities, that are more sensitive to $1/R$, should be simultaneously 
considered.
In this respect,  the sizable downward 
shift of  the zero ($\hat s_0$) in the 
forward-backward asymmetry $A_{\rm FB}$ in $B\to X_s \mu^+\mu^-$  and 
the suppression of $Br(B\to X_s\gamma)$ by roughly $20\%$ at $1/R=300\gev$ 
appear to be most interesting \cite{Buras:2003mk}. 

 As the most recent analysis of the $B\to X_s \gamma$ decay at the NNLO level
results in its SM branching ratio being by more than one $\sigma$ below the
experimental values, the model in question is put therefore under considerable
pressure and the values of $1/R$ as low as 300 GeV appear rather improbable from
the present perspective \cite{Haisch:2007vb}. A decrease of the experimental
error without a significant change of its central value and a better
understanding of non-perturbative effects in the $B\to X_s\gamma$ decay could
result in $1/R\approx \ord (1 {\rm TeV})$ and consequently very small new physics
effects in $K \to \pi \nu \bar \nu$ decays in this model.

\subsection{Models with Lepton-Flavour Mixing}
In the presence of flavour mixing in
the leptonic sector,  the transition $K_L\to\pi^0\nu_i\bar\nu_j$, with 
$i\not=j$ could receive significant CP-conserving contributions \cite{Grossman:1997sk}. 
Subsequently
this issue has been analyzed in \cite{Perez:1999kw,Perez:2000fx} and 
in \cite{Grossman:2003rw}.
Here we summarize briefly the main findings of these papers.

In \cite{Perez:1999kw,Perez:2000fx} the effect of light sterile right-handed neutrinos 
leading to scalar and tensor dimension-six operators has been 
analyzed. As shown there, the effect of these 
operators is negligible, if the right-handed neutrinos interact with 
the SM fields only through their Dirac mass terms.

Larger effects are expected from the operators 
\be
O^{ij}_{sd}=(\bar s\gamma_\mu d) (\bar\nu^i_L\gamma^\mu\nu^j_L), 
\ee
that for $(i\not=j)$  create a neutrino pair which is not a CP eigenstate.
As shown in \cite{Grossman:2003rw} the condition for a non-vanishing $\klpn$ 
rate in this 
case is rather strong. One needs either CP violation in the quark sector 
or a new effective interaction that violates both quark and lepton
universality. One finds then the following pattern of effects:
\begin{itemize}
\item
If the source of  universality breaking is confined to mass matrices,  
the effects of lepton-flavour mixing get washed out in the
$K\to\pi\nu\bar\nu$ rates after the sum over the neutrino flavours has 
been done. There are in principle detectable effects of lepton mixing 
only in cases where there are two different lepton-flavour mixing
matrices, although they cannot be large.
\item
In models in which  simultaneous violation of quark and lepton universality
proceeds entirely through Yukawa couplings, the CP conserving effects in 
$K\to\pi\nu\bar\nu$ are suppressed by Yukawa couplings. As explicitly 
shown in \cite{Grossman:2003rw} even in the MSSM with flavour violation and large 
$\tan\beta$ these types of effects are negligible.
\item
In exotic scenarios, such as R-parity violating supersymmetric models, 
lepton flavour mixing could generate sizable CP-conserving contributions 
to $\klpn$ and generally in $K\to\pi\nu\bar\nu$ rates.
\end{itemize}

\subsection{Other Models}
There exist other numerous analyses of $K\to\pi\nu\bar\nu$ decays within 
various extensions of the SM. 
For completeness we briefly describe them here.

In \cite{Carlson:1996jh} the rate for $\klpn$ has been calculated in several 
extensions of the 
SM Higgs 
sector, including the Liu-Wolfenstein two-doublet model of spontaneous CP-
violation and the Weinberg three doublet model. 
It has been concluded that although in the usual two Higgs doublet model, 
with CP-violation governed by the CKM matrix, some measurable effects 
could be seen, in models in which CP-violation arises either entirely or 
predominantly from the Higgs sector the decay rate is much smaller than in 
the SM.

The study of $K\to\pi\nu\bar\nu$ in models with four generations, 
extra vector-like quarks  and isosinglet down quarks 
 can be found
in \cite{Huang:2000xe,Hung:2001wy,Hawkins:2002qb,Yanir:2002cq,Aguilar-Saavedra:2002kr,Hattori:1998mv}. 
In particular in four generation models \cite{Huang:2000xe,Yanir:2002cq,Hattori:1998mv} due to
 three additional  mixing angles and
two additional  complex phases, $Br(\klpn)$ can be enhanced by 1-2 orders of 
magnitude 
with respect to the SM expectations and also $Br(\kpn)$ can be significantly 
enhanced. Unfortunately, due to many free parameters, the four generation 
models are not very predictive. A new analysis of 
$K\to\pi\nu\bar\nu$ in a model with an extra isosinglet down quark appeared 
in \cite{Deshpande:2004xc}. Putting all the available constraints on the parameters of 
this model, the authors conclude that $Br(\kpn)$ can still be enhanced up 
to the present experimental central value, while $Br(\klpn)$ can reach 
$1\cdot 10^{-10}$.

The decays $K\to\pi\nu\bar\nu$ have also been investigated in a seesaw 
model for quark masses \cite{Kiyo:1998zm}. In this model there are scalar 
operators $(\bar s d)(\bar \nu_{\tau}\nu_{\tau})$, resulting from LR box
diagrams, that make the rate for $\klpn$ non-vanishing even in the CP
conserving limit and in the absence of lepton-flavour mixing. But the 
enhancement of $Br(\klpn)$ due to these operators is at most of order $30\%$ 
even for $M_{W_R}=500\gev$ with a smaller effect in $Br(\kpn)$. 

The effects of the electroweak symmetry breaking on rare $K$ and $B$ decays, 
including $K\to\pi\nu\bar\nu$, in the presence of new strong dynamics, have
been worked out in \cite{Buchalla:1995dp,Burdman:1997qn}. Deviations from the SM in 
$K\to\pi\nu\bar\nu$ have been shown to be correlated with the ones in $B$ 
decays \cite{Burdman:1997qn}.

The implications of a modified effective $Zb\bar b$ vertex on $K\to
\pi\nu\bar\nu$, in connection
with the small disagreement between the SM and the measured asymmetry
$A^b_{FB}$ at LEP, have been discussed in \cite{Chanowitz:2001bv,Chanowitz:1999jj}. While 
the predictions are rather uncertain, an enhancement of $Br(\kpn)$ by 
a factor of two, towards the central experimental value, is possible.

Enhancement of both $K\to\pi\nu\bar\nu$ branching ratios up to $50\%$ 
has been found in a five dimensional split fermions scenario \cite{Chang:2002ww} 
and the decay $\kpn$ turns out to be the best for providing the 
constraints on the bulk SM in the Randall-Sundrum scenario \cite{Burdman:2002gr}. 

\subsection{Summary}
We have seen in this and the previous section that many scenarios of new physics
allow still for significant enhancements of both $Br(\kpn)$ and $Br(\klpn)$:
$Br(\kpn)$ can still be enhanced by factors of 2-3 and $Br(\klpn)$ could be 
by an order of magnitude larger than expected within the SM. While for 
obvious reasons most of the papers concentrate on possible enhancements of 
both branching ratios, their suppressions in several scenarios are still
possible. This is in particular the case of the MSSM with MFV and in several
models in which CP violation arises from the Higgs sector.

Because most models contain several free parameters, definite predictions for
$K\to\pi\nu\bar\nu$ can only be achieved  by considering simultaneously as
many processes as possible so that these parameters are sufficiently
constrained.

\section{Comparison with Other Decays}\label{sec:decays}
After this exposition of $\kpn$ and $\klpn$ decays in the SM and its most 
studied extensions we would like to briefly compare the potential of these two clean 
rare decays in extracting the CKM parameters and in testing the SM and 
its extensions with other  
prominent $K$ and $B$ decays for which a rich literature exists.
A subset of relevant references will be given below.

In the $K$ system, the most investigated in the past are the parameters 
$\varepsilon_K$ and the ratio $\varepsilon'/\varepsilon$ that describe 
respectively the 
indirect and direct CP violation in $K_L\to \pi\pi$ decays and 
the rare decays $K_L\to\mu^+\mu^-$ and $K_L\to\pi^0 e^+ e^-$. 
None of them can compete in the theoretical 
cleanness with the decays considered here but some of them are still useful.

While $K_L\to \mu^+ \mu^-$ and $\varepsilon^\prime /\varepsilon$ suffer from
large hadronic uncertainties, the case of the decays $K_L \to \pi^0 \mu^+ \mu^-$
and $K_L \to \pi^0 e^+ e^-$ is much more promising. They provide an interesting and complementary window to $|\Delta S|=1$ SD
transitions. While the latter is theoretically not as clean as the
$K\to \pi \nu \bar \nu$ system, it is sensitive to different types of
SD operators. The $K_L \to \pi^0 \ell^+\ell^-$ decay amplitudes have
three main ingredients: i)~a clean direct-CP-Violating (CPV) component
determined by SD dynamics; ii)~an indirect-CPV term due
to $K^0$--$\overline{K^0}$ mixing; iii)~a LD CP-Conserving (CPC) component
due to two-photon intermediate states. Although generated by very
different dynamics, these three components are of comparable size and
can be computed (or indirectly determined) to good accuracy within
the SM \cite{Buchalla:2003sj, Isidori:2004rb}. In the presence of
non-vanishing NP contributions, the combined measurements of $K\to \pi
\nu \bar \nu$ and $K_L \to \pi^0 \ell^+\ell^-$ decays provide a unique
tool to distinguish among different NP models.

Most advanced analyses of these decays within the SM can be found in
\cite{Buchalla:2003sj, Isidori:2004rb, Friot:2004yr}, where further references
to earlier literature can be found. We would like also to mention the recent
analyses of these decays in the context of the MSSM \cite{Isidori:2006qy} and
other NP scenarios \cite{Mescia:2006jd}, in
particular in the LHT model \cite{Blanke:2006eb}.

The situation with $B$ decays is very different. First of all there are many 
more channels than in $K$ decays, which allows to eliminate or reduce many 
hadronic uncertainties by simultaneously considering several decays and 
using flavour symmetries. Also the fact that now the $b$ quark mass 
is involved in the effective theory allows  to calculate
hadronic amplitudes in an expansion in the inverse power of the $b$ quark mass 
and invoke 
related heavy quark effective theory, heavy quark expansions, QCD
factorization for non-leptonic decays, perturbative QCD approach and others. 
During the last years considerable advances in this field have been made 
\cite{Battaglia:2003in}. 
While in semi-leptonic tree level decays this progress allowed to decrease
the errors on the elements $|V_{ub}|$ and $\vcb$ \cite{Battaglia:2003in}, 
in the case of prominent 
radiative decays like $B\to X_s\gamma$ and $B\to X_s l^+ l^-$, these 
methods allowed for a better estimate of hadronic uncertainties. 
In addition during last decade and in this decade theoretical uncertainties
in these decays 
have been considerably reduced
through the computations of NLO and in certain cases NNLO 
QCD corrections \cite{Misiak:2006zs,Buchalla:1995vs,Buras:1998ra,Fleischer:2002ys,Fleischer:2004xw,Ali:2003te,Hurth:2003vb,Nir:2001ge,Buchalla:2003ux}.

In the case of non-leptonic decays, various strategies for the determination 
of the angles of the unitarity triangle have been proposed. 
 Excellent reviews of these strategies are   
\cite{Cavoto:2007fp,Fleischer:2002ys,Fleischer:2004xw}. See also \cite{Buras:2005xt,Buras:2003jf,Buras:2004sc} and \cite{Ali:2003te,Hurth:2003vb,Nir:2001ge,Buchalla:2003ux}.
These strategies generally 
use simultaneously several decays and are based on plausible dynamical
assumptions that can be furthermore tested by invoking still other decays. 

There is no doubt that these methods will give us considerable insight 
into flavour and QCD dynamics but it is fair to say that most of them 
cannot match the $K\to \pi\nu\bar\nu$ decays with respect to the theoretical 
cleanness. On the other hand there exist a number of strategies for the
determination of the angles and also sides of the unitarity triangle that 
certainly can compete with the $K\to\pi\nu\bar\nu$ complex and in certain 
cases are even slightly superior to it, provided corresponding measurements 
can be made precisely. 

Yet, the present status of FCNC processes in the $B_d$-system indicates that the
new physics in this system enters only at a subleading level. While certain
departures from the SM are still to be clarified, this will not be easy in
particular in the case of non-leptonic decays.

More promising from the point of view of the search of new physics is the
$B_s$-system. While the measurement of $\Delta M_s$ did not reveal large
contributions from NP, the case of the CP asymmetry $S_{\psi \phi}$ and of the
branching ratios $Br(B_{d,s}\to \mu^+\mu^-)$ could be very different as they all
are very strongly suppressed within the SM. The experiments at LHC will
undoubtly answer the important question, whether these observables signal NP
beyond the SM.  Even more detailed investigations will be available at
a Super-B machine.

\section{Subleading Contributions to \boldmath{$K\to\pi\nu\bar\nu$}}
\label{sec:longdist}

In this section we discuss briefly the subleading
contributions to the decays $\kpn$ and $\klpn$ that we have neglected 
so far. More detailed discussions and explicit calculations have been 
presented in {\cite{Rein:1989tr,Hagelin:1989wt,Lu:1994ww,Fajfer:1996tc,Geng:1996kd,Ecker:1987hd,Buchalla:1998ux,Isidori:2005xm}}.
These effects can be potentially interesting especially when the NNLO
calculation anticipated in Section \ref{sec:num} is actually performed
and $Br(\kpn)$ and $Br(\klpn)$ are measured with an accuracy of 5$\%$.

Accordingly, we begin with the discussion of $\kpn$, where there can be,
in principle, two additional contributions to the branching ratio:
\begin{itemize}
\item Effects through soft $u$ quarks in the penguin loop that induce an on
shell $K^+ \to \pi^+ Z^0 \to \pi^+ \nu\bar\nu$ transition as well
as similar processes induced by $W-W$ exchange. These are long
distance effects and addressed
{\cite{Rein:1989tr,Hagelin:1989wt,Lu:1994ww,Fajfer:1996tc,Geng:1996kd,Ecker:1987hd}}
in chiral perturbation theory. 
 
\item Higher dimensional operators contributing to the
  OPE in the charm sector {\cite{Falk:2000nm}}.
\end{itemize}
Most recently, both effects have been investigated in detail in
{\cite{Isidori:2005xm}}, paying in particular attention to the
cancellation of the renormalization scale dependence between both contributions.

Therefore, we follow here {\cite{Isidori:2005xm}} in a more elaborate
discussion of both effects in detail:
In particular, concerning the effects of higher dimensional operators,
the results of {\cite{Falk:2000nm}} have been fully confirmed. These contributions have to be considered
only in the charm sector, if one assumes a natural scaling of
$~ M_K^2/m_q^2$ in the Wilson coefficients. The scaling of the Inami-Lim
functions then leads to an overall scaling of $M_K^2/M_W^4$, which is
independent of the quark masses. The top contribution is then simply
suppressed by CKM factors. 

Going to dimension eight, one finds two operators that appear when
expanding the penguin and box diagrams:
\begin{eqnarray}
O^l_1&=& \bar s
\gamma^{\nu}(1-\gamma^5)d(i\partial)^2\left(\bar\nu_l\gamma_{\nu}(1-\gamma^5)\nu_l\right)
 \nonumber\\
O^l_2&=&
\bar s\gamma^{\nu}(1-\gamma^5)(iD)^2 d~\bar\nu_l\gamma_{\nu}
(1-\gamma^5)\nu_l + \nonumber\\ &&
 2\bar s\gamma^{\nu}(1-\gamma^5)(iD^{\mu}) 
 d~\bar\nu_l\gamma_{\nu}(1-\gamma^5)(\partial_{\mu})\nu_l + \nonumber\\&&
\bar s
\gamma^{\nu}(1-\gamma^5)d~\bar\nu_l\gamma_{\nu}(1-\gamma^5)(i\partial)^2\nu_l
\,,
\end{eqnarray}
where $D^{\mu}$ is the covariant derivative involving the gluon field.
The coefficients of these operators are determined by matching of the
diagrams in Fig \ref{lightloops}, where one finds that the neutral coupling in the
left diagram generates $O^l_1$ while the charged coupling in the right
diagram are responsible for $O^l_2$. These coefficients are given in
\cite{Falk:2000nm, Isidori:2005xm}.
\begin{figure}[htb]
\centering
\epsfig{file=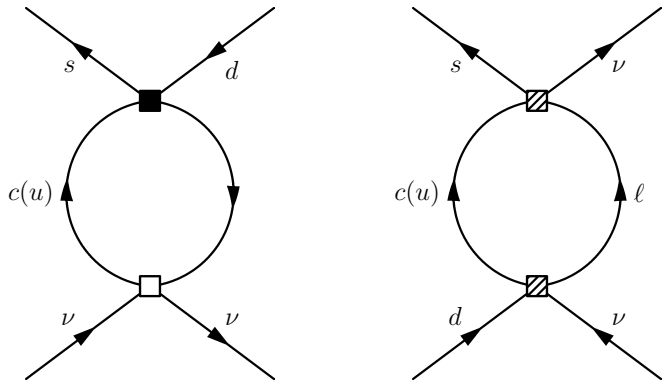,height=5cm}
\caption{One loop diagrams with light quarks that generate higher
  dimensional operators. (from \cite{Isidori:2005xm})
 \label{lightloops}}
\end{figure}
While the matrix element of $O^l_1$ can be rather reliably estimated
and gives a negligible contribution compared to the leading dimension
six terms, the matrix element of $O^l_2$ is harder to estimate, due to the
gluon appearing in the covariant derivative. A numerical estimate is
performed using the Lorentz structure and parametrizing the remaining
ignorance of hadronic effects by a bag factor, which is determined by
matching onto the genuine long distance contributions and demanding
that the renormalization scale dependence should cancel. Further 
progress can be achieved through lattice calculations \cite{Isidori:2005tv}

While the discussion so far is rather straightforward, the genuine
long distance effects from $u$ quark loops have received much more attention
\cite{Rein:1989tr,Hagelin:1989wt,Lu:1994ww,Fajfer:1996tc,Geng:1996kd,Isidori:2005xm}.
Again, we follow \cite{Isidori:2005xm}, where the most recent and
complete discussion is given. In particular, it is shown that previous
calculations missed several terms that are necessary to obtain the
correct matching between short and long distance components in the amplitude.

In order to address these effects, on begins with the chiral effective
$\Delta S = 1$ Hamiltionan (see for example \cite{D'Ambrosio:1996nm}
for a review). From the chiral transformation properties, one finds
that this Hamiltonian consists of pieces that transform as ($8_L,1_R$)
and ($27_L,1_R$) under the chiral symmetry group $SU(3)_L\times
SU(3)_R$. Experimentally, one finds that the octet piece is enhanced
(this corresponds to the usual $\Delta I=1/2$ rule) so that the
($27_L,1_R$) can be neglected. To lowest order in the chiral
expansion and using only the octet contribution there is then one
operator that contributes:
\begin{equation}
{\cal L}^{(2)}_{ \left|  \Delta S \right| =1 }  = 
G_{8} F^{4} \left\langle \lambda_{6} D^{\mu}U^{\dagger} D_{\mu}U\right\rangle,
\label{eq:LW2}
\end{equation}
where $G_8 \approx 9\times 10^{-6} \gev^{-2}$, $U$ is the
conventional representation of the pseudoscalar meson fields and
$\langle \rangle$ implies a trace.
Using the Hamiltonian thus obtained, one finds that the leading order
diagrams in CHPT (Fig {\ref{CHpt}}) cancel
\cite{Ecker:1987hd,Lu:1994ww}. However, to be consistent, there are additional operators to be included since the
$SU(2)_L$ generators are broken and, for an effective chiral Lagrangian,
also non gauge invariant operators with the correct representation
must be added (this is not necessary for the $K \to \pi \gamma$ vertex
{\cite{Ecker:1987hd}}).
Including these operators also leads to the same parametric
renormalization scale dependence as in the short distance part of the amplitude.
Then, the complete chiral Lagrangian is given by {\cite{Isidori:2005xm}}
\begin{equation}
{\cal L}^{(2)}_{ \left|  \Delta S \right| =1} = 
G_{8} F^{4} \langle \lambda_{6}  \left[  D^{\mu}U^{\dagger} D_{\mu}U 
- 2 i g_Z Z_{\mu} U^{\dagger} D^{\mu} U \left( Q - \frac{a_1}{6}\right) \right] \rangle,
\end{equation}
where $a_1$ is related to the coupling of the $Z$ to the $U(1)_L$
charge \cite{Lu:1994ww}.
One finds then that the $O(p^2)$ terms do not cancel for the charged
($K^+ \rightarrow \pi^+ Z$) amplitude:
\begin{equation}
{\cal A}( K^+ \to \pi^+ \nu \bar\nu )_Z = \frac{G_F}{\sqrt{2}} G_8 F^2 
\left[ 4 p^\mu \right] 
\sum_l \bar{\nu}_l \gamma_\mu (1-\gamma_5) \nu_l
\label{eq:dec_M}
\end{equation}
In \cite{Isidori:2005xm} this calculation is extended to $O(p^4)$ which
involves several one loop diagrams. The final contributions come from
$W-W$ exchange diagrams \cite{Hagelin:1989wt,Isidori:2005xm} which
correspond, on the short distance side, to the contributions from
$O^l_2$ and accordingly should cancel the respective renormalization
scale dependence.
We give here just the tree level result {\cite{Isidori:2005xm}}:
\begin{equation}
{\cal A}(K^+ \to \pi^+\nu\bar\nu)_{WW} =   G^2_F  F^2 \lambda
\sum_{l = e,\mu} 2~p^\mu \bar{\nu}_l \gamma_\mu (1-\gamma_5) \nu_l
\end{equation}
Summing up all contributions, one can include all subleading effects
discussed in this section by shifting the value of $P_c(X)$:
\begin{equation}
 P^{(6)}_c \to  P^{(6)}_c +\delta P_{c,u} \qquad \qquad \delta P_{c,u} = 0.04 \pm 0.02,
\label{eq:PC2}
\end{equation}
which implies a shift of roughly $6\%$ in the branching ratio.

\begin{figure}[htb]
\centering
\epsfig{file=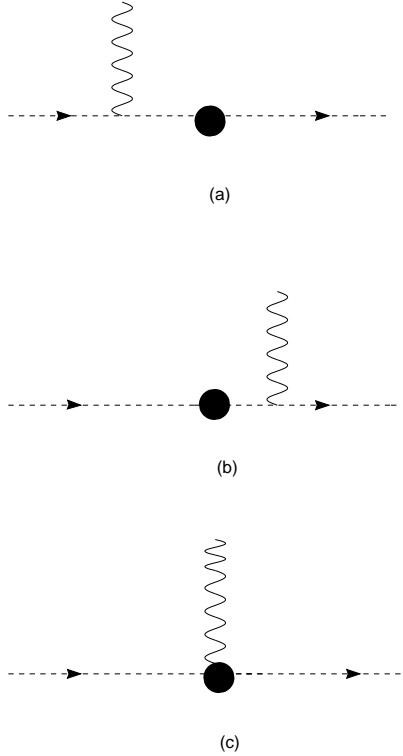,height=10cm}
\caption{Leading order chiral perturbation theory diagrams
  contributing to a $K^+\to \pi^+ Z^0$ vertex (from \cite{Lu:1994ww}). The
  dashed lines denote the pion and kaon, while the wavy line denotes
  the $Z^0$, and the dot indicates the insertion of a flavor changing
  effective vertex.
 \label{CHpt}}
\end{figure}

Let us now turn to $\klpn$. Long distance contributions here are
mostly equivalent to CP conserving effects and have been
comprehensively studied in \cite{Buchalla:1998ux}. As for $\kpn$, there are
effects from soft up quarks, that are treated in chiral perturbation
theory, and higher dimensional operators in the charm sector, which 
are actually short distance effects. It is found that they are
suppressed by several effects, reinforcing the theoretically clean
character of this decay. Let us briefly describe these effects.

The contributions from soft up quarks in the penguin loops have been
studied in {\cite{Buchalla:1998ux}} and in {\cite{Geng:1996kd}}. As
is the case for $\kpn$, the leading diagrams appear at one loop
order. They
are calculated explicitely by Buchalla and Isidori
{\cite{Buchalla:1998ux}}, who find, taking into account also phase space
suppression, that the CP conserving long distance contributions are
suppressed by approximately a factor of $10^{-5}$ compared to the
dominant top contribution. 

The next contribution that can be important are then higher dimensional
operators in the OPE. As \cite{Buchalla:1998ux} studies only CP conserving
contributions only one operator
that is antisymmetric in neutrino momenta, survives from the expansion
of the box diagrams (contributions from $Z^0$-penguins also drop out for
the same reason):
 \begin{equation}\label{hcpc}
H_{CPC}=-\frac{G_F}{\sqrt{2}}\frac{\alpha}{2\pi\sin^2\Theta_W}
\lambda_c\, \ln\frac{m_c}{\mu}\, \frac{1}{M^2_W}T_{\alpha\mu}
\bar\nu({\overleftarrow{\partial^\alpha}} -\partial^\alpha)\gamma^\mu
(1-\gamma_5)\nu~,
\end{equation}
\begin{equation}\label{tam}
T_{\alpha\mu}=\bar s{\overleftarrow D}_\alpha\gamma_\mu(1-\gamma_5)d-
\bar d\gamma_\mu(1-\gamma_5)D_\alpha s~.
\end{equation}
There arise now several suppression factors: First, there is
the naive suppression of the operator scaling, which is estimated to
be $ \mathcal{O}(\lambda_c M_K^2 /{\rm Im}\lambda_t M_W^2) \approx 10\%$
compared to the leading top contribution. 
Here, the smallness of $M_K/M_{W}$ is compensated by
the ratio of CKM
factors $\lambda_c/{\rm Im}\lambda_t$. 

The suppression is more severe when the matrix elements are 
calculated, since the leading order $K_L-\pi^0$ matrix element in chiral perturbation theory is found to be:
\begin{equation}\label{mefin}
\langle\pi^0(p)|T_{\alpha\mu}|K_L(k)\rangle=
-\frac{i}{2}[(k-p)_\alpha (k+p)_\mu+\frac{1}{4}m^2_K g_{\alpha\mu}]~,
\end{equation}
which vanishes when multiplied with the leptonic current in the
operator due to the equations of motion and the negligible neutrino
masses.
 The chiral suppression of the NLO ($p^4$) terms leads to an additional
 reduction of higher dimensional operator contributions by about 
  $m_K^2/(8\pi^2f_{\pi}^2)\approx 20\%$.
Finally, one has to take into account also phase space effects, which
further suppress these terms.

Estimating the $\mathcal{O}(p^4)$ matrix elements and performing the phase space
calculations, the authors of {\cite{Buchalla:1998ux}} find that short distance CP
conserving effects are suppressed by a factor of $10^{-5}$ compared to the 
dominant top
contribution and conclude that they are ''safely negligible, by a comfortably
large margin''.

It is then fair to say, from the present perspective, that long
distance effects are rather well under control especially in $\klpn$,
but also in $\kpn$, where the contributions and its uncertainty can
now be rather reliably quantified and included in numerical
analyses. This is gratifying, since the NNLO
calculation is available and are of the same order of magnitude.

\section{Conclusions and Outlook}\label{sec:concl}

In the present review we have summarized the present status of the rare decays 
$\kpn$ and $\klpn$, paying in particular attention to theoretical and 
parametric 
uncertainties. Our analysis reinforced the importance of these decays in 
testing the SM and its extensions. We have pointed out that the clean 
theoretical character of these decays remains valid in essentially all 
extensions of the SM, whereas this is often not the case for non-leptonic 
two-body $B$ decays used to determine the CKM parameters through CP 
asymmetries and/or other strategies. 
Here, in extensions of the SM  in which new operators and new weak 
phases are present, the mixing induced asymmetry $a_{\phi K_S}$ 
and other similar asymmetries can suffer 
from potential 
hadronic uncertainties that make the determination of the relevant 
parameters  problematic unless the hadronic matrix element can be 
calculated with sufficient precision. In spite of advances in 
non-perturbative 
calculations of non-leptonic amplitudes for $B$ decays
\cite{Beneke:2003zv,Beneke:1999br,Keum:2000ph,Keum:2000wi,Bauer:2001yt,Bauer:2002uv,Bauer:2002nz,Stewart:2003gt,Beneke:2002ph,Beneke:2002ni},
 we are still far away 
from precise calculations of non-leptonic amplitudes from first principles.
On the other hand the branching ratios for $\kpn$ and $\klpn$ 
can be parametrized in essentially 
all extensions of the SM by a single complex function 
$X$ (real in the case of MFV models) that can be calculated in perturbation
theory in any given extension of the SM.

There exists, however, a handful of strategies in the $B$ system that
similarly to $K\to\pi\nu\bar\nu$, are very clean. Moreover, in contrast to
$K\to\pi\nu\bar\nu$, there exist strategies involving $B$ decays that allow not only a
theoretically clean determination of the UT but also one free from new physics
pollution.

Our main findings are as follows:
\begin{itemize}
\item
Our present predictions for the branching ratios read
{ \begin{equation}\label{SMkpf}
Br(\kpn)_{\rm SM}=
(8.1 \pm 1.1)\cdot 10^{-11},
\ee
\be\label{SMklf}
 Br(\klpn)_{\rm SM}=
(2.6 \pm 0.3)\cdot 10^{-11} .
\end{equation}}
This is an accuracy of $\pm14\%$ and $\pm 12\%$, 
respectively. 
\item   
Our analysis of theoretical uncertainties in $K\to\pi\nu\bar\nu$, 
that come almost 
exclusively from the charm contribution to $K^+\to \pi^+\nu\bar\nu$, 
reinforced the importance 
of the recent NNLO calculation of this contribution \cite{BUGOHA,Buras:2006gb}. 
Indeed the $\pm 18\%$ 
uncertainty in $P_c(X)$ coming dominantly from the scale uncertainties 
and the value of $m_c(m_c)$, translates into an uncertainty of 
$\pm 7.0\%$ in 
the determination of $\vtd$, $\pm0.04$ in the determination of 
$\sin2\beta$ and $\pm 10\%$ in the prediction for $Br(\kpn)$. 
The NNLO analysis reduced the uncertainty in $P_c$ to $12 \%$ and
further progress on the determination of 
$m_c(m_c)$ could reduce the error in $P_c(X)$ down to $\pm 5\%$, 
implying the 
reduced error in $\vtd$ of $\pm 2\%$, in $\sin2\beta$  of 
$\pm 0.011$ and $\pm 3\%$ in $Br(\kpn)$.
\item
Further progress on the determination of the CKM parameters, that in the next
few years
will dominantly come from BaBar, Belle and 
Tevatron and later from LHC and BTeV, should allow eventually 
the predictions for $Br(\kpn)$ and $Br(\klpn)$ with the uncertainties of 
roughly $\pm 5\%$ or better. It should be emphasized that this
accuracy cannot be matched by any other rare decay branching ratio 
in the field of meson decays.
\item
We have analyzed the impact of precise measurements of $Br(\kpn)$ and 
$Br(\klpn)$ on the unitarity triangle and other observables of interest, 
within the SM. 
In particular we have analyzed the accuracy with which $\sin 2\beta$ and 
the angle $\gamma$ could be extracted from these decays. Provided 
both branching ratios can be measured with the accuracy of $\pm 5\%$, 
an error on $\sin 2\beta$ of  $\pm 0.038$ could be achieved.  The determination 
of $\gamma$ requires an accurate measurement of $Br(\kpn)$ and the reduction 
of the errors in $P_c(X)$ and $\vcb$. 
With a measurement better than $\pm 5\% $ of $Br(\kpn)$ and the reduction of the 
errors in $P_c(X)$ and $\vcb$ anticipated,
$\gamma$ could 
be measured with an error of  $\pm 5^\circ$.
\item
We have emphasized that the simultaneous investigation of the
$K\to\pi\nu\bar\nu$  
complex, the  mass differences $\Delta M_{d,s}$ and the angles $\beta$ and 
$\gamma$ from 
clean strategies in two body $B$ decays, should allow to disentangle 
different new physics contributions to various observables and 
determine new parameters of the extensions of the SM. The $(R_t,\beta)$,
 $(R_b,\gamma)$, $(\beta,\gamma)$ and $(\bar\eta,\gamma)$
strategies for UT when combined with $K\to\pi\nu\bar\nu$ decays are very 
useful in 
this goal. This is in particular the case for the $(R_b,\gamma)$ strategy 
that is 
related to the reference unitarity triangle \cite{Goto:1995hj,Cohen:1996sq,Barenboim:1999in,Grossman:1997dd}. A graphical 
representation of these investigations is given in Fig.~\ref{bsuplot}.
\item
We have presented a new "golden relation" between $\beta$, $\gamma$ and 
$Br(\klpn)$, 
given in (\ref{newrel}), that with improved values of $m_t$ and 
$Br(\klpn)$ should allow 
very clean test of the SM one day. 
Another new relation is the one between $\beta$, $\gamma$ and $Br(\kpn)$, 
that is given in (\ref{AJBNEW}). Although not as clean as the golden 
relation in 
(\ref{newrel}) because of the presence of $P_c$, it should play a useful 
role in future investigations.
\item
We have presented the results for both decays in
models with minimal flavour violation and in several 
scenarios with new complex phases in $Z^0$ penguins and/or 
$B^0_d-\bar B^0_d$ mixing. 
We have reviewed the results for $Br(\kpn)$ and $Br(\klpn)$ in a number of 
specific 
extensions of the SM. In particular we have discussed LHT, $Z^\prime$ and 
supersymmetry with MFV, more general supersymmetric models 
with new complex phases, models with universal extra dimensions and 
models with lepton-flavour mixing.
Each of these models has some characteristic predictions for the 
branching ratios in question, so that it should be possible to 
distinguish between various alternatives. Simultaneous investigations of 
other observables should be very helpful in this respect. 
In some of these scenarios the departures from the SM expectations are still
allowed to be spectacular.
\item
Finally we have compared the usefulness of $K\to\pi\nu\bar\nu$ decays 
in testing various models with the one of other decays. While in the 
$K$ system $K\to\pi\nu\bar\nu$ decays have no competition, there is 
a handful of $B$ decays and related strategies that are also theoretically 
very clean. It is precisely the comparison between the results of these 
clean strategies in the $B$ system with the ones obtained one day from 
$K\to\pi\nu\bar\nu$ decays that will be most interesting. 
\item 
In spite of an impressive agreement of the SM with the available data, 
large departures from the SM expectations in $B_s$ decays are still
possible. However, even if future Tevatron and LHC data
would not see any significant new physics effect in these decays, this 
will not imply necessarily that new physics is not visible in $\klpn$, $\kpn$ 
and $K_L\to\pi^0\ell^+\ell^-$. On the contrary, as seen in particular in the
case of the LHT model \cite{Blanke:2006eb}, there are scenarios in which the effects in $B$-physics
are tiny, while 
large departures in these three decays will still be possible. It  may 
then be that in the end, it will be $K$ physics and not $B$ physics 
that will offer the best information about the new phenomena at 
very short distance scales, in accordance with the arguments in \cite{Bryman:2005xp,Grinstein:2006cg}
\end{itemize}

We hope we have convinced the reader that the very clean rare decays 
$\kpn$ and $\klpn$ deserve a prominent status in the field of flavour and 
CP violation and 
that precise measurements of their branching ratios are of utmost importance.
Let us hope that our waiting for these measurements will not be too long.

\noindent
\section{Acknowledgments}
\noindent
 We would like to thank Steve Kettell, Laur Littenberg and J\"urgen 
Engelfried for information
about the future prospects for $K\to\pi\nu\bar\nu$ experiments and Steve 
Kettell and Gino Isidori for comments on the manuscript.  Further we would
like to thank Frederico Mescia and Christopher Smith for providing information
on details of $\kappa_+$ and $\kappa_L$.
F.S. would like to thank the IFAE, Barcelona, where final revisions and updates have been performed.
The work presented here was supported in part by the German 
Bundesministerium f\"ur
Bildung und Forschung under the contract 05HT4WOA/3, 05HT6WOA  and the 
DFG Project Bu.\ 706/1-2.

\bibliographystyle{apsrmp}
\bibliography{rmpbib}
\newpage
\end{document}